\documentclass[fleqn,usenatbib]{mnras}

\usepackage{txfonts}
\usepackage{footnote}
\usepackage[T1]{fontenc}
\usepackage{ae,aecompl}
\usepackage[flushleft]{threeparttable}
\usepackage{longtable,threeparttablex,booktabs,threeparttable}
\usepackage{longtable}
\usepackage[flushleft]{threeparttable}
\usepackage{longtable,threeparttablex,booktabs,threeparttable}
\usepackage{graphicx}
\usepackage{wrapfig}
\usepackage{pdflscape}
\usepackage{rotating}
\usepackage{epstopdf}
\usepackage{supertabular}
\hypersetup{draft}

\usepackage{graphicx}	
\usepackage{amssymb}	
\usepackage{multirow,multicol}
\newcommand{\kms}{km\,s$^{-1}$} 
\newcommand{\hi}{H{\sc i}}
\newcommand{\hii}{H{\sc i}~21-cm}
\title[H{\sc i} 21\,cm absorption in compact AGNs]{A Giant Metrewave Radio Telescope survey for associated H{\sc i} 21\,cm absorption in the Caltech-Jodrell Flat-spectrum sample.}

\author[Aditya \& Kanekar]{J. N. H. S. Aditya$^{1}$$^{,2}$\thanks{adityaj@iucaa.in},
Nissim~Kanekar$^2$\thanks{Swarnajayanti Fellow; nkanekar@ncra.tifr.res.in}\\
$^{1}$Inter-University Centre for Astronomy and Astrophysics, Pune 411007, India\\
\newline
$^{2}$National Centre for Radio Astrophysics, Tata Institute of Fundamental Research, Pune 411007, India}

\date{Accepted XXX. Received YYY; in original form ZZZ}

\pubyear{2018}

\begin{document}
\label{firstpage}
\pagerange{\pageref{firstpage}--\pageref{lastpage}}
\maketitle

\begin{abstract}
We report a Giant Metrewave Radio Telescope (GMRT) survey for associated H{\sc i} 21-cm absorption 
from 50 active galactic nuclei (AGNs), at $z \approx 0.04 - 3.01$, selected from the Caltech-Jodrell 
Bank Flat-spectrum (CJF) sample. Clean spectra were obtained towards 40 sources, yielding two new 
absorption detections, at $z = 0.229$ towards TXS~0003+380 and $z = 0.333$ towards TXS~1456+375, besides 
confirming an earlier detection, at $z = 1.277$ towards TXS~1543+480. There are 92 CJF sources, at 
$0.01 \lesssim z \lesssim 3.6$, with searches for associated H{\sc i} 21-cm absorption, by far the 
largest uniformly-selected AGN sample with searches for such absorption. We find weak ($\approx 2\sigma$) 
evidence for a lower detection rate of H{\sc i} 21-cm absorption at high redshifts, with detection 
rates of $28^{+10}_{-8}$\% 
and $7^{+6}_{-4}$\% in the low-$z$ ($z < z_{\rm med}$) and high-$z$ ($z > z_{\rm med}$) sub-samples,
respectively. We use two-sample tests to find that the strength of the H{\sc i} 21-cm 
absorption in the AGNs of our sample depends on both redshift and AGN luminosity, with a lower 
detection rate and weaker absorption at high redshifts and high ultraviolet/radio AGN luminosities. 
Unfortunately, the luminosity bias in our sample, with high-luminosity AGNs arising at high redshifts,
implies that it is not currently possible to identify whether redshift evolution or AGN luminosity
is the primary cause of the weaker absorption in high-$z$, high-luminosity AGNs. We find 
that the strength of H{\sc i} 21-cm absorption does not depend on AGN colour, 
suggesting that dust extinction is not the main cause of reddening in the CJF sample.

\end{abstract}

\begin{keywords}
galaxies: active - quasars: absorption lines - galaxies: high redshift - radio
lines: galaxies
\end{keywords}


\section{Introduction}


Neutral hydrogen (\hi) is an important constituent of the gas in the environments of active galactic 
nuclei (AGNs). For radio-loud AGNs, the \hii\ transition allows the presence of such gas to be 
discerned, and its kinematics studied, via \hii\ absorption studies against the AGN radio 
continuum. Such ``associated'' \hii\ absorption studies allow a detailed probe of physical 
conditions in AGN environments \citep[see, e.g., ][for a recent review]{morganti18}, and their 
evolution with redshift. For example, one can test whether the strength of the \hii\ absorption 
depends on the nature of the AGN (e.g. core-dominated or lobe-dominated, high-luminosity or 
low-luminosity, flat spectrum or steep spectrum, etc), providing information on the distribution 
and the excitation of neutral gas in different AGN environments. One might also test whether the
strength of \hii\ absorption depends on redshift: weaker \hii\ absorption in typical AGNs over 
some range of redshifts might indicate a paucity of gas in the AGN environments at this epoch, 
shedding light on the gas accretion process and the fuelling of AGNs. The \hii\ absorption velocity, 
relative to the AGN redshift, also contains information on local conditions, with redshifted 
(relative to the AGN) absorption indicating the presence of inflowing gas, and blueshifted 
absorption, a signature of gas outflows. If the associated \hii\ absorption systematically 
arises at higher velocities than the AGN redshift, it would suggest that neutral gas is 
predominantly flowing towards typical AGNs at this redshift, which could fuel the 
nuclear activity \citep[e.g.][]{vangorkom89}. Conversely, mostly blueshifted \hii\ absorption 
would imply a predominance of gas outflows, that might result in shutting down of the AGN activity 
\citep[e.g.][]{vermeulen03,gupta06}. Similarly, narrow \hii\ lines with velocity spreads 
$\lesssim 100$ km~s$^{-1}$ and relatively small velocity offsets from the AGN systemic redshift 
are likely to indicate gas clouds with low velocity dispersion, perhaps rotating in a circumnuclear disk 
\citep[e.g.][]{dwaraka95,conway99,gereb15}. Broad absorption features, with widths $\approx 200-300$~\kms\ 
suggest the presence of unsettled gas, possibly interacting with the active nucleus. Finally, the broadest lines, 
with widths $\gtrsim 500$~\kms, are likely to arise due to interactions between the neutral gas and the radio 
jets in powerful radio sources \citep[e.g.][]{morganti03,morganti05,mahony13}.

A further important advantage of \hii\ absorption studies in probing AGN environments stems from the 
fact that the radio emission is often extended, stemming from both the AGN core and structures such as 
jets, lobes, hotspots, etc. This allows the exciting possibility of using very long baseline interferometry 
(VLBI) techniques to map the \hii\ absorption against the extended radio continuum, on scales of $\approx 10-1000$~pc, 
to determine the spatial structure of the absorbing gas, and connections with the radio continuum
\citep[e.g.][]{mundell95,carilli98c,peck98,peck99,conway99,beswick02,morganti04,labiano06,struve12,morganti13}.  
Such VLBI absorption mapping studies are critical to identify the location of the absorbing gas 
in the AGN environment.

Since the first detection of associated \hii\ absorption, by \citet{roberts70} in NGC5128, a large number of 
searches for associated \hii\ absorption have been carried out, using a variety of radio telescopes. More than 
400 AGNs have so far been searched for associated \hii\ absorption, with more than seventy-five detections 
\citep[e.g.][]{dickey86,vangorkom89,carilli98,moore99,gallimore99,morganti01,morganti05,pihlstrom03,vermeulen03,gupta06,chandola11,allison12,gereb15,maccagni17,aditya18}.  
Molecular absorption has also been detected in a handful of these systems, indicating the presence of dense gas 
\citep[e.g.][]{gardner76,wiklind94,wiklind96,kanekar02,kanekar08c}. However, the vast majority of the detections 
of \hii\ absorption are at low redshifts, $z < 0.25$ \citep[e.g.][]{vangorkom89,gereb15,maccagni17}: the current 
sample of associated \hii\ absorbers is dominated by the 59 detections of \citet{gereb15} and \citet{maccagni17}, 
at $0.05 < z < 0.25$. Further, most searches for redshifted \hii\ absorption have been in highly heterogeneous 
samples, making it hard to differentiate between the causes for the presence or absence of absorption 
(e.g. redshift, AGN type, AGN luminosity, etc.). We have hence been using the Giant Metrewave Radio 
Telescope (GMRT) survey for redshifted associated \hii\ absorption towards a uniformly-selected AGN 
sample, distributed over a wide redshift range. Earlier results from this survey were reported in 
\citet{aditya16} and \citet{aditya17}. 


\section{The AGN targets: The Caltech-Jodrell Flat-Spectrum sample}
\label{sec:cjfsample}

Our primary goal was to carry out a large survey for \hii\ absorption from AGNs at high redshifts, $z \gtrsim 1$, 
to study redshift evolution in AGN environments. Of course, an important goal was to significantly increase the 
number of known associated \hii\ absorbers at high redshifts. We also aimed to target a uniformly-selected AGN 
sample, to minimize the heterogeneity in our AGN environments. Earlier studies at low redshifts had found evidence 
for a higher detection rate of \hii\ absorption ($\approx 40$\%) in more compact radio sources, the 
GHz-Peaked Spectrum (GPS) sources and the Compact Steep Spectrum (CSS) sources, 
than in more extended radio galaxies ($\approx 15$\%; e.g. \citealp{vermeulen03,gupta06}). Similarly, \citet{pihlstrom03} 
found that GPS sources have typically higher \hii\ optical depths than even the somewhat larger CSS sources. Foreground 
\hi\ clouds of a given size would obscure a larger fraction of the radio emission from a compact source than from 
an extended source: lower observed integrated \hii\ optical depths towards extended radio sources can thus be naturally 
explained by such covering factor effects. At a fixed observed \hii\ optical depth sensitivity, the likelihood 
of detecting \hii\ absorption is hence higher towards compact radio sources. To maximize our chances of detections 
of \hii\ absorption, which would allow us to trace the distribution and kinematical properties of neutral gas 
in high-$z$ AGN environments, we hence chose source compactness, i.e. an inverted or a flat low-frequency radio 
spectrum \citep[e.g.][]{kellermann81,shu91,odea98}, as the main criterion in sample selection.

We aimed to ensure homogeneity in our target AGNs by selecting them from a near-complete, flux-density 
limited sample, the Caltech-Jodrell Bank Flat-spectrum (CJF) sample \citep[][]{pearson88,henstock95,taylor96}. 
The CJF sample includes all radio sources with (1)~4.85~GHz flux density $\geq 350$~mJy, (2)~flat radio spectra 
between 1.4 and 4.85~GHz, with $\alpha_{\rm 1.4 \; GHz}^{\rm 4.85 \; GHz} \geq -0.5$, and (3)~declination 
$\delta > 35^\circ$ \citep[][]{taylor96}, with a total of 293 sources, at redshifts $0 \lesssim z \lesssim 4.0$.



At the beginning of our survey, 29 sources of the CJF sample had been earlier searched for associated \hii\ 
absorption, mostly at $z < 1$, with a detection rate of $\approx 40$\% \citep[e.g.][]{vermeulen03,gupta06}. We 
aimed to search for associated \hii\ absorption from the remaining 82 CJF sources for which the redshifted \hii\ 
line frequency lies in the GMRT 327~MHz, 610~MHz and 1420~MHz observing bands (which cover the frequency ranges 
$300-360$~MHz, $570-660$~MHz, and $1000-1450$~MHz, respectively). Finally, we were unable to observe 8 of the above 
82 CJF sources, five at $1.1 < z < 1.5$ (CJ2~1534+501, 7C~1550+5815, CJ2~1308+471, S5~0454+84, and TXS~1851+488) and 
three at $3.0 < z < 3.6$ (S4~0636+68, BZQ~J1526+6650, and B3~1839+389), due to their low flux densities that 
would have required prohibitively large integration times to achieve a good \hii\ optical depth sensitivity. 


Our initial GMRT observations of 24~AGNs of the CJF sample were presented, and the results discussed, by 
\citet{aditya16}, and an additional detection of \hii\ absorption at $z \approx 1.223$ towards TXS~1954+513 
was presented and discussed by \citet{aditya17}. In the present paper, Section~\ref{sec:obs} describes the 
new GMRT observations, data analysis, and results for 50 AGNs of our target sample, Section~\ref{sec:detect} 
discusses the individual detections of \hii\ absorption, while Section~\ref{sec:discuss} discusses the 
dependence of the strength of the \hii\ absorption on various AGN properties.


\section{Observations, Data Analysis, and Results}
\label{sec:obs}

\subsection{The GMRT observations and data analysis}
\label{sec:gmrt_obs}

The GMRT 327~MHz, 610~MHz, and 1420~MHz receivers were used to carry out a search for 
associated \hii\ absorption from the 49 AGNs of the CJF sample without earlier searches for 
such absorption. We also observed the AGN TXS~1543+480, at $z = 1.277$, to confirm the 
detection of \hii\ absorption by \citet[][]{curran13}. The observations were carried out 
over 2014 April to 2015 August, in GMRT Cycles 26 (proposal 26\_052, PI: Kanekar) and 28 
(proposal ID: 28\_089, PI: Aditya), using the GMRT Software Backend (GSB) as the correlator.
21 sources were observed with the 1420~MHz receivers, 28 with the 610~MHz receivers, and
one with the 327~MHz receivers. Bandwidths of 16.7~MHz or 33.3~MHz were used for all 
observations, sub-divided into 512 channels, and centred at the redshifted \hii\ line 
frequency, with two circular polarizations. This yielded velocity resolutions of, respectively, 
$\approx 8-34$~\kms, and velocity coverages of $\approx 3600 - 17500$~\kms, in the different 
observing bands. The velocity coverage was sufficient to detect wide (velocity spread $\approx 1000$~\kms) 
\hii\ absorption, while the velocity resolution allowed excellent sensitivity to even relatively 
narrow ($\approx 20$~km s$^{-1}$) spectral components. 

A standard flux density calibrator (3C48, 3C147 or 3C286) was observed at the start or end 
of each observing run. Observations of the target source were split into $\approx 30-40$~minute 
scans, bracketed by short ($\approx 6$ minutes) scans on a nearby phase calibrator. The flux and/or 
phase calibrators were also used to calibrate the system passband; no additional observations of 
bandpass calibrators were carried out. The on-source observing times were $\approx 1-1.5$~hours for 
sources observed with the 610~MHz and 1420~MHz receivers, and $\approx 3$~hours for the lone source
observed with the 327~MHz receivers. 

Our initial GMRT observations yielded tentative detections of \hii\ absorption in two sources, 
TXS~0003+380 and TXS~1456+375, besides confirming the detection of \hii\ absorption in TXS~1543+480 
\citep[][]{curran13}. The first two sources were re-observed with the GMRT to confirm the 
reality of the absorption. These observations again used the GSB as the backend, but with a bandwidth 
of 4.17~MHz, sub-divided into 512 channels, and centred at the frequency of the putative \hii\ absorption 
feature. This provided a higher velocity resolution of $\approx 2.2$~\kms, albeit with a narrower 
velocity coverage $\approx 1175$~\kms. The observational details are 
summarized in Table~\ref{table:obs}.
 
All data were analysed using the Astronomical Image Processing System \citep[AIPS; ][]{greisen03} package, 
following standard procedures for data editing, gain and bandpass calibration, self-calibration, and imaging
\citep[see, e.g.,][]{aditya16}. These procedures yielded a continuum image of the field at the observing 
frequency and the final spectral line cube (after subtracting out all continuum emission). The \hii\ spectrum 
was extracted from the spectral cube by taking a cut along the frequency axis at the location of the target 
AGN. The final spectrum was then obtained by subtracting out a second-order polynomial baseline, fit to 
line- and RFI-free regions, to compensate for any residual bandpass effects.

\setcounter{table}{0}
\begin{table*}
\footnotesize
\caption{The 50 target sources, in order of increasing redshift.
\label{table:obs}}
\begin{center}
\begin{tabular}{|lcccccccc|}

\hline
Source &  $ z $ & $\nu_{\rm 21-cm}$ & $S_{\nu}$$^{b}$ &  $\Delta v$ & Beam & $\Delta S$ & $\int \tau dv$$^{c}$ & $\rm N_{HI}$$^d$ \\
       &        &    MHz        &    mJy    & km $s^{-1}$ & $'' \times ''$ &  mJy      &    km $s^{-1}$ &  $\times 10^{20}~{\rm cm}^{-2}$    \\
\hline
     
      TXS 0344+405$^{C}$ & 0.039 & 1367.08 & $24.3 $    & 28.6     & $7.0 \times 6.4   $ & 1.7  & $< 14$            & $< 26$ \\ 
 TXS 0344+405$^{\rm SE}$ & 0.039 & 1367.08 & $82.4 $    & 28.6     & $7.0 \times 6.4   $ & 1.8  & $< 4.9$           & $< 8.9$ \\ 
 TXS 0344+405$^{\rm NW}$ & 0.039 & 1367.08 & $89.1 $    & 28.6     & $7.0 \times 6.4   $ & 2.0  & $< 5.5$           & $< 10$ \\ 
            TXS 0733+597 & 0.041 & 1364.46 & $502.8$    & 28.6     & $3.9 \times 2.7   $ & 1.5  & $< 0.67$          & $< 1.2$ \\ 
              S5 2116+81 & 0.084 & 1310.33 & $137.2$    & 14.9     & $7.4 \times 3.6   $ & 1.2  & $< 1.4$           & $< 2.5$ \\ 
            TXS 1418+546 & 0.153 & 1231.92 & $595.6$    & 31.7     & $4.1 \times 2.9   $ & 2.4  & $< 1.5$           & $< 2.7$ \\ 
              S4 0749+54 & 0.200 & 1183.67 & $533.3$    & 32.9     & $3.6 \times 2.9   $ & 2.7  & $< 1.8$           & $< 2.1$ \\ 
            TXS 0003+380 & 0.229 & 1155.74 & $547.3$    & 2.1$^a$  & $4.5 \times 3.5   $ & 2.6  & $1.943 \pm 0.057$ & $3.54 \pm 0.11$ \\ 
              S5 1356+47 & 0.230 & 1154.80 & $544.8^e$  & 16.9     &  RFI                & --   & --                & -- \\
            TXS 0010+405 & 0.255 & 1131.79 & $358.9$    & 34.5     & $3.1 \times 2.5   $ & 2.3  & $< 1.4$           & $< 2.5$ \\ 
            TXS 1719+357 & 0.263 & 1124.63 & $262.4$    & 34.7     & $4.5 \times 3.4   $ & 1.4  & $< 1.2$           & $< 2.2$ \\ 
            B3 0251+393  & 0.289 & 1101.94 & $229.8^e$  & 8.9      & RFI                 & --   & --                & -- \\
            TXS 0716+714 & 0.300 & 1092.62 & $834.8^e$  & 17.9     & RFI                 & --   & --                & -- \\
            TXS 1700+685 & 0.301 & 1091.78 & $372.9^e$  & 8.9      & RFI                 & --   & --                & -- \\
            S5 1928+73   & 0.302 & 1090.94 & $4058.9^e$ & 17.9     & RFI                 & --   & --                & -- \\
         JVAS J1010+8250 & 0.322 & 1074.40 & $532.1^e$  & 18.2     & RFI                 & --   & --                & -- \\
            TXS 0424+670 & 0.324 & 1072.81 & $719.1$    & 36.4     & $8.6 \times 4.1   $ & 2.9  & $< 0.89$          & $< 1.6$ \\ 
             B3 1456+375 & 0.333 & 1065.57 & $148.4$    & 2.3$^a$  & $9.1 \times 3.8   $ & 1.6  & $3.834 \pm 0.079$ & $6.98 \pm 0.15$ \\ 
              S5 2007+77 & 0.342 & 1058.42 & $660.0$    & 36.9     & $4.8 \times 4.0   $ & 24.3 & $< 9.9$           & $< 18$ \\ 
            TXS 0035+367 & 0.366 & 1039.82 & $581.2$    & 37.5     & $5.6 \times 4.9   $ & 1.2  & $< 0.51$          & $< 0.93$ \\ 
            TXS 0954+658 & 0.368 & 1038.30 & $1104.8$   & 37.6     & $5.6 \times 4.2   $ & 20.1 & $< 5.8$           & $< 11$ \\ 
            CJ2 0925+504 & 0.370 & 1036.79 & $310.4 $   & 37.7     & $4.1 \times 3.6   $ & 1.6  & $< 1.2$           & $< 2.2$ \\ 
            TXS 0110+495 & 0.389 & 1022.61 & $615.4 $   & 38.2     & $4.8 \times 4.0   $ & 1.6  & $< 0.43$          & $< 0.78$ \\ 
            TXS 1030+415 & 1.117 &  670.95 & $636.6 $   & 29.1     & $7.2 \times 4.0   $ & 2.2  & $< 0.69$          & $< 1.3$ \\ 
            TXS 0249+383 & 1.122 &  669.37 & $877.0^e$  & 29.2     & RFI                 & --   & --                & -- \\
              S5 1044+71 & 1.150 &  660.65 & $1522.9$   & 29.5     & $10.6 \times 3.9  $ & 2.9  & $< 0.38$          & $< 0.69$ \\ 
             8C 1305+804 & 1.183 &  650.67 & $1334.2^e$ & 30.1     & RFI                 & --   & --                & -- \\
            TXS 1105+437 & 1.226 &  638.09 & $406.5 $   & 30.6     & $10.5 \times 6.5  $ & 1.3  & $< 0.70$          & $< 1.3$ \\ 
            TXS 1015+359 & 1.228 &  637.52 & $702.7 $   & 30.5     & $8.3 \times  4.8  $ & 1.3  & $< 0.43$          & $< 0.78$ \\ 
            TXS 1432+422 & 1.240 &  634.10 & $286.1 $   & 30.8     & $5.9 \times 5.4   $ & 1.2  & $< 0.95$          & $< 1.7$ \\ 
             S5 1150+81  & 1.250 &  631.29 & $1788.1$   & 30.9$^a$ & $15.7 \times 6.1  $ & 8.8  & $< 1.1$           & $< 2.0$ \\ 
            TXS 1020+400 & 1.254 &  630.17 & $1496.5$   & 30.9     & $13.2 \times 6.8  $ & 3.6  & $< 0.51$          & $< 0.93$ \\ 
              S5 1039+81 & 1.260 &  628.49 & $726.5 $   & 31.1$^a$ & $16.1 \times 5.8  $ & 5.2  & $< 1.5$           & $< 2.8$ \\ 
      TXS 1543+480$^{f}$ & 1.277 &  623.81 & $861.5 $   & 31.3$^a$ & $8.0 \times 5.3   $ & 2.1  & $9.56 \pm 0.36$   & $17.43 \pm 0.66$ \\ 
            TXS 1656+571 & 1.281 &  622.71 & $1456.3$   & 31.3     & $23.4 \times 12.6 $ & 4.9  & $< 0.74$          & $< 1.4$ \\ 
            TXS 0833+416 & 1.301 &  617.29 & $432.2 $   & 31.6$^a$ & $7.5 \times 4.7   $ & 1.8  & $< 0.63$          & $< 1.1$ \\ 
            TXS 2138+389 & 1.306 &  615.96 & $612.5^e$  & 15.9     & RFI                 & --   & --                & -- \\
            TXS 2319+444 & 1.310 &  614.89 & $378.4 $   & 31.7     & $9.2 \times 5.3   $ & 1.3  & $< 0.65$          & $< 1.9$ \\ 
            TXS 1240+381 & 1.318 &  612.77 & $547.2 $   & 31.8     & $6.6 \times 4.6   $ & 1.6  & $< 0.63$          & $< 1.1$ \\ 
            TXS 2007+659 & 1.325 &  610.92 & $631.1 $   & 31.9     & $10.4 \times 4.6  $ & 2.3  & $< 0.76$          & $< 1.4$ \\ 
         JVAS J2236+7322 & 1.345 &  605.71 & $270.7 $   & 32.2     & $13.9 \times 6.4  $ & 1.3  & $< 0.97$          & $< 1.8$ \\ 
            TXS 1342+663 & 1.351 &  604.17 & $231.8 $   & 32.3     & $11.1 \times 6.3  $ & 1.1  & $< 0.98$          & $< 1.8$ \\ 
            TXS 1739+522 & 1.375 &  598.06 & $1006.7$   & 32.6     & $8.9 \times 4.9   $ & 2.9  & $< 0.58$          & $< 1.1$ \\ 
            TXS 1442+637 & 1.380 &  596.80 & $656.6 $   & 32.7$^a$ & $9.2 \times 4.3   $ & 5.4  & $< 1.3$           & $< 2.3$ \\ 
            TXS 1010+350 & 1.410 &  589.37 & $520.1 $   & 33.1     & $9.5 \times 6.6   $ & 1.8  & $< 0.67$          & $< 1.2$ \\ 
            TXS 2229+695 & 1.413 &  588.64 & $291.4 $   & 33.1     & $10.3 \times 3.5  $ & 2.7  & $< 1.7$           & $< 3.1$ \\ 
            TXS 0145+386 & 1.442 &  581.65 & $207.4 $   & 33.6     & $9.3 \times 6.5   $ & 1.3  & $< 1.2$           & $< 2.2$ \\ 
         JVAS J2311+4543 & 1.447 &  580.46 & $156.4 $   & 33.6     & $13.6 \times 5.7  $ & 1.7  & $< 2.2$           & $< 3.9$ \\ 
              S5 1058+72 & 1.460 &  577.40 & $1499.7$   & 33.8     & $14.0 \times 4.6  $ & 2.3  & $< 0.31$          & $< 0.57$ \\ 
             B3 1746+470 & 1.484 &  571.82 & $187.9 $   & 34.1     & $9.9 \times 8.5   $ & 4.2  & $< 4.5$           & $< 8.2$  \\ 
            TXS 0859+681 & 1.499 &  568.39 & $458.4^e$  & 34.3     & RFI                 & --   & --                & --       \\
            TXS 1427+543 & 3.013 &  353.95 & $2402.7$   & 27.6     & $16.6 \times 8.0  $ & 7.4  & $< 0.44$          & $< 0.81$ \\ 

\hline
\hline
\end{tabular}
\end{center}
\begin{tablenotes}
\item[a]$^{a}$~This spectrum has not been Hanning-smoothed and resampled.
\item[b]$^{b}$~Flux densities are measured using the task JMFIT.
\item[c]$^{c}$~The upper limits on the integrated \hi\ optical depth assuming a line FWHM of 100~km~s$^{-1}$.
\item[d]$^{d}$~The \hi\ column density is calculated for an assumed spin temperature of 100 K.
\item[e]$^{e}$~The flux density of the source at the redshifted \hi\ line frequency was estimated 
by interpolating between the 1.4~GHz flux density (from the FIRST or NVSS surveys; 
\citealp[][]{becker95}; \citealp[][]{condon98}) and the 325 or 365 MHz flux density 
(from the WENSS and Texas surveys, respectively; \citealp[][]{rengelink97}; 
\citealp[][]{douglas96}) in the literature.  
\item[f]$^{f}$~The detection of associated \hii\ absorption towards this source was originally reported by \citet[][]{curran13}.
\item[]~For TXS 0344+405, ``C'' corresponds to the core, ``SE'' to the south-east lobe, 
and NW to the north-west lobe.
\end{tablenotes}
\end{table*}

\subsection{Results}
\label{sec:results}

\begin{figure*}
\includegraphics[scale=0.28]{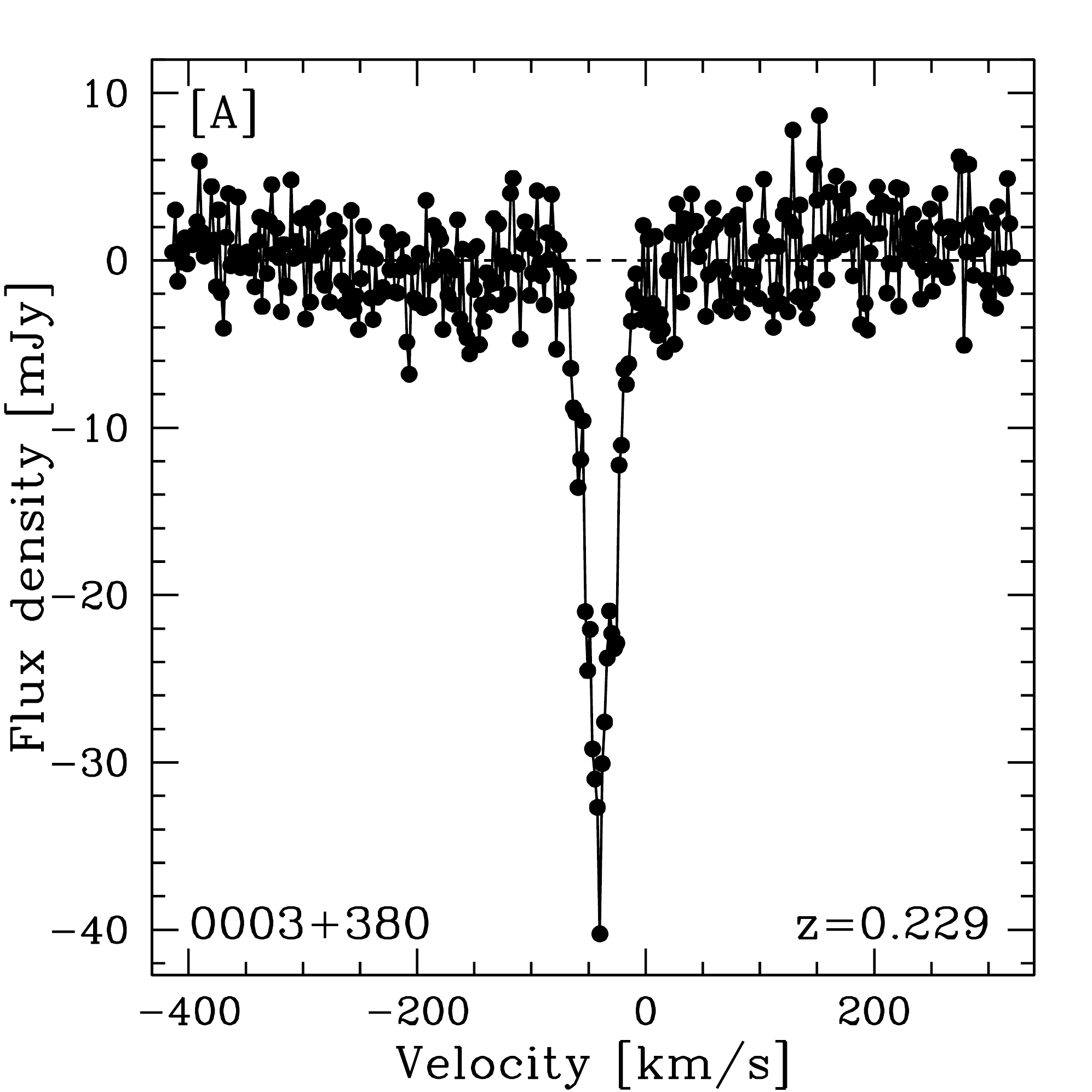}
\includegraphics[scale=0.28]{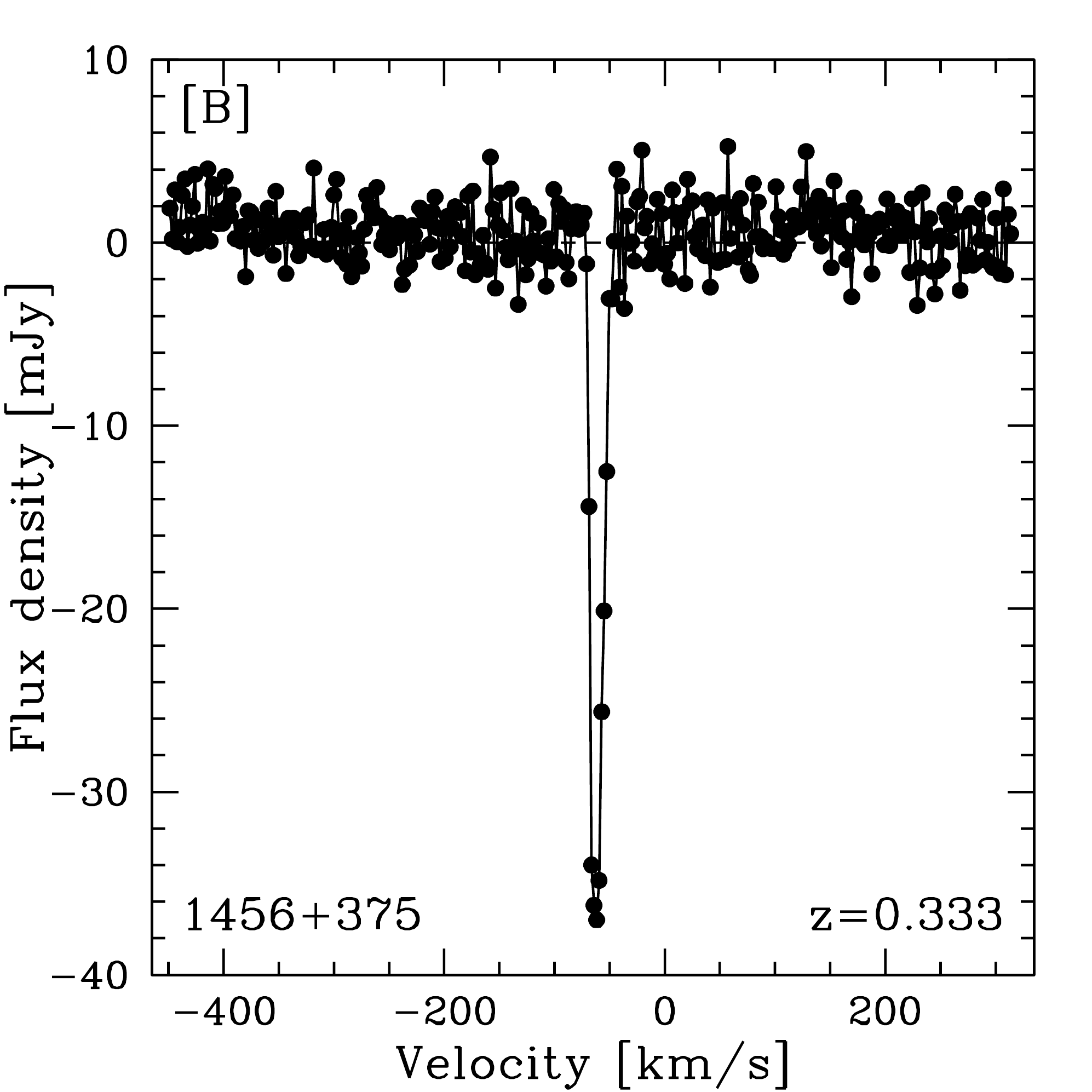}
\includegraphics[scale=0.28]{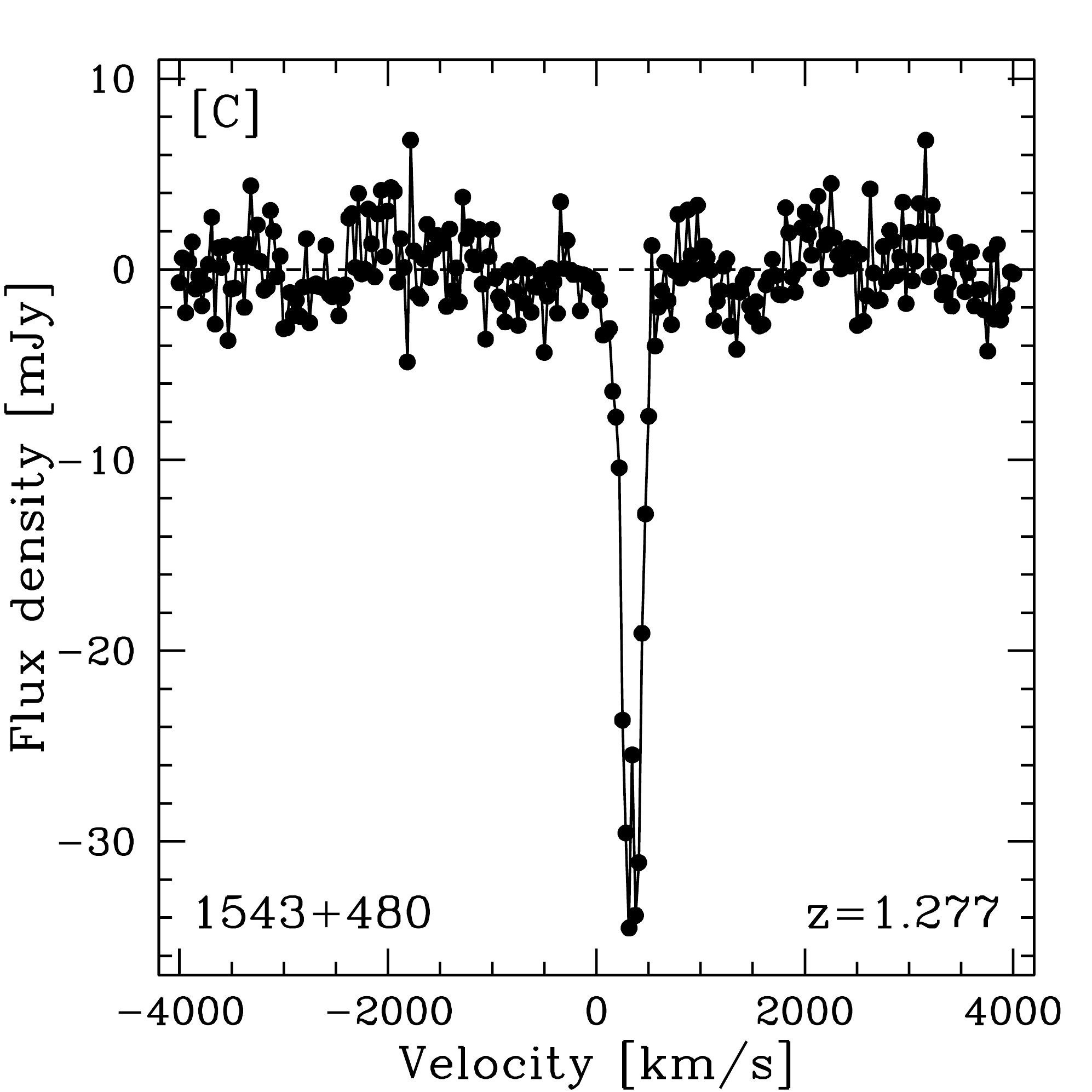}
\caption[]{The GMRT \hii\ absorption spectra towards [A]~TXS~0003+380 at $z = 0.229$,
[B]~TXS~1456+375 at $z = 0.333$, and [C]~TXS 1543+480, at $z = 1.277$. In all panels, the 
bottom axis contains velocity, in \kms, relative to the AGN redshift (listed in each panel).}
\label{fig:detect}
\end{figure*}

For ten sources, half at $z \approx 0.29-0.32$, the GMRT data were severely affected by RFI around the 
redshifted \hii\ line frequency and it was not possible to obtain reliable RFI-free spectra. In most cases, 
the continuum data themselves were badly affected by the RFI, rendering the flux density estimates unreliable. 
These sources are indicated by the label ``RFI'' in Table~\ref{table:obs}.

Except for TXS~0344+405, all the AGNs of our sample are compact in the GMRT continuum images, unresolved by 
the GMRT synthesized beam. The AGN flux densities at the redshifted \hii\ 
line frequency were measured by fitting a single-Gaussian model to a small region of the image plane around 
the target, using the AIPS task {\sc jmfit}. The measured flux densities are listed in Table~\ref{table:obs}.
We note that the {\sc jmfit} measurement errors are $< 1$~mJy in all cases; however, the errors on the flux densities 
are dominated by systematic effects, $\approx 10$\% (from our experience) for the GMRT at these frequencies. 
TXS~0344+405 shows extended radio structures, consisting of a central core and two extended emission 
components, with a typical Fanaroff-Riley II morphology \citep[][]{fanaroff74}. Since our aim is to search for 
neutral gas in the environments of compact AGNs, we have considered the \hii\ absorption spectrum against only 
the core of TXS~0344+405 in the later analysis. 

We obtained two new detections of associated \hii\ absorption, at $z = 0.229$ towards TXS~0003+380 and $z = 0.333$ 
towards TXS~1456+375, and also confirmed the detection of \hii\ absorption at $z = 1.277$ towards TXS~1543+480. 
The \hii\ absorption spectra towards these AGNs are shown in Fig.~\ref{fig:detect}, with flux density (in mJy) 
plotted against velocity (in \kms), relative to the AGN redshift. For the new detections, the \hii\ absorption 
spectra are from the confirming runs, at finer velocity resolution. The velocity-integrated \hii\ optical depths 
for these sources are $1.943 \pm 0.057$~\kms\ (TXS~0003+380), $3.834 \pm 0.079$~\kms\ (TXS~1456+375), and
$9.56 \pm 0.36$~\kms\ (TXS~1543+480). We note that our measured integrated \hii\ optical depth for TXS~1543+480 
is in excellent agreement with the estimate of $9.69 \pm 0.53$~\kms\ by \citet{curran13}.

For 37 AGNs, the spectra were found to be consistent with noise, with no evidence for \hii\ absorption; these are 
shown in Fig.~\ref{fig:nondetect} in the Appendix, in order of increasing redshift. The shaded regions in the 
spectra indicate velocity ranges that were corrupted by RFI. The figure includes three spectra for TXS~0344+405, 
towards the core and the two extended radio lobes, none of which show \hii\ absorption.  

For the non-detections, we obtained $3\sigma$ upper limits to the velocity-integrated \hii\ optical depth by 
assuming a line full-width-at-half-maximum (FWHM) of 100~\kms, and measuring the root-mean-square (RMS) 
optical depth noise from RFI-free channels after smoothing each spectrum to approximately the same resolution.
For two sources, TXS~0954+658 and S5~2007+77, a weak ripple was found to be present across the observing band, 
probably due to time variability in the antenna bandpass shapes that could not be calibrated out. Attempts to 
excise these ripples by data editing were unsuccessful. Since the peak-to-peak amplitude of the ripple is 
relatively low, the spectra may still be used to constrain the presence of strong \hii\ absorption. Hence, 
for these two sources, we conservatively estimated the upper limit to the \hii\ optical depth to be 
$3~\times$ the peak-to-peak spread in the ripple in the optical depth spectrum.

Table~\ref{table:obs} summarizes the observational details and results from our GMRT observations, with 
the sources ordered by increasing redshift. The columns of the table are (1)~the AGN name, 
(2)~the AGN redshift, $z$ (3)~the redshifted \hii\ line frequency, $\nu_{\rm 21 cm}$, in MHz,
(4)~the AGN flux density, $S_{\nu}$, at the redshifted \hii\ line frequency, in mJy, (5)~the velocity 
resolution $\Delta V$ of the final \hii\ spectrum, in \kms, (6)~the FWHM of the synthesized beam, in $'' \times ''$, 
(7)~the RMS noise $\Delta S$ on the final spectrum at the velocity resolution of column~(5), in mJy, 
(7)~the integrated \hii\ optical depth $\int \tau dv$ in \kms, or, for non-detections, the $3\sigma$ upper 
limit to $\int \tau dv$, assuming a line FWHM of $100$~km s$^{-1}$, (8)~the \hi\ column density N$_{\rm HI}$ in 
cm$^{-2}$, or, for non-detections, the $3\sigma$ upper limit to N$_{\rm HI}$, assuming a spin temperature 
($\rm T_s$) of 100~K. For the ten sources affected by RFI, the sixth column contains ``RFI'', and the remaining 
columns indicate that no information is available. In most cases, the velocity resolution quoted in column~(5) 
is after Hanning-smoothing and resampling the spectrum.

\section{The detections of \hii\ absorption}
\label{sec:detect}

In this section, we discuss the detections of \hii\ absorption and the implications of 
our results for conditions in the AGN environment. 


\subsection{The $z=0.229$ absorber towards TXS~0003+380}
\label{sec:0003+380}

The radio emission of TXS~0003+380 is unresolved in the GMRT continuum image at $\approx 1156$~MHz, with 
no evidence of any extended structure. A prominent core is present in the milli-arcsec (mas) scale VLBI map at 2.3~GHz, 
along with a weak signature of a jet projected towards the south-west \citep[][]{fey00}. More than 95\% of 
the 2.3~GHz flux density arises from the core, with a total of only $\approx 35$~mJy arising from the components 
in the radio jet \citep[see Table~2 of ][]{fey00}. The AGN spectrum is very flat at low radio frequencies, 
$\lesssim 5$~GHz, with flux densities of $\approx 500-800$~mJy over the wide frequency range $74 - 5000$~MHz. 
The radio flux density is known to vary rapidly, on a timescale of $2 - 20$~days \citep[][]{heeschen84}.
Such `flickering' may arise due to new plasma ejections, when the radiation is highly beamed towards the 
observer \citep[][]{ghisellini93}. Optical and X-ray studies of this source also find evidence for strong 
blazar characteristics \citep[e.g.][]{massaro09}.

The GMRT \hii\ absorption spectrum towards TXS~0003+380 is shown in Fig.~\ref{fig:detect}[A]. The absorption 
has a peak line depth of $\approx 40$~mJy, comparable to the total flux density ($\approx 35$~mJy) in the 
jet components at 2.3~GHz. This indicates that the absorption is likely to arise against the radio core.  
The absorption line is relatively narrow, with a full width between 20\% points of only $\approx 45$~\kms. 
The detected \hii\ absorption implies an \hi\ column density of $\rm (3.54 \pm 0.11) \times (T_s/100\;K) 
\times 10^{20}$~cm$^{-2}$. While the \hii\ absorption appears to be blueshifted by $\approx 50$~\kms\ from 
the AGN redshift \citep[$z= 0.229$;][]{stickel94}, we note that the latter authors do not quote an error on 
their redshift estimate.

\subsection{The $z = 0.333$ absorber towards TXS~1456+375}
\label{sec:1456+375}

TXS~1456+375 is unresolved in the GMRT 1065~MHz continuum image, and is also extremely compact in 
the VLBI 5~GHz image of \citet[][]{helmboldt07}, with a linear size $\lesssim 10$~mas. The AGN 
spectrum is flat at low frequencies, with spectral index  $\alpha = -0.02$ between 408~MHz and 1.4~GHz,
steepening slightly to $\alpha = -0.27$ between 1.4~GHz and 5~GHz \citep[][]{helmboldt07}.
The source has been classified as a blazar in the literature, based on its optical and near-infrared 
properties \citep[e.g.][]{massaro09, davenport15}. Our line of sight is hence likely to lie close to 
the direction of any jet emission. In addition, the AGN spectrum rises steeply with increasing 
wavelength in the optical and near-infrared bands, with R$-$K$=5.14$~mag, causing it to be classified as 
a ``red'' quasar \citep[][]{glikman12}.

The GMRT \hii\ absorption spectrum towards TXS~1456+375 is displayed in Fig.~\ref{fig:detect}[B]. 
The \hii\ absorption is only marginally offset from the AGN redshift ($z = 0.33343 \pm 0.00017$; 
\citealp[][]{schneider05}), with the deepest absorption blueshifted by $\approx 60$~\kms\ (consistent 
with the AGN redshift within $\approx 2\sigma$ significance). The \hii\ absorption is very narrow,
with a width between 20\% points of only $\approx 20$~\kms, and no evidence of extended absorption.
The \hi\ column density inferred from the \hii\ absorption is $\rm (6.98 \pm 0.15) \times (T_s/100\;K) 
\times 10^{20}$~cm$^{-2}$. The small angular extent of the 5~GHz radio emission implies that the 
detected \hii\ absorption probably arises against the radio core.


\subsection{The $z= 1.277$ absorber towards TXS~1543+480}
\label{sec:1543+480}

TXS~1543+480 is unresolved in our GMRT 624~MHz continuum image, but shows two clear lobes, with 
similar flux densities, in the 5~GHz VLBI image \citep[][]{helmboldt07}. The authors classify this
source as a candidate Compact Symmetric Object (CSO). CSOs have symmetric parsec-scale structure, 
dominated by steep-spectrum extended emission on both sides of the core. The core is usually faint, 
or even undetected, in such sources. It is plausible that the two radio components detected in the 
5~GHz VLBI image of TXS~1543+480 arise from parsec-scale lobes at the ends of VLBI-scale jets, with 
the core remaining undetected. Multi-frequency VLBI studies are needed to measure the spectral indices 
of the mas-scale radio components, and to test whether both have steep spectra, as expected for a CSO. 
At present, only 5~GHz VLBI images are available for TXS~1543+480 in the literature and it hence remains a 
candidate CSO \citep[][]{helmboldt07}. 

If TXS~1543+480 is indeed a CSO, its 624~MHz flux density is likely to be dominated by the emission from the 
two radio lobes, as these would be expected to have steep spectra. The detected \hii\ absorption is hence likely to arise
against one or both of the VLBI radio lobes. The \hii\ absorption is quite extended, with a width between 
20\% points of $\approx$~330~\kms. Such wide absorption could arise either due to absorption against 
both radio lobes, or due to disturbed gas that is interacting with the AGN jets. 

We have earlier reported a similar case of associated \hii\ absorption towards another CSO, TXS~1245-197
at $z = 1.275$ \citep[][]{aditya18}. It is intriguing that both TXS~1543+480 and TXS~1245-197 show wide 
\hii\ absorption, with widths of $\gtrsim 300$~\kms\ between 20\% points. Earlier studies of the kinematic 
ages of CSOs have been used to argue that CSOs are small due to their youth, and not because they 
are ``frustrated'' sources that reside in a dense environment, unable to grow to large sizes 
\citep[][]{owsianik98,owsianik98a,taylor00}. However, our detections of wide \hii\ absorption from CSO 
environments support the notion that the radio jets in CSOs may indeed be evolving in a dense gaseous 
environment, with their expansion constrained by the surrounding medium.

\section{Discussion}
\label{sec:discuss}

\subsection{A uniformly-selected flat-spectrum sample}
\label{sec:sample}

Searches for redshifted associated \hii\ absorption have so far been carried out in more than 400 AGNs, 
with $\gtrsim 75$ detections of \hii\ absorption \citep[e.g.][]{vangorkom89,vermeulen03,gupta06,curran10,gereb15,maccagni17}. 
The vast majority of both searches and detections are at low redshifts, $z < 1$, where the typical 
detection fraction is $\approx 30$\% \citep[e.g.][]{pihlstrom03,gupta06,maccagni17}. The situation is 
very different at high redshifts, $z > 1$, with searches for associated \hii\ absorption in only $\approx 25$~AGNs 
\citep[e.g.][]{gupta06,curran13} and just four detections in the literature, prior to our survey. The implied 
detection rate of \hii\ absorption is $\approx 16^{+13}_{-8}$\%, where the errors are from Poisson statistics 
\citep[][]{gehrels86}. While the detection rate of \hii\ absorption at $z > 1$ is only half that at low redshifts,
the difference between the two is not statistically significant due to the large uncertainty in the high-$z$ 
value, simply due to the fact that few high-$z$ AGNs have hitherto been targetted in \hii\ absorption studies.
An important part of the present survey was simply to increase the number of searches for redshifted \hii\ 
absorption at $z > 1$. If high-$z$ AGNs yield a detection fraction similar to those at low redshifts, the survey 
would then yield a large sample of associated \hii\ absorbers, suitable for detailed kinematic studies of AGN 
environments. Conversely, if the detection fraction remains low at $z > 1$, this would be evidence for redshift 
evolution in AGN environments.

Further, most studies of associated \hii\ absorption \citep[e.g.][]{vermeulen03,gupta06,curran13,maccagni17} have 
targetted highly heterogeneous AGN samples at all redshifts. This heterogeneity makes it difficult to distinguish 
between possible redshift evolution in the AGN environment and differences in the AGN samples at different redshifts. For 
example, \citet[][]{gupta06} carried out an analysis of 96 AGNs, mostly at low to intermediate redshifts, $z < 1$. 
They found little evidence for redshift evolution in either the detection rates of \hii\ absorption or in the 
distribution of the \hii\ optical depths. However, their sample included a range of AGN types, with 21 large radio 
galaxies, 13 flat-spectrum sources, 35 CSS sources, and 27 GPS sources. The heterogeneity of the 
sample makes it difficult to reliably interpret the observational data.

An alternative explanation for the tentative result that the strength of associated \hii\ absorption may be weaker 
at high redshifts stems from the luminosity bias in most AGN samples used in such studies. High-$z$ AGN samples 
typically contain more objects with higher rest-frame UV and radio luminosities. \citet[][]{curran08} suggest 
that the high AGN luminosity in the UV and/or radio wavebands could lead to a lower \hii\ optical depth, either 
by ionizing the \hi\ (and thus reducing the \hi\ column density) or by altering the hyperfine level populations 
(and thus increasing the spin temperature). \citet[][]{curran08} hence argued that the high luminosities of 
high-$z$ AGNs may be the primary cause of the low detection rate of associated \hii\ absorption at high redshifts
\citep[see also][]{curran13}. However, we note that the AGN samples of \citet{curran08} and \citet{curran13} were 
also highly heterogeneous, containing all the AGNs that had been searched for \hii\ absorption in the literature.

The present survey for \hii\ absorption aims to address the above issues by carrying out a search for \hii\ absorption
in a large and uniformly-selected sample of AGNs, selected from the CJF catalogue, to investigate the dependence of the 
\hii\ absorption strength and detectability on redshift and AGN properties (e.g. luminosity, colour, spectral index, etc). 
In our pilot study \citep[][]{aditya16}, we combined 23 new searches for redshifted \hii\ absorption with 29 searches 
from the literature (all in AGNs from the CJF sample), to find tentative evidence (at $\approx 3\sigma$ significance) 
that the strength of associated \hii\ absorption depends on both redshift and AGN luminosity, with weaker \hii\ absorption
obtained at both high redshifts and higher AGN luminosities. 

In the present work, we have completed our GMRT \hii\ absorption survey of AGNs of the CJF sample, targetting nearly all 
CJF sources whose redshifted \hii\ line frequencies lie within the GMRT's legacy 327, 610, and 1420-MHz bands. Including 
39 sources with usable data from this study, our full sample consists of 92 flat-spectrum AGNs, which includes 63 sources 
from our GMRT observations and 29 from the literature. The sample contains 16 detections of \hii\ absorption [including the
three confirmed detections of this paper towards TXS~0003+380, TXS~1456+375, and TXS~1543+480, the confirmed detection towards 
TXS~1954+513 \citep{aditya17}, and our tentative detection towards TXS~0604+728 \citep{aditya16}], and 76 non-detections, 
yielding upper limits to the \hii\ optical depth. This is by far 
the largest sample of uniformly-selected AGNs that have been searched for associated \hii\ absorption. The sample covers a 
large redshift range, $0.01 \lesssim z \lesssim 3.6$, with more than half of the sample ($\approx 50$ sources) at $z > 1$. 
   
The 92 CJF  sources of our full sample are listed in order of increasing redshift, in Table~\ref{table:cjf}. The columns of 
this table are (1)~the AGN name, (2)~the AGN redshift, (3)~the velocity-integrated \hii\ optical depth in \kms, or, for 
non-detections of \hii\ absorption, the 3$\sigma$ upper limit to the \hii\ optical depth, assuming a Gaussian profile 
with a line FWHM of 100~\kms, (4)~the rest-frame 1216~\AA\ AGN luminosity $L'_{UV} \equiv Log[L_{UV}/(W\;Hz^{-1})]$, 
inferred by interpolating between measured luminosities in UV and/or optical wavebands (see below), (5)~the rest-frame 
1.4~GHz AGN luminosity $L'_{\rm 1.4\;GHz} \equiv {\rm Log}[L_{\rm 1.4\;GHz}/({\rm W\;Hz^{-1}})]$, (6)~the AGN spectral 
index close to the 
redshifted \hii\ line frequency, $\rm \alpha_{\rm 21-cm}$, (7)~the AGN colour (R$-$K) between the R- and K-bands, 
(8)~the reference for the search for \hii\ absorption, and (9)~references for the estimates of UV, optical, and 
near-infrared (NIR) luminosities, which were used to infer $L'_{\rm UV}$ and (R$-$K).

The rest-frame 1216~\AA\ AGN luminosity was estimated following the procedure of \citet[][]{curran10}. We first determined 
the flux density $F_{\rm UV}$ at $1216\times(1 + z)$~\AA\ for each AGN, using a power-law spectrum to interpolate between 
measured flux densities at two nearby optical and/or UV wavebands from the literature. The luminosity at rest-frame 1216~\AA\ 
was then inferred from the expression ${L_{UV}} = {4\pi D_{\rm AGN}^{2} F_{\rm UV}}/(1+z)$, where $D_{\rm AGN}$ is the 
AGN's luminosity distance. For two AGNs, TXS~0424+670 and TXS~1020+400, the flux density is known only at a single optical 
waveband, quite distant from the redshifted 1216~\AA\ wavelength. These systems hence do not have a listed rest-frame 
1216~\AA\ luminosity in Table~\ref{table:cjf}.

The radio spectral indices of the AGNs were computed from their flux densities at the redshifted \hii\ line frequency 
and a nearby frequency at which a flux density estimate was available in the literature. For all sources at $z > 1$, 
the second frequency was 1.4~GHz, from the FIRST or NVSS surveys \citep{becker95,condon98}. For sources at $z < 1$,
the second frequency was either 365~MHz \citep[the Texas survey;][]{douglas96} or 325~MHz \citep[the WENSS survey;][]{rengelink97}. 
Finally, the (R$-$K) colour could only be inferred for 58 AGNs of the full sample; 
the remaining sources do not have K-band information in the literature.

In the following sections, we will examine the dependence of the \hii\ detection fraction and the distribution of integrated 
\hii\ optical depth on redshift, radio spectral index, AGN radio and UV luminosities, and the (R$-$K) colour, for the 
full sample of 92 flat-spectrum sources.

\setcounter{table}{1}
\begin{table*}
\footnotesize
\caption{The 92 CJF sources of the full sample with searches for redshifted \hii\ absorption, 
	listed in order of increasing redshift. 63 sources are from the present survey, 39 from this paper 
	and 24 from \citet{aditya16,aditya17}, while 29 are from the literature. See main text for discussion.
\label{table:cjf}}
\begin{center}
\begin{tabular}{|lcccccccc|}
\hline \\
	AGN        &  $z$  & $\int \tau dv$     & $L'_{\rm UV}$$^a$~\tnote{a} & $L'_{\rm 1.4\;GHz}$ & $\alpha_{\rm 21-cm}$ & (R$-$K)$^b$~\tnote{b} & Refs.$^c$~\tnote{c} & Refs.$^d$~\tnote{d} \\
		   &       & \kms               &       &       &       &     &  (21-cm) & (UV,Opt.,NIR)        \\
\hline
 TXS 1146+596      & 0.011 & $5.3 \pm 1.8$      & 20.30 & 23.09 & 0.26  & 2.3  & 3   & 1--4     \\
 TXS 0316+413      & 0.018 & $1.3$              & 20.61 & 25.13 & 1.05  & -0.2 & 4   & 1,3,4    \\
 B3 0651+410       & 0.022 & $<0.82$            & 18.69 & 23.35 & 0.60  & -1.0 & 5   & 1,3,4    \\
 TXS 1101+384      & 0.030 & $<0.63$            & 21.28 & 24.18 & 0.17  & -1.7 & 6   & 4,5      \\
 TXS 1744+557      & 0.030 & $<1.2$             & 19.53 & 24.02 & 0.22  & -3.1 & 7   & 1,3,4    \\
 TXS 1652+398      & 0.034 & $<2.4$             & 21.58 & 24.56 & -0.12 & -1.3 & 6   & 1,3,4    \\
     TXS 0344+405  & 0.039 & $<6.3$             & 21.34 & 23.23 & 1.70  & 0.6  & 1   & 1,3,4    \\
     TXS 0733+597  & 0.041 & $<0.67$            & 19.69 & 24.24 & -0.28 & -1.8 & 1   & 1,3,4    \\
 TXS 1254+571      & 0.042 & $33.9 \pm 3.8$     & 20.51 & 24.06 & -0.57 & 6.1  & 14  & 2--4     \\
 TXS 1807+698      & 0.051 & $<$1.6             & 20.91 & 25.00 & 1.05  & 0.2  & 6   & 1,3,4    \\
 TXS 0402+379      & 0.055 & $0.98 \pm 0.11$    & 19.88 & 25.31 & -0.29 & 3.6  & 9   & 3,6      \\
 TXS 1144+352      & 0.063 & $<1.1$             & 20.26 & 24.75 & 0.54  & -0.9 & 7   & 1,3,4    \\
 TXS 2200+420      & 0.069 & $<0.95$            & 20.30 & 25.78 & -0.14 & 2.0  & 6   & 7--10    \\
       S5 2116+81  & 0.084 & $<1.4$             & 21.25 & 24.31 & -0.67 & 1.5  & 1   & 1,3,4    \\
     TXS 1418+546  & 0.153 & $<1.5$             & 21.71 & 25.48 & 0.58  & 4.7  & 1   & 1,4,11   \\
 TXS 1946+708      & 0.101 & $15.8 \pm 4.6$     & 18.60 & 25.35 & -0.33 & 1.3  & 10  & 3,4      \\
 TXS 0309+411      & 0.134 & $<0.92$            & 18.48 & 25.23 & 0.08  & 1.5  & 7   & 3,4      \\
       S4 0749+54  & 0.200 &  $<1.2$            & 20.71 & 25.68 & 0.35  & 2.5  & 1   & 1,4,11   \\
 IVS B1622+665     & 0.201 & $<1.7$             & 20.00 & 25.24 & 0.53  & 2.9  & 5   & 3,4      \\
 S5 1826+79        & 0.224 & $<15$              & 21.28 & 25.57 & 0.59  & 2.1  & 11  & 1,3,4    \\
 TXS 2021+614      & 0.227 & $<0.21$            &  --   & 26.43 & 0.07  & 3.3  & 11  & 3,4      \\
     TXS 0003+380  & 0.229 & $1.943 \pm 0.057$  & 20.62 & 25.80 & 0.49  & 3.6  & 1   & 1,3,4    \\
 TXS 2352+495      & 0.238 & $1.7$              & 20.84 & 26.46 & -0.09 & 2.9  & 11  & 3,4,12   \\
 TXS 0831+557      & 0.241 & $0.58$             & 21.39 & 27.03 & -0.15 & 0.3  & 11  & 1,3,4    \\
     TXS 0010+405  & 0.255 & $<1.4$             & 21.24 & 25.72 & 1.37  & 2.2  & 1   & 1,3,4    \\
     TXS 1719+357  & 0.263 & $<1.2$             & 21.90 & 25.61 & 0.74  & 3.2  & 1   & 1,3,4    \\
 TXS 1943+546      & 0.263 & $2.9$              & 20.85 & 26.46 & -0.45 & 1.4  & 11  & 1,3,6    \\
     TXS 0424+670  & 0.324 & $<0.89$            &  --   & 26.24 & -0.46 &  --  & 1   &  --      \\
      B3 1456+375  & 0.333 & $3.834 \pm 0.079$  & 20.65 & 25.58 & -0.41 & 5.1  & 1   & 1,3,7    \\
       S5 2007+77  & 0.342 & $<9.9 $            & 22.03 & 26.25 & 0.09  & 3.5  & 1   & 3,7      \\
     TXS 0035+367  & 0.366 & $<0.51$            & 21.56 & 26.26 & -0.17 & 2.4  & 1   & 1,3,4    \\
     TXS 0954+658  & 0.368 & $<5.8 $            & 21.96 & 26.55 & 0.01  & 3.3  & 1   & 1,3,4    \\
     CJ2 0925+504  & 0.370 & $<1.2 $            & 22.75 & 26.00 & 0.38  & 2.3  & 1   & 1,3,7    \\
     TXS 0110+495  & 0.389 & $<0.43$            & 21.09 & 26.34 & -0.01 & 2.4  & 1   & 1,3,4    \\
 TXS 1031+567      & 0.459 & $<0.76$            & 19.90 & 26.96 & -0.20 &  --  & 11  & 2,13     \\
 TXS 1355+441      & 0.646 & $19$               & 20.60 & 26.94 & -0.34 &  --  & 11  & 1,2,13   \\
 S4 0108+38        & 0.669 & $46 \pm 7$         & 21.63 & 26.95 & 1.16  &  --  & 12  & 1,14,15  \\
 TXS 1504+377      & 0.672 & $27.20 \pm 0.04$   & 20.30 & 26.95 & -0.21 &  --  & 15  & 2        \\
 TXS 0923+392      & 0.695 & $<0.54$            & 23.47 & 27.49 & -0.38 & 1.7  & 11  & 1,4,15   \\
 TS5 0950+74       & 0.695 & $<1.4$             & 21.65 & 27.15 & 0.92  &  --  & 11  & 3        \\
 TXS 1642+690      & 0.751 & $<0.69$            & 22.66 & 27.34 & 0.03  &  --  & 11  & 3        \\
 S4 1843+35        & 0.764 & $<6.0$             & 24.66 & 27.55 & -0.03 &  --  & 11  & 3        \\
     TXS 1030+415  & 1.117 & $<0.69$            & 23.32 & 27.30 & -0.61 & --   & 1   & 2        \\
 TXS 0600+442      & 1.136 & $<0.71$            & --    & 27.61 & -0.37 & --   & 2   & --       \\
       S5 1044+71  & 1.150 & $<0.38$            & 23.26 & 27.70 & -0.97 & 4.7  & 1   & 1,3,4    \\
 TXS 2356+390      & 1.198 & $<0.42$            & 22.36 & 27.36 & -1.16 & --   & 2   & 1,3      \\
 TXS 0821+394      & 1.216 & $<0.44$            & 23.55 & 27.97 & -0.75 & 3.0  & 2   & 1--4     \\
     TXS 1954+513  & 1.223 & $0.698 \pm 0.036$  & 23.23 & 27.75 & 0.06  & 2.4  & 16  & 3,4      \\
     TXS 1105+437  & 1.226 & $<0.70$            & 23.04 & 27.18 & -0.38 & --   & 1   & 1,2      \\
     TXS 1015+359  & 1.228 & $<0.43$            & 24.31 & 27.42 & -0.02 & --   & 1   & 1        \\
     TXS 1432+422  & 1.240 & $<0.95$            & 22.62 & 27.04 & 0.10  & --   & 1   & 1,3,4    \\
 TXS 0945+408      & 1.249 & $<0.30$            & 23.93 & 27.89 & -0.46 & 2.5  & 2   & 1,3,4    \\
       S5 1150+81  & 1.250 & $<1.1 $            & 23.54 & 27.84 & -0.33 & 3.3  & 1   & 1,3,4    \\
     TXS 1020+400  & 1.254 & $<0.51$            &  --   & 27.77 & -0.78 & 3.1  & 1   & 3,4      \\
       S5 1039+81  & 1.260 & $<1.5 $            & 23.90 & 27.46 & 0.01  & 2.9  & 1   & 3,4      \\
\hline
\hline
\end{tabular}
\end{center}
\end{table*}

\setcounter{table}{1}
\begin{table*}
\caption{(contd.)
		\label{cjfcontd}}
\begin{center}
\begin{tabular}{|lcccccccc|}
\hline \\
	AGN        &  $z$  & $\int \tau dv$     & $L'_{\rm UV}$$^a$~\tnote{a} & $L'_{\rm 1.4\;GHz}$ & $\alpha_{\rm 21-cm}$ & (R$-$K)$^b$~\tnote{b} & Refs.$^c$~\tnote{c} & Refs.$^d$~\tnote{d} \\
		   &       & \kms               &       &       &       &     &  (21-cm) & (UV,Opt.,NIR)        \\
\hline
 TXS 0641+392      & 1.266 &  $<0.68$           & 22.64 & 27.22 & 0.79  & --   & 2   & 3        \\
 TXS 0537+531      & 1.275 &  $<0.32$           & 22.84 & 27.46 & 0.01  & 2.7  & 2   & 1,3,4    \\
 TXS 1543+480      & 1.277 & $9.69 \pm 0.53$    & 22.08 & 27.36 & -0.23 & 6.3  & 13  & 1,2,4    \\
     TXS 1656+571  & 1.281 &  $<0.74$           & 23.56 & 27.77 & -0.54 & --   & 1   & 1,3      \\
 TXS 0707+476      & 1.292 &  $<0.30$           & 23.86 & 27.63 & -0.15 & 1.4  & 2   & 1,3,4    \\
     TXS 0833+416  & 1.301 &  $<0.63$           & 24.06 & 27.26 & -0.02 & 0.2  & 1   & 1,3,4    \\
     TXS 2319+444  & 1.310 &  $<0.65$           & 22.41 & 27.20 & -0.05 & --   & 1   & 1,3      \\
 TXS 0248+430      & 1.311 & $< 1.4$            & 23.10 & 27.84 & 0.22  & -0.9 & 3   & 1,3,4,16 \\
     TXS 1240+381  & 1.318 &  $<0.63$           & 23.53 & 27.37 & 0.01  & 2.9  & 1   & 1,3,4    \\
 TXS 0850+581      & 1.318 &  $<0.37$           & 23.53 & 27.63 & -0.26 & --   & 2   & 1        \\
     TXS 2007+659  & 1.325 &  $<0.76$           & 22.51 & 27.44 & -0.26 & --   & 1   & 1,3      \\
 S5 2353+81        & 1.344 &  $<0.87$           & 22.30 & 27.41 & -0.53 & 2.9  & 2   & 1,3,6    \\
  JVAS J2236+7322  & 1.345 &  $<0.97$           & 22.64 & 27.08 & -0.01 & --   & 1   & 1,3      \\
     TXS 1342+663  & 1.351 &  $<0.98$           & 22.74 & 27.02 & 1.23  & 4.3  & 1   & 1--4     \\
 TXS 0035+413      & 1.353 &  $<0.76$           & 23.12 & 27.39 & 0.35  & --   & 2   & 1        \\
     TXS 1739+522  & 1.375 &  $<0.58$           & 23.73 & 27.67 & 0.51  & 3.1  & 1   & 1,3,4    \\
     TXS 1442+637  & 1.380 &  $<1.3 $           & 23.90 & 27.51 & 0.05  & 2.0  & 1   & 1,3,4    \\
 TXS 1030+611      & 1.401 &  $<0.87$           & 23.56 & 27.53 & -0.37 & 5.3  & 2   & 1--4     \\
     TXS 1010+350  & 1.410 &  $<0.67$           & 23.94 & 27.41 & -0.43 & --   & 1   & 1        \\
     TXS 2229+695  & 1.413 &  $<1.7 $           & 24.47 & 27.15 & 0.64  & --   & 1   & 3,13     \\
 TXS 0820+560      & 1.418 &  $<0.34$           & 23.69 & 27.87 & -0.26 & --   & 2   & 1        \\
 TXS 0805+410      & 1.418 &  $<0.69$           & 23.32 & 27.40 & -0.07 & --   & 2   & 1,2      \\
 TXS 0804+499      & 1.436 &  $<0.95$           & 23.45 & 27.54 & 0.31  & --   & 2   & 1,2      \\
     TXS 0145+386  & 1.442 &  $<1.2 $           & 23.54 & 27.02 & 0.58  & 1.7  & 1   & 1,3,4    \\
 TXS 0917+624      & 1.446 &  $<0.79$           & 23.22 & 27.70 & 0.23  & 3.2  & 2   & 1--4     \\
  JVAS J2311+4543  & 1.447 &  $<2.2 $           & 22.82 & 26.90 & 0.76  & --   & 1   & 1,3      \\
       S5 1058+72  & 1.460 &  $<0.31$           & 24.27 & 27.89 & -0.21 & 1.9  & 1   & 1,3,4    \\
 TXS 0859+470      & 1.470 &  $<0.43$           & 23.54 & 28.17 & -0.23 & --   & 2   & 2        \\
 TXS 2253+417      & 1.476 &  $<0.71$           &  --   & 27.82 & 0.30  & --   & 2   & --       \\
      B3 1746+470  & 1.484 &  $<4.5 $           & 23.43 & 27.00 & 0.54  & --   & 1   & 1,3      \\
 TXS 0340+362      & 1.484 &  $<3.0 $           & 22.75 & 27.19 & 0.35  & --   & 2   & 3        \\
     TXS 1427+543  & 3.013 &  $<0.44$           & 23.77 & 28.66 & -0.60 & --   & 1   & 2        \\
 TXS 0800+618      & 3.033 &  $<1.2 $           & 23.67 & 28.22 & -0.05 & --   & 2   & 1,3      \\
 TXS 0642+449      & 3.396 &  $<2.9 $           & 24.47 & 27.86 & 0.72  & 2.9  & 2   & 3,4,17   \\
 TXS 0620+389      & 3.469 &  $<0.20$           & 24.28 & 28.50 & -0.16 & 2.4  & 2   & 3,14,18  \\
 TXS 0604+728      & 3.530 &  $4.29 \pm 0.28$   & 23.96 & 28.64 & -0.38 & --   & 2   & 1,3      \\
 TXS 0749+426      & 3.589 &  $<0.76$           & 24.60 & 28.14 & 0.16  & 2.4  & 2   & 2,3,18   \\
\hline
\hline
\end{tabular}
\end{center}
\begin{tablenotes}
\item[]~Notes to the table:
\item[a]$^{a}$The inferred rest-frame 1216 \AA\ AGN luminosity, obtained by extrapolating from measurements in two nearby optical 
and/or ultraviolet bands. For two AGNs (indicated by a ``--" in this column), the UV luminosity could not be obtained as the AGN 
flux density is only available at a single optical waveband in the literature. 
\item[b]$^b$For sources with ``--" entries, the flux density is not known in the K-band; the (R$-$K) colour hence could not be obtained.
\item[c]$^c$References for the associated \hii\ absorption searches : (1)~This paper; (2)~\citet{aditya16}; (3)~\citet[][]{gupta06}; 
(4)~\citet[][]{young1973}; (5)~\citet[][]{orienti06}; (6)~\citet[][]{vangorkom89}; (7)~\citet[][]{chandola13}; (8)~\citet[][]{dickey82}; 
(9)~\citet[][]{morganti09}; (10)~\citet[][]{peck99}; (11)~\citet[][]{vermeulen03}; (12)~\citet[][]{carilli98}; 
(13)~\citet[][]{curran13}; (14)~\citet[][]{gallimore99}; (15)~\citet[][]{kanekar08c}; (16)~\citet[][]{aditya17}.
\item[d]$^d$References for the ultraviolet, optical, and infrared luminosity measurements, which were used to obtain the 
rest-frame 1216~\AA\ UV luminosity (following the procedure of \citealt[][]{curran10}), and the (R$-$K) colour: 
(1)~\citet[][]{bianchi14}; (2)~\citet[][]{abazajian09}; (2)~\citet[][]{monet03}; (4)~\citet[][]{cutri03}; 
(5)~\citet[][]{massaro04}; (6)~\citet[][]{cutri13}; (7)~\citet[][]{chen05b}; (8)~\citet[][]{howard04}; (9)~\citet[][]{odell78}; 
(10)~\citet[][]{raiteri09}; (11)~\citet[][]{urry00}; (12)~\citet[][]{zacharias04}; (13)~\citet[][]{veron10}; (14)~\citet[][]{souchay15};
(15)~\citet[][]{healey08}; (16)~\citet[][]{rao06}; (17)~\citet[][]{fedorov11}; (18)~\citet[][]{kuhn04}.
\end{tablenotes}
\end{table*}

\subsection{Redshift evolution}
\label{sec:redshift}

\begin{figure}
\includegraphics[scale=0.45]{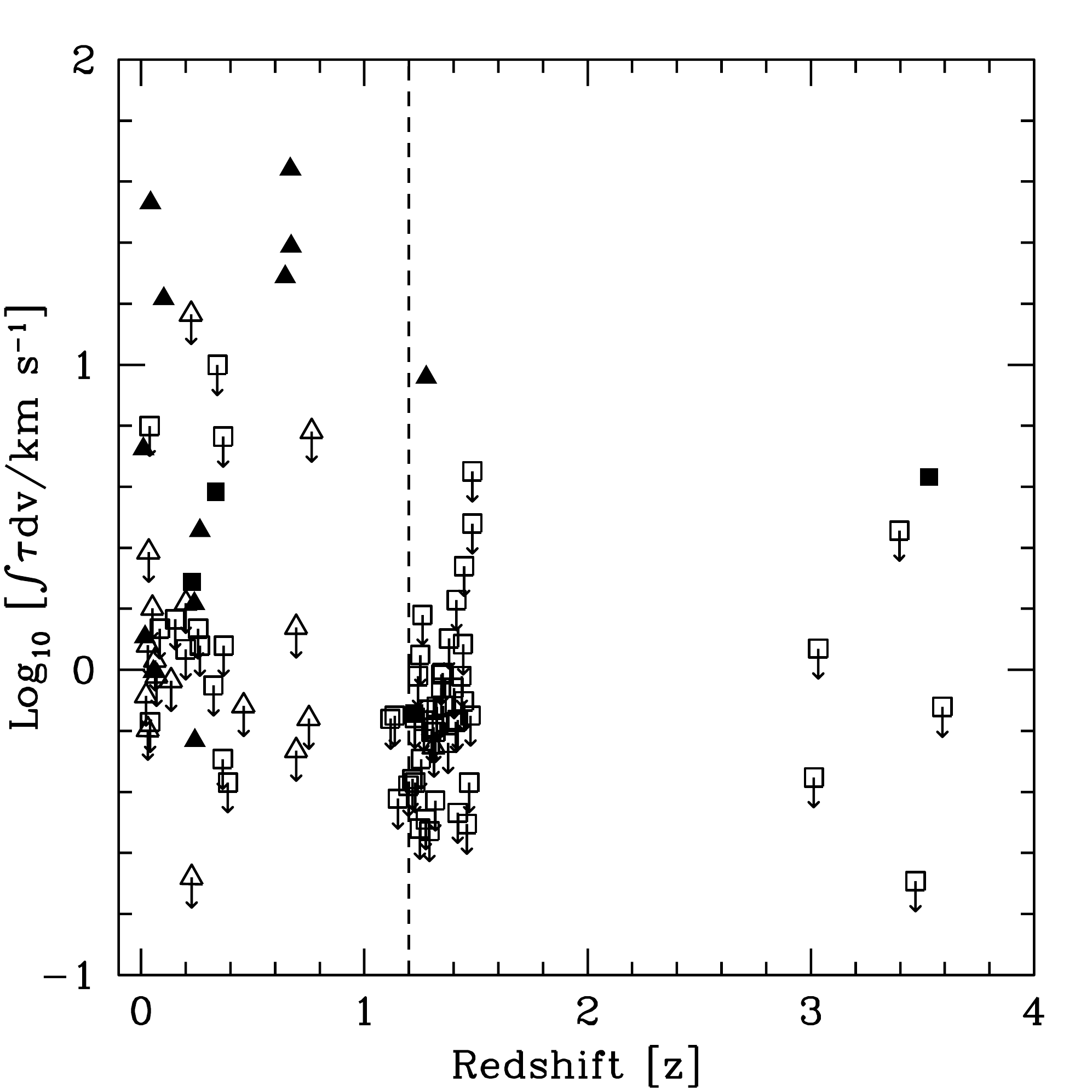}
\caption[]{The velocity-integrated \hii\ optical depth of the 92 CJF sources of the sample, 
plotted as a function of redshift. The 63 sources observed in the present GMRT survey are shown 
by squares, while the 29 sources from the literature are represented by triangles. 
Filled symbols indicate detections of \hii\ absorption, while open symbols indicate 
upper limits on the \hii\ optical depth. The dashed vertical line indicates the median redshift 
of the sample, $z_{\rm med} = 1.2$.
\label{fig:tau_vs_z}}
\end{figure}

\begin{figure}
\includegraphics[scale=0.45]{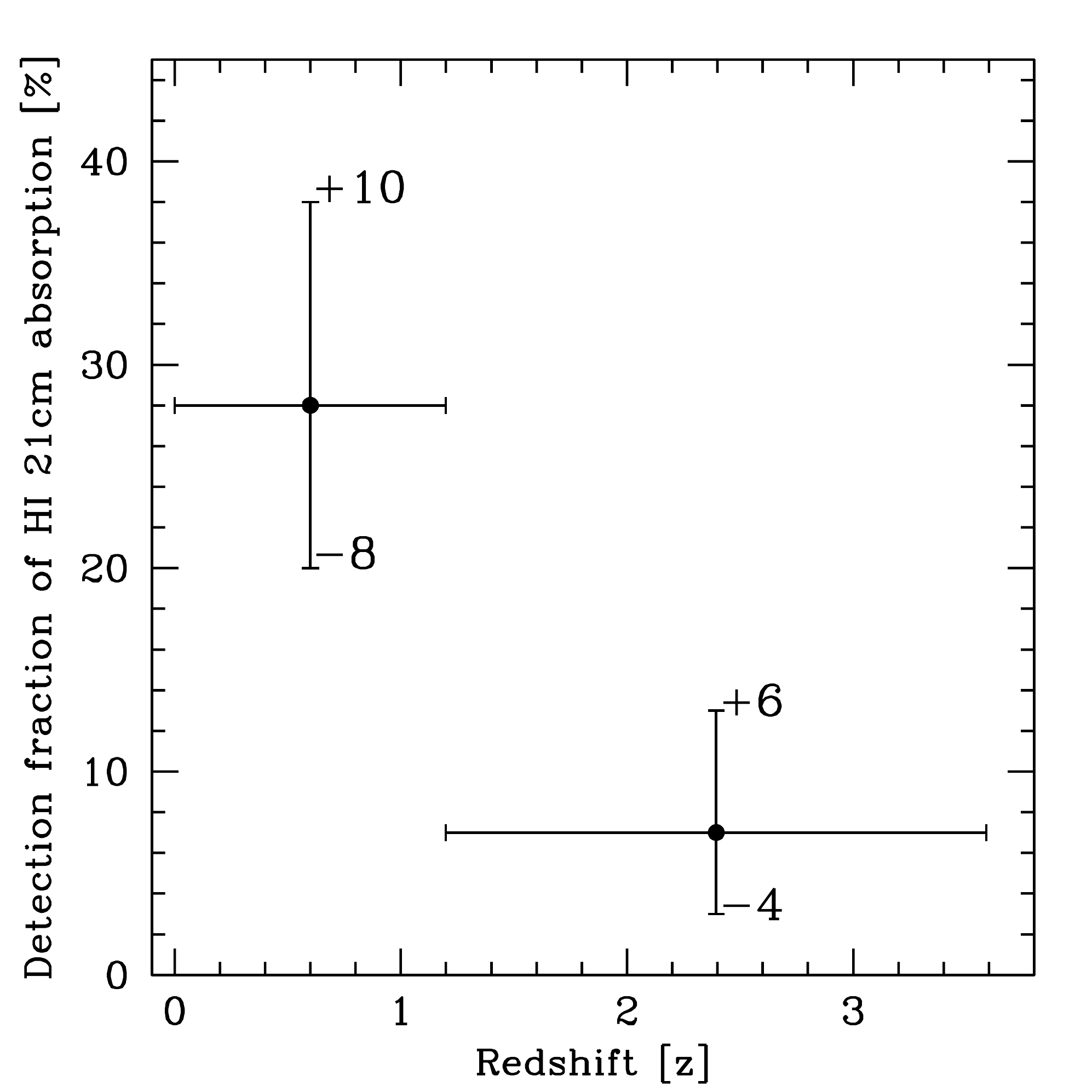}
\caption[]{The detection rate of \hii\ absorption for the low-$z$ and high-$z$ sub-samples,
separated at the median redshift. 
\label{fig:det_frac}}
\end{figure}

Figure~\ref{fig:tau_vs_z} plots the velocity-integrated \hii\ optical depth, in logarithmic units, against 
AGN redshift, for the full sample of 92 sources. It is clear that our GMRT observations, especially at 
$1.1 \lesssim z \lesssim 1.5$, are sufficiently sensitive to detect \hii\ opacities lower than those 
of most of the detections of \hii\ absorption. Further, most \hii\ detections are concentrated 
at low redshifts, $z < 1$, with 13 detections at $z < 1$, and just 3 detections at $z > 1$ 
\citep[including the tentative detection at $z \approx 3.530$ towards TXS~0604+728;][]{aditya16}. 
The median redshift of the sample is $z_{\rm med} = 1.200$. Dividing the sample at this redshift into 
low-$z$ and high-$z$ sub-samples, the former has 13 detections and 33 non-detections of \hii\ absorption, 
whereas the latter has 3 detections (one of which is tentative) and 43 non-detections. The detection 
rates of \hii\ absorption (see Fig.~\ref{fig:det_frac}) are $28^{+10}_{-8}\%$ and $7^{+6}_{-4}\%$ 
\citep[again estimating the $1\sigma$ errors from Poisson statistics;][]{gehrels86} for the 
$z < z_{\rm med}$ and $z > z_{\rm med}$ sub-samples, respectively; the high-$z$ detection rate is even lower, 
$4^{+6}_{-3}\%$, when the tentative detection towards TXS~0604+728 is excluded from the sample. The 
high-$z$ sub-sample thus has a lower \hii\ detection rate, albeit only at $\approx 2.1\sigma$ 
significance.

In addition to the \hii\ detection rates, we tested whether the distribution of the strength of 
the \hii\ absorption in AGN environments varies with redshift. For this purpose, we used survival 
analysis, in the {\sc asurv} package \citep[][]{isobe86}, to correctly include the upper limits on 
the \hii\ opacity. Within {\sc asurv}, the Peto-Prentice generalized Wilcoxon test finds that the 
null result that the velocity-integrated \hii\ optical depths of the low-$z$ and high-$z$ AGN 
sub-samples (again separated at $z_{\rm med}$) are drawn from the same distribution is ruled out at 
$3\sigma$ significance (increasing to $3.4\sigma$ significance when the tentative detection towards 
TXS~0604+728 is excluded from the sample). We thus find statistically significant evidence for 
redshift evolution in the strength of associated \hii\ absorption in a uniformly-selected AGN 
sample, with lower-redshift AGNs showing both a higher detection rate of \hii\ absorption and 
significantly higher integrated \hii\ optical depths.

The above redshift dependence of the strength of associated \hii\ absorption could stem from 
a variety of reasons: (1)~less neutral hydrogen in high-$z$ AGN environments, implying lower 
\hi\ column densities, (2)~higher gas spin temperatures in high-$z$ environments, as has been 
seen in ``intervening'' galaxies towards AGNs, the damped Lyman-$\alpha$ absorbers 
\citep[e.g.][]{kanekar03,kanekar14}, or (3)~lower covering factors in the high-$z$ AGN sample, 
yielding a lower {\it observed} \hii\ optical depth. Unfortunately, in the case of associated \hii\ 
studies, we do not have direct estimates of the \hi\ column density, and hence cannot separate
between the first two possibilities, a low \hi\ column density or a high gas spin temperature. 
We will initially consider the low covering factor hypothesis to account for the low detection rate 
of \hii\ absorption, before investigating whether AGN conditions might yield either a low gas content 
or a high spin temperature.


\subsection{Covering factor issues: The radio spectral index}
\label{sec:tau_alpha}

A possible cause for the lower strength of associated \hii\ absorption in high-$z$ AGNs 
is that the gas covering factor is systematically lower in the high-redshift systems.
In such a scenario, the observed difference between the \hii\ absorption properties of the 
low-$z$ and high-$z$ sub-samples would arise not due to changes in the properties (e.g. 
\hi\ column density or spin temperature) of neutral hydrogen in AGN environments, 
but due to differences in the structure of the radio emission in the high-$z$ and low-$z$
AGNs of the sample. Specifically, if the low-frequency radio emission of high-$z$ AGNs arises
primarily from extended structure (albeit still unresolved on GMRT baselines), which is not 
occulted by foreground gas clouds, the \hii\ optical depth estimated via the GMRT observations 
could be significantly lower than the true optical depth. For example, the radio emission 
of the low-$z$ AGNs at the redshifted \hii\ line frequency might predominantly arise from 
a compact radio core, while that of the high-$z$ AGNs might arise from either the radio jet 
or radio lobes.

The simplest way of testing the above scenario is to measure the fraction of radio 
emission arising from the AGN core at, or close to, the redshifted \hii\ line frequency,
via high-resolution VLBI imaging studies \citep[e.g.][]{kanekar09a,kanekar14}. This 
core fraction then gives a lower limit to the covering factor, under the assumption that 
the foreground gas clouds are likely to cover the core. Unfortunately, VLBI observations 
at frequencies $\lesssim 1$~GHz are technically challenging and are hence not available 
for most of the AGNs of our sample.

An alternative approach to addressing the covering factor issue is based on the fact
that the compact emission from the core tends to undergo synchrotron self-absorption 
and hence typically has an inverted or flat spectrum, while extended radio emission from the radio 
jet or the lobes tends to have a steep spectrum. If the radio emission of the high-$z$ AGNs 
at the redshifted \hii\ line frequency is dominated by the extended structure, yielding 
a low covering factor, one would expect these AGNs to have a systematically steeper spectral 
index at the \hii\ line frequency than the AGNs of the low-$z$ sub-sample. We note that, 
although our target AGNs have been uniformly chosen from the CJF sample, with flat spectral 
indices, $\alpha \geq -0.5$ \citep[][]{taylor96}, the CJF spectral index criterion is based 
on the AGN flux densities at two relatively high frequencies, 1.4~GHz and 4.85~GHz. It is 
hence possible that the radio emission at the low redshifted \hii\ line frequency is 
dominated by steep-spectrum extended structure, rather than by the flat- or inverted-spectrum 
radio core.

We will hence use the AGN's radio spectral index $\alpha_{\rm 21-cm}$ at the redshifted \hii\ 
frequency as a proxy for the compactness of the AGN. A flat or an inverted spectrum near
the redshifted \hii\ line frequency ($\alpha_{\rm 21-cm} \gtrsim 0$) would indicate a core-dominated 
source, and a relatively high covering factor ($f \approx 1$), whereas a steep spectrum 
($\alpha_{\rm 21-cm} \lesssim -0.7$) would indicate extended radio structure and a possibly low 
covering factor ($f \ll 1$).  If the measured \hii\ optical depths are found to depend on 
the spectral index, or if the 
high-$z$ AGN sub-sample has a systematically steeper spectral index than the low-$z$ 
sample, it would suggest that a low covering factor may be the cause of the lower observed 
\hii\ optical depths at high redshifts.

\begin{figure*}
\includegraphics[scale=0.4]{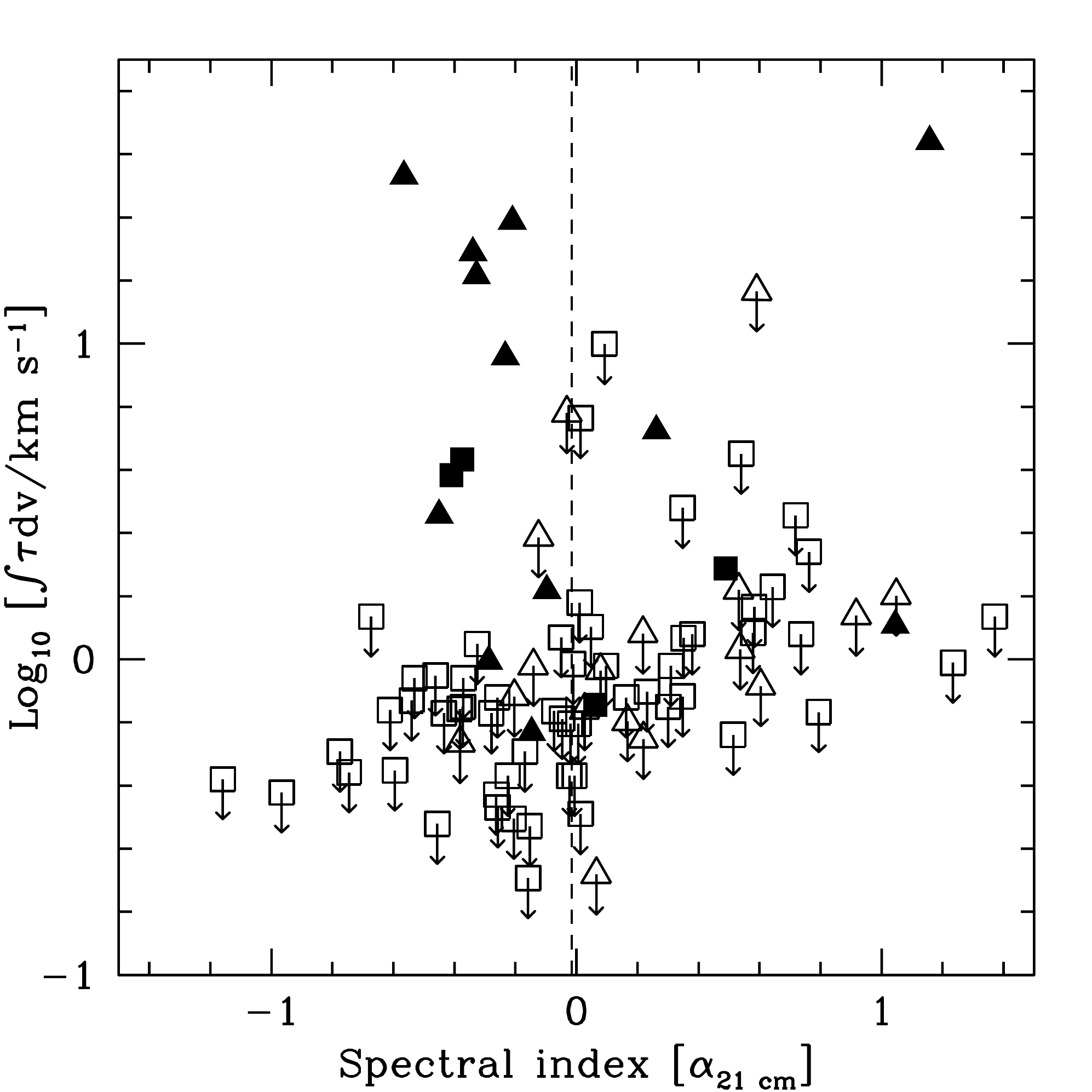}
\includegraphics[scale=0.4]{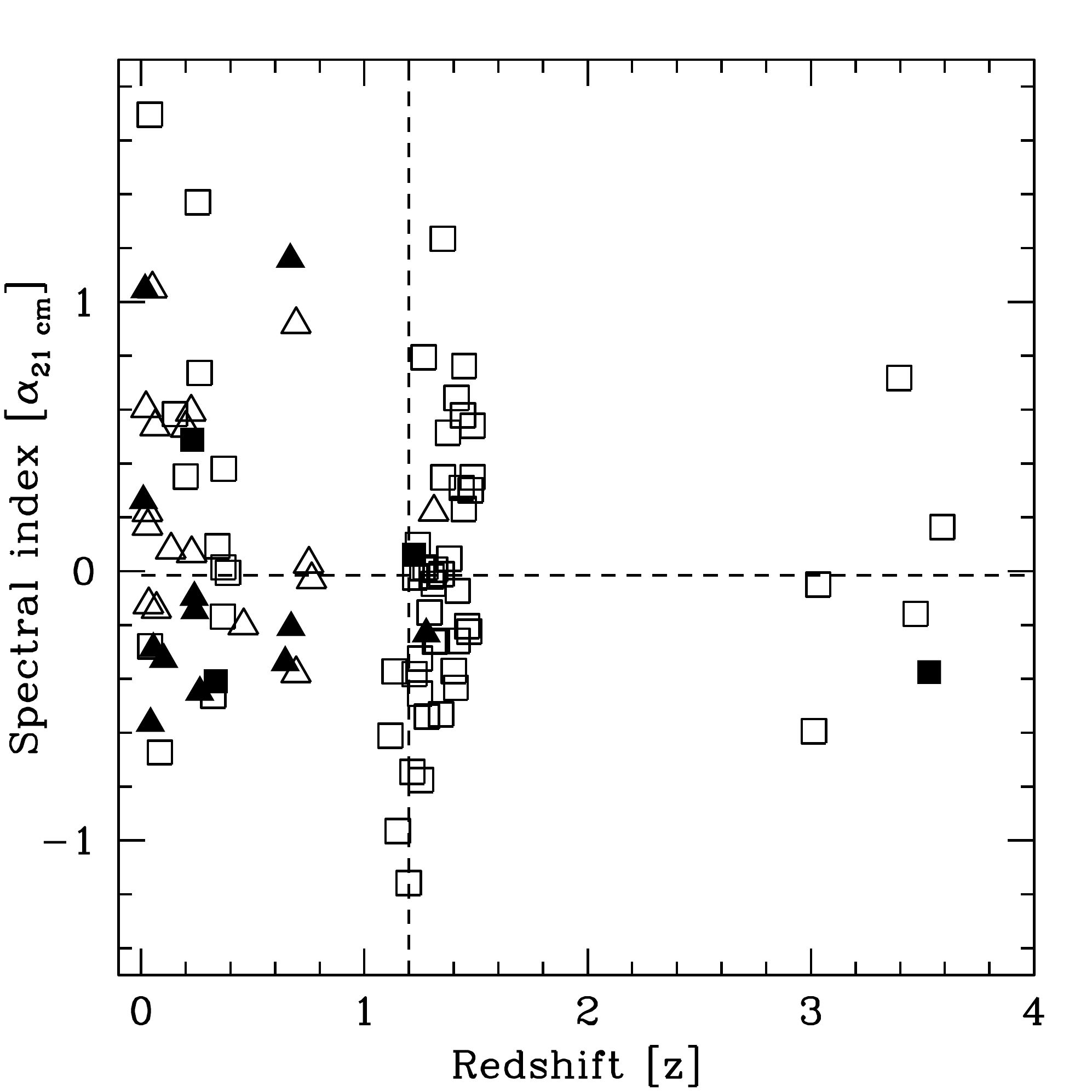}
\caption{[A]~Left panel: The integrated \hii\ optical depth of the 92 CJF sources plotted against the 
	spectral index, $\alpha_{\rm 21-cm}$, around the redshifted \hii\ line frequency; the dashed vertical 
	line indicates the median spectral index, $\alpha_{\rm 21-cm,med} = -0.015$. 
	[B]~Right panel: The spectral index $\alpha_{\rm 21-cm}$, plotted against AGN redshift; the dashed 
	vertical line indicates the median redshift, $z_{\rm med} = 1.2$, while the dashed horizontal line indicates 
	the median spectral index, $\alpha_{\rm 21-cm} = -0.015$. In both panels, the 63 AGNs of our survey are 
	shown as squares, while the 29 literature sources are shown as triangles. Filled symbols represent 
	detections, while open symbols represent upper limits on the integrated \hii\ optical depth. See
	main text for discussion.
\label{fig:alpha}}
\end{figure*}  

Figure~\ref{fig:alpha}[A] shows the integrated \hii\ optical depths of the 92 CJF sources 
of our sample plotted against $\alpha_{\rm 21-cm}$. While there are clearly a few sources with 
$\alpha_{\rm 21-cm} \lesssim -0.5$, most of the AGNs are seen to have flat radio spectra, $\alpha_{\rm 21-cm} \approx  0$.
Indeed, the median spectral index of the sample, $\alpha_{\rm 21-cm,med} = -0.015$, is very close to zero, 
indicating that the sample is dominated by compact objects, with $\alpha_{\rm 21-cm} > -0.5$. A Peto-Prentice two-sample 
test finds that the distributions of the \hii\ optical depths of the two sub-samples, separated at the 
median $\alpha_{\rm 21-cm}$, are consistent (within $\approx 1.3\sigma$ significance) with the null hypothesis 
of being drawn from the same underlying distribution. We thus find no evidence for a dependence of the 
strength of the associated \hii\ absorption on the AGN spectral index.

Figure~\ref{fig:alpha}[B] shows the low-frequency AGN spectral index $\alpha_{\rm 21-cm}$ plotted against
redshift; no trend is apparent in the figure. A Gehan-Wilcoxon two-sample test comparing the distributions 
of $\alpha_{\rm 21-cm}$ values of the low-$z$ and the high-$z$ sub-samples, separated at $z_{\rm med} = 1.2$, finds 
that the data are consistent (within $\approx 1.2\sigma$ significance) with the null hypothesis of being 
drawn from the same underlying distribution. We thus find no evidence for a systematic difference between 
the spectral indices of the AGNs of the high-$z$ and the low-$z$ sub-samples. 
  
In summary, we find no statistically significant evidence either that the strength of the \hii\ absorption
depends on the low-frequency AGN spectral index or that the low-frequency spectral index varies systematically 
with redshift. Both of these would have been expected if low covering factors are the cause of the weaker 
\hii\ absorption observed in the high-$z$ AGN sub-sample. We hence conclude that it is unlikely that the 
observed differences in the \hii\ absorption properties of the low-$z$ and high-$z$ AGN sub-samples 
can be explained by covering factor issues.

\subsection{The AGN colour: Evidence for dust reddening?}
\label{sec:dust}

\begin{figure}
\includegraphics[scale=0.4]{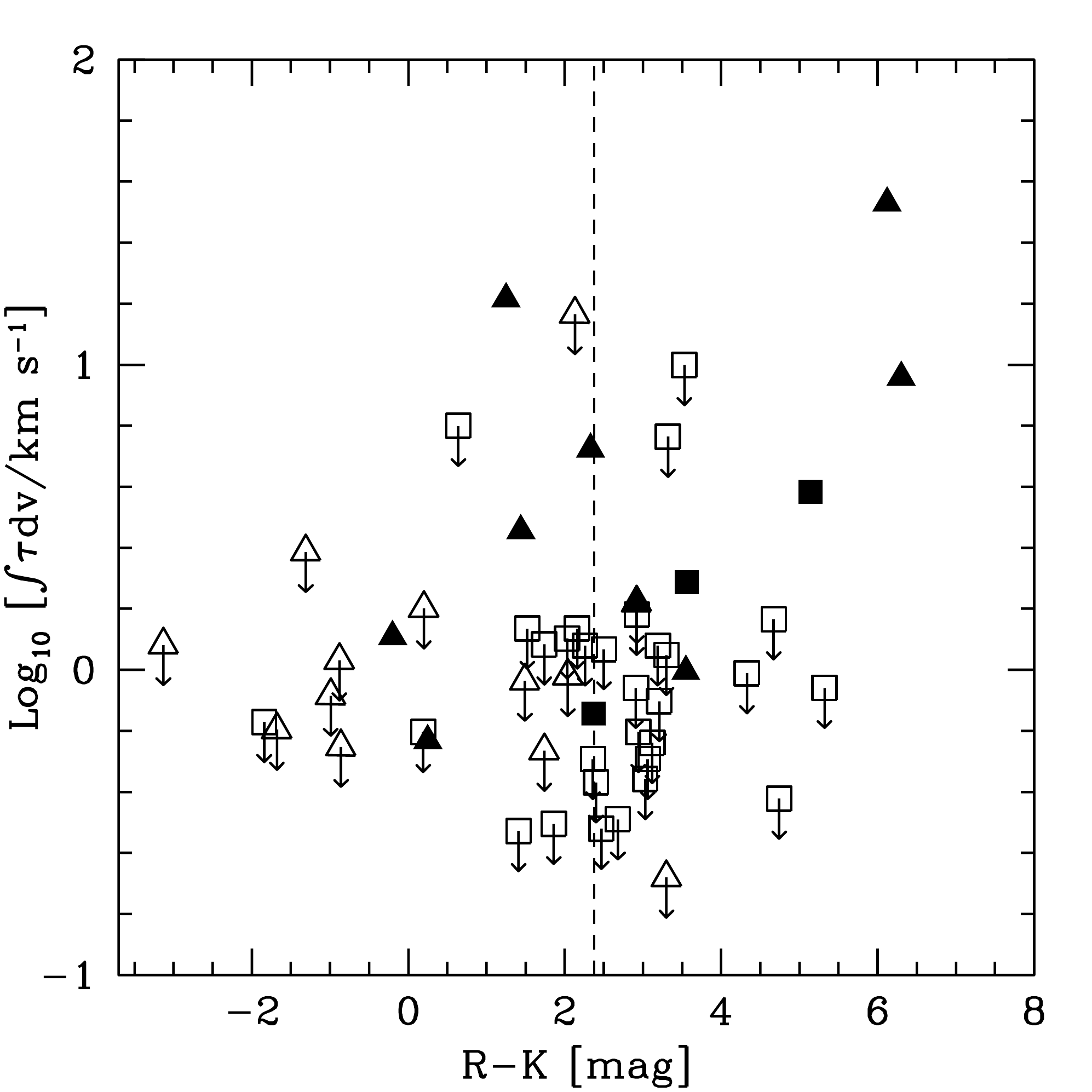}
\caption[]{The integrated \hii\ optical depth, in logarithmic units, plotted against the 
(R$-$K) colour for the 58 AGNs of the sample with NIR photometry. The dashed vertical line indicates 
the median colour, $\rm (R$-$K) = 2.38$. The sources from our survey and the literature are shown as 
squares and triangles, respectively. Detections of \hii\ absorption are shown as filled symbols 
and non-detections as open symbols. See main text for discussion.
	\label{fig:tau_rk}}
\end{figure}

For red quasars, the high extinction at optical wavebands is believed to be caused by dust extinction 
\citep[e.g.][]{webster95}. However, it has also been suggested in the literature that not all red 
quasars are dusty systems \citep[e.g.][]{benn98}. Earlier \hii\ studies have yielded ambiguous 
results: for example, \citet[][]{carilli98} detected strong \hii\ absorption in four out of five 
red quasars at intermediate redshifts ($z \approx 0.7$), suggesting a high \hi\ column consistent
with the dust obscuration hypothesis. Strong molecular absorption has also only been seen in very red 
AGNs \citep[e.g.][]{wiklind94,wiklind96,kanekar02,kanekar05}. However, later \hii\ absorption studies have 
found no significant correlation between the detectability of associated \hii\ absorption and the AGN colour 
\citep[e.g.][]{curran08,aditya16}, suggesting that the reddening may have other causes besides 
dust obscuration. Our detection of associated \hii\ absorption towards TXS~1456+375 is consistent with 
the dust-reddening hypothesis.

It has long been known that the Galactic extinction at 5500~\AA\ is proportional to the total 
hydrogen column density, with $A_{v}=6.29 \times 10^{-22} \left[ N_{\rm HI} + 2N_{\rm H_2} \right]$ 
\citep[e.g.][]{savage77}. This is understood as arising due to reddening produced by the large amounts 
of dust associated with a high hydrogen column. \citet{webster95} hence suggested that ``red'' quasars, 
with a steeper spectral index between the optical and NIR wavebands than typical quasars, are likely to 
acquire their redder colours due to dust extinction. Consistent with this hypothesis, \citet{carilli98}
found that 80\% of red quasars showed either associated or intervening \hii\ absorption, while only 
11\% of optically-selected Mg{\sc ii} absorbers showed radio absorption. Similarly, all five of the known 
redshifted radio molecular absorbers at $z > 0.2$ have background quasars with extremely red 
colours, $\rm (R$-$K) > 4$ \citep{wiklind94,wiklind95,wiklind96,wiklind97,kanekar02,kanekar05,curran06b,curran08}.
Our detection of associated \hii\ absorption towards TXS~1456+375 is consistent with the dust-reddening hypothesis.

However, it has also been noted in the literature that not all red quasars appear to be dusty systems
\citep[e.g.][]{benn98}. Indeed, recent searches for associated \hii\ absorption in red AGNs have
not been very successful \citep[e.g.][]{yan16}, while no significant correlation has been found between 
the detectability of associated \hii\ absorption and the AGN colour \citep[e.g.][]{curran08,aditya16}, 
suggesting that the reddening may have other causes besides dust obscuration. We note that the sample of 
\citet{carilli98} was very small (five systems), and limited to low redshifts, $z \lesssim 0.7$.

In this section, we test the hypothesis that associated \hii\ absorption is more likely to 
arise in red AGNs, by examining our sample for a correlation between the strength of the 
\hii\ absorption with AGN (R$-$K) colour. Unfortunately, NIR photometry was available for 58 AGNs 
of the sample and so the present analysis is restricted to these 58 systems.

At the outset, we note that the choice of a flat-spectrum sample is likely to be biased against 
the reddest sightlines. In standard unification schemes \citep[e.g.][]{urry95}, a flat AGN spectrum 
is expected to mostly arise from sightlines closer to orthogonal to the obscuring torus. Such sightlines 
are unlikely to be affected by extinction from dust in the torus, and so we do not expect very 
large (R$-$K) values, $\rm (R$-$K) \gtrsim 5$, in our sample. 

Fig.~\ref{fig:tau_rk} shows the integrated \hii\ optical depth plotted versus the (R$-$K) colour 
for the 58 AGNs with NIR photometry. Of the 12 AGNs with detections of \hii\ absorption,
five have relatively red colours, (R$-$K)~$> 3$. Further, three of these five systems lie at the top 
right of the figure, indicating both red colours and high integrated \hii\ optical depths; indeed,
these three AGNs have $(R$-$K) > 5$, comparable to the colours of the AGNs that show molecular 
absorption! At least for these systems, the red colours are likely to arise due to the presence 
of high columns of gas and associated dust at the AGN redshift. However, we also note 
that there are four AGNs that show \hii\ absorption at relatively low (R$-$K) values, 
(R$-$K)~$\lesssim 1.3$, and that one of these has the second-highest integrated \hii\ optical 
depth of the full sample.

To test the dependence of the strength of the \hii\ absorption on the (R$-$K) colour, we 
divided the sample of 58 systems at the median (R$-$K) value, $\rm (R$-$K)= 2.38$, and carried 
out two-sample tests on the high-(R$-$K) and low-(R$-$K) sub-samples. A Peto-Prentice two-sample 
test for censored data finds that the null hypothesis that the sub-samples are drawn from the 
same underlying distribution is rejected at only $1.4\sigma$ significance. The two sub-samples 
are thus consistent with being drawn from the same distribution, and we find no statistically 
significant evidence for a dependence of the strength of \hii\ absorption on the (R$-$K) colour 
of the AGN. Thus, while three of the four highest integrated \hii\ optical depths do arise 
from the reddest AGNs, our results from the CJF sample do not provide support for the 
dust-reddening hypothesis.

\subsection{Effects of the AGN luminosity}
\label{sec:uv_radio_lum}

\begin{figure*}
\includegraphics[scale=0.4]{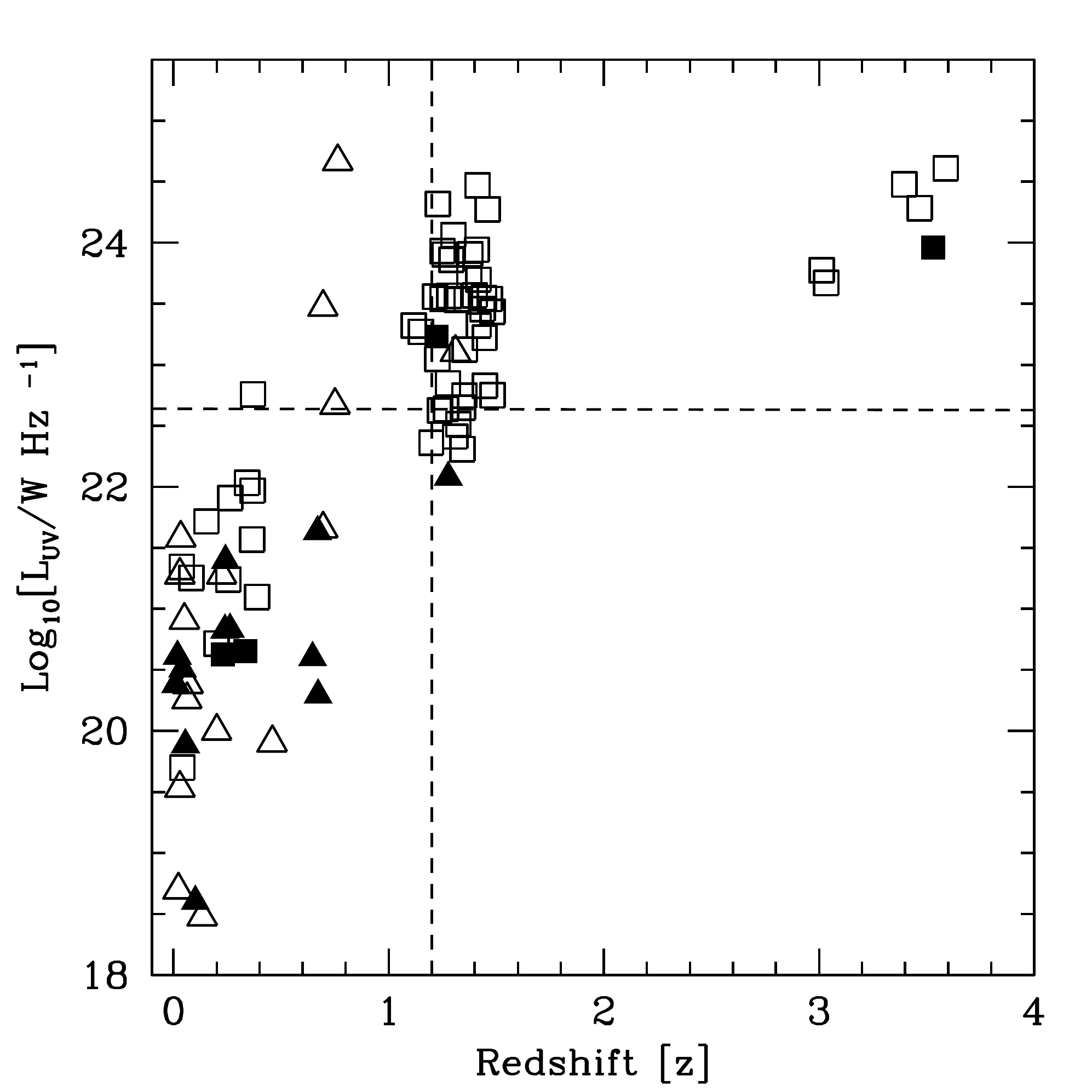}
\includegraphics[scale=0.4]{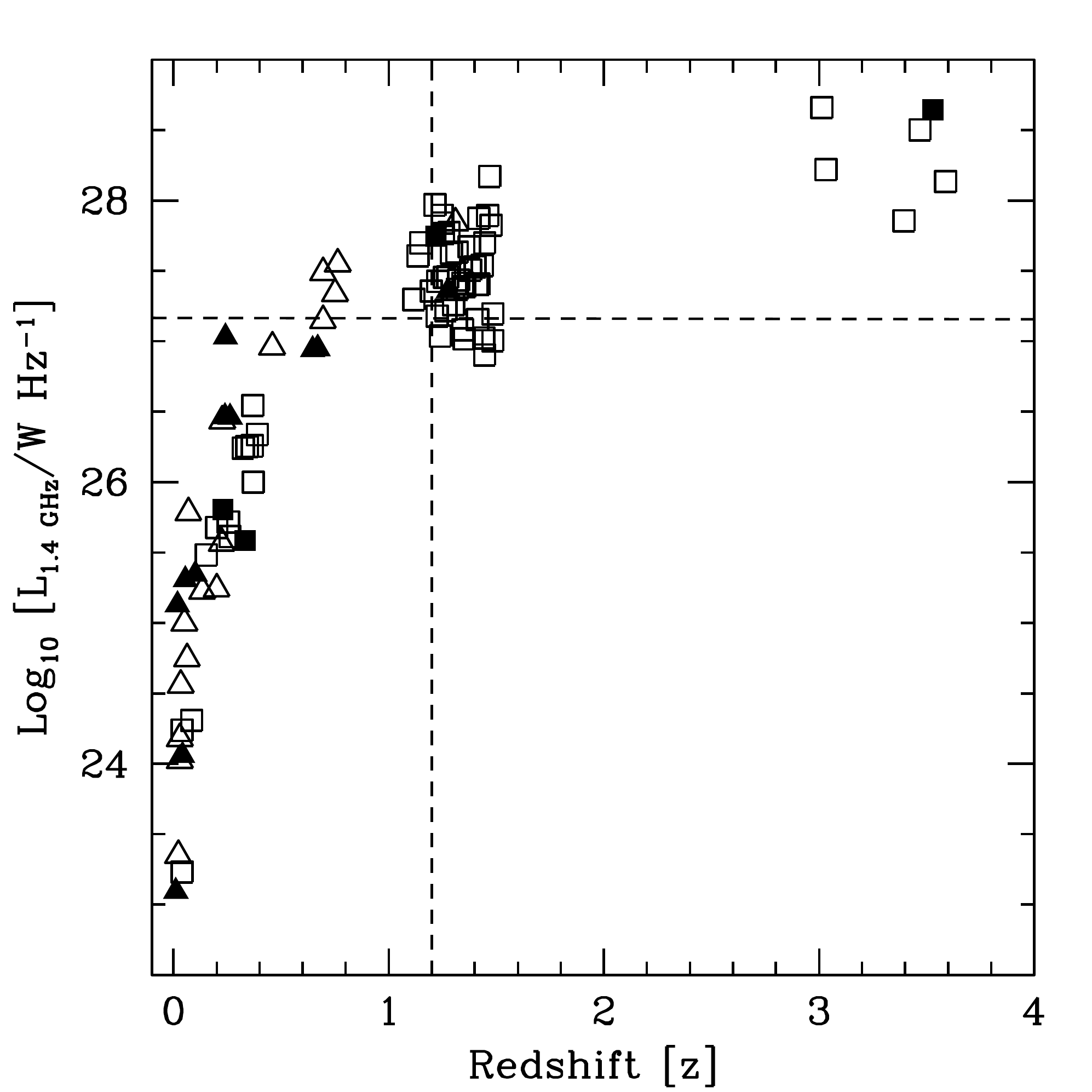}
	\caption{The AGN luminosities of the CJF sources of the sample in [A]~(left panel) at rest-frame 
	 UV 1216~\AA\ (87 sources) and [B]~(right panel) rest-frame radio~1.4~GHz (92 sources), plotted,
	 in logarithmic units, against the AGN redshift. The dashed vertical lines indicate the median 
	 redshift, $z_{\rm med}$, while the dashed horizontal lines indicate the median UV 1216~\AA\ and radio 1.4~GHz 
	 luminosities ($L_{UV,med} = 10^{22.64}$~W~Hz$^{-1}$ and $L_{1.4\;GHz,med} = 10^{27.17}$~W~Hz$^{-1}$, 
	 respectively). The squares and triangles represent, respectively, the sources from our survey and the 
	 literature. Filled and open symbols represent, respectively, \hii\ detections and upper limits on the 
	 integrated \hii\ optical depth. See text for discussion.}
\label{fig:lum_z}
\end{figure*}

As mentioned briefly in Section~\ref{sec:sample}, a high AGN luminosity can adversely affect the 
strength of the associated \hii\ absorption \citep[e.g.][]{curran08,curran13}. This is because a
high UV luminosity can cause ionization of the nearby neutral hydrogen, thus reducing the \hi\ 
column density, and can also affect the spin temperature \citep[which is coupled to the colour 
temperature of the Lyman-$\alpha$ radiation field; ][]{wouthuysen52,field58,field59}. Similarly, a high 
AGN rest-frame 1.4~GHz luminosity can raise the gas spin temperature \citep[][]{field58,field59}. 
Both a decrease in the \hi\ column density and a raising of the spin temperature would reduce 
the strength of the \hii\ absorption. Thus, if the higher-redshift AGNs of our sample tend to have 
higher UV and/or radio luminosities, this would provide an alternative explanation of the weaker 
\hii\ absorption in the high-$z$ sub-sample.

Figs.~\ref{fig:lum_z}[A] and [B] show, respectively, the rest-frame UV 1216~\AA\ luminosity and the 
rest-frame 1.4~GHz radio luminosity of the sources of the CJF sample plotted against redshift. In both 
panels, the vertical dashed line is at the median redshift of the sample, while the horizontal dashed line
is at the median luminosity. It is clear that the high-$z$ AGNs of the sample typically lie close to
or above the median luminosity (in both UV and radio wavebands), while the low-$z$ AGNs mostly lie 
below the median luminosity. Again dividing the sample at $z_{\rm med} = 1.2$, a Gehan-Wilcoxon test finds 
that the null hypothesis that the rest-frame UV and radio luminosities of the two sub-samples are drawn from 
the same distribution is rejected, respectively, at $\approx 7.7\sigma$ and $\approx 7.8\sigma$ significance. 
As such, it is clear that the sample contains a strong bias towards higher UV and radio luminosities at 
high redshifts. We note, in passing, that the rest-frame UV $1216$~\AA\ and radio 1.4~GHz luminosities 
of the sources in our sample are strongly correlated (at $\approx9\sigma$ significance, via a Kendall-tau 
test).

\begin{figure*}
\includegraphics[scale=0.4]{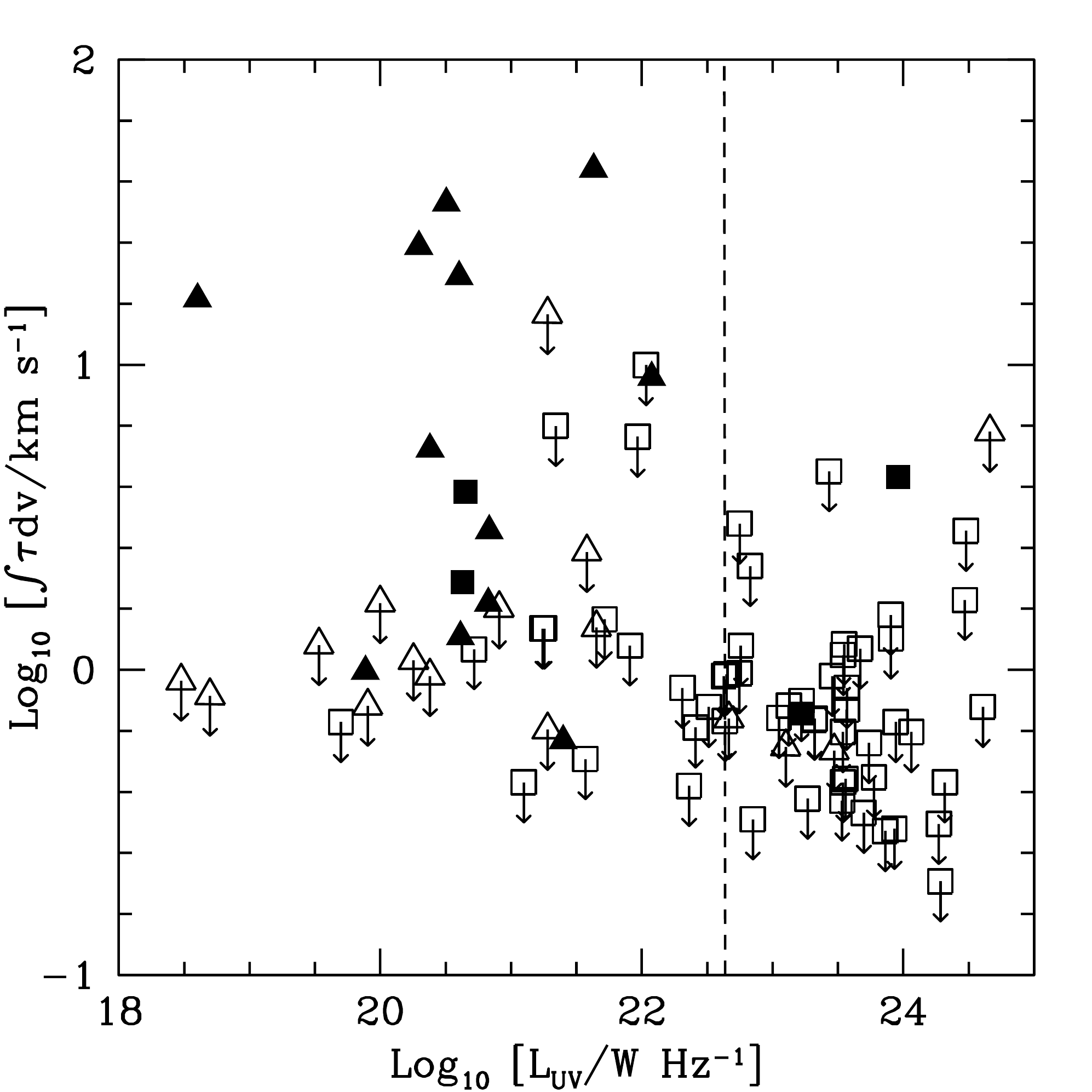}
\includegraphics[scale=0.4]{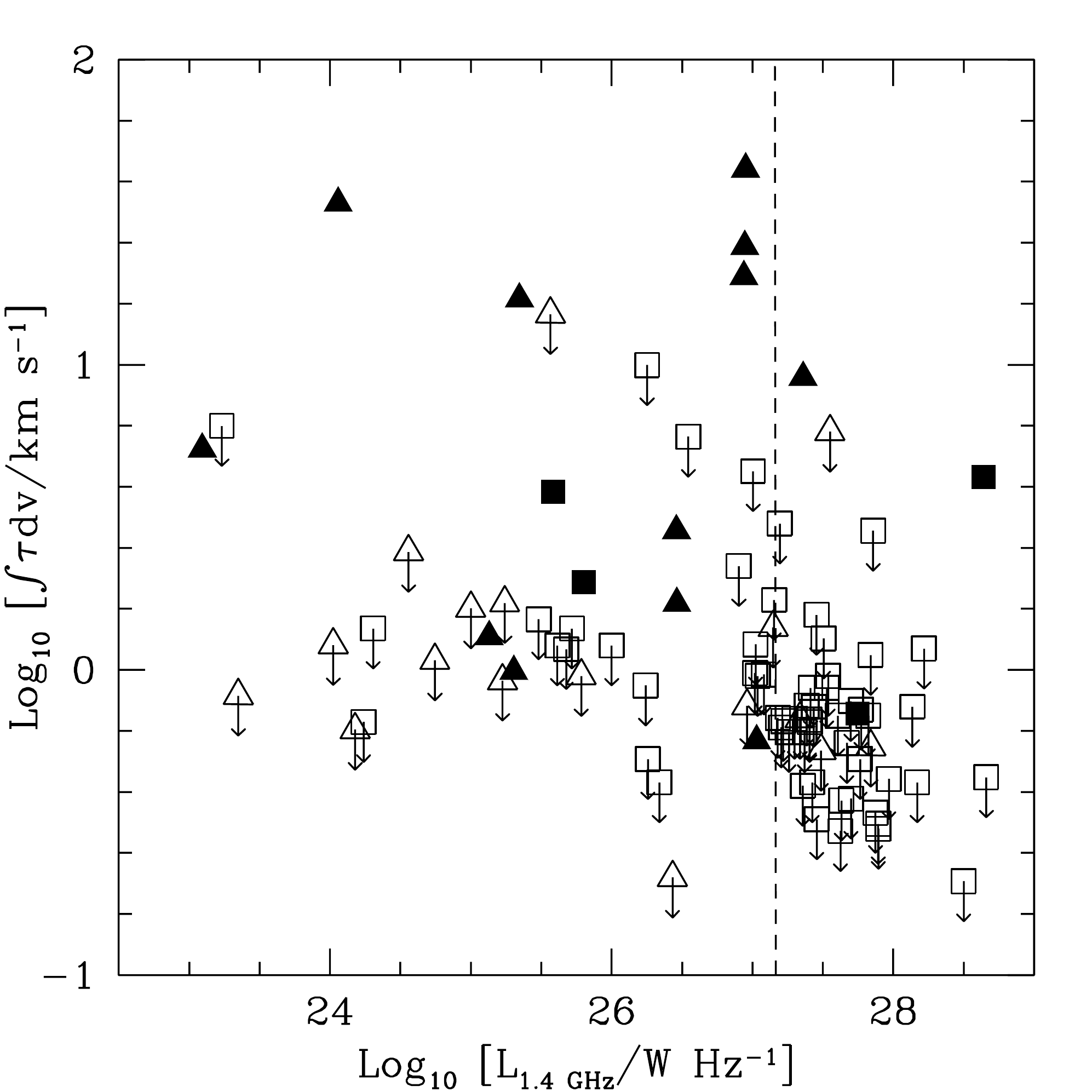}
\caption[]{The integrated \hii] optical depths of the AGNs of the our sample plotted 
	against [A]~(left panel) the rest-frame UV 1216~\AA\ luminosity and [B]~(right panel) the 
	rest-frame radio 1.4~GHz luminosity, with all quantities in logarithmic units. The dashed vertical 
	lines indicate the median UV 1216~\AA\ and radio 1.4~GHz luminosities 
	($L_{UV,med} = 10^{22.64}$~W~Hz$^{-1}$ and $L_{1.4\;GHz,med} = 10^{27.17}$~W~Hz$^{-1}$, respectively).
	The sources from our survey and the literature are represented by squares and triangles, respectively. 
	Filled and open symbols represent, respectively, detections and upper limits on the \hii\ optical depth.
\label{fig:tau_lum}}
\end{figure*}

It thus appears that the apparent redshift evolution in the strength of the associated \hii\
absorption might also arise due to differences in the UV and/or radio luminosities of the AGNs 
of the low-$z$ and the high-$z$ sub-samples. Figs.~\ref{fig:tau_lum}[A] and [B] plot the 
integrated \hii\ optical depth versus, respectively, the rest-frame UV 1216~\AA\ luminosity 
and the rest-frame radio 1.4~GHz luminosity (with all quantities in logarithmic units).
The dashed vertical lines in the two figures indicate the median luminosities, 
$L_{UV,med} = 10^{22.64}$~W~Hz$^{-1}$ and $L_{1.4\;GHz,med} = 10^{27.17}$~W~Hz$^{-1} $, 
respectively. It is clear that most of the detections of \hii\ absorption lie in the low-luminosity 
halves of the two figures; further, the integrated \hii\ optical depths towards low-luminosity AGNs 
appear higher than the typical $3\sigma$ upper limits on the integrated \hii\ optical depths towards 
AGNs with high luminosities.  Formally, a Peto-Prentice two-sample test (for censored data) finds 
that the null hypotheses that the \hii\ optical depth distributions of the low-luminosity and 
high-luminosity sub-samples are drawn from the same distribution is rejected at, respectively, 
$\approx 3.6\sigma$ and $\approx 3.3\sigma$ significance, for the rest-frame UV $1216$~\AA\ and 
rest-frame radio 1.4~GHz luminosities.

We thus find statistically significant evidence for a dependence of the strength of associated \hii\ 
absorption in the CJF sample on both redshift and AGN luminosity in the rest-frame UV 1216~\AA\ and radio 
1.4~GHz wavebands, but not on the low-frequency radio spectral index or the (R$-$K) colour. Weaker \hii\ 
absorption is obtained at higher redshifts and higher radio and UV luminosities. Unfortunately, there is 
a strong correlation between AGN luminosity and redshift in the AGNs of our sample (see Figs.~\ref{fig:lum_z}[A] 
and [B]), due to which it is not currently possible to break the degeneracy between redshift and luminosity,
and identify the primary cause, if any, for the differences in the strength of the \hii\ absorption. 
Searches for \hii\ absorption in either a low-luminosity AGN sample at high redshifts, or a 
high-luminosity sample at low redshifts would be required to break the present degeneracy.

\section{Summary}
\label{sec:summary}

We have used the GMRT to carry out a search for associated \hii\ absorption in 50 flat-spectrum 
AGNs, selected from the CJF sample. The data on ten AGNs were rendered unusable by RFI. 
We obtained new detections of \hii\ absorption in two sources, at $z = 0.229 $ towards 
TXS~0003+380 and $z =0.333$ towards TXS~1456+375, and also confirmed an earlier detection 
\citep[by][]{curran13} of \hii\ absorption at $z = 1.277$ towards TXS~1543+480. The measured 
velocity-integrated \hii\ optical depths towards the above three AGNs lie in the range 
$\int \tau d{\rm V} \approx 1.9 - 9.6$~\kms, implying \hi\ column densities of 
$\approx (3.5 - 17.5) \times 10^{20}$~cm$^{-2}$, for an assumed spin temperature of $100$~K and 
covering factor of unity. For the remaining 37 AGNs, the $3\sigma$ upper limits on the integrated 
\hii\ optical depth range from $0.3 - 14$~\kms, with a median value of $\approx 0.97$~\kms.

Our full sample of CJF sources with searches for redshifted \hii\ absorption consists of 
92 AGNs, 63 from our survey \citep[including 24 sources from][]{aditya16,aditya17} and 29 from 
the literature. This is currently the largest sample of uniformly-selected AGNs with 
searches for associated \hii\ absorption, with 16 \hii\ detections and 76 non-detections,
at redshifts $0.01-3.6$, and with a median redshift of $z_{\rm med} = 1.2$.  

We find that both the strength and the detectability of \hii\ absorption appear higher at 
low redshifts, $z < z_{\rm med}$. The detection rate of \hii\ absorption is $28^{+10}_{-8}$\% 
for the low-$z$ AGN sub-sample (with $z < 1.2$), but only $7^{+6}_{-4}$\% for the high-$z$ 
sub-sample (with $z > 1.2$). While the difference in detection rates has only $\approx 2.1\sigma$ 
significance, a Peto-Prentice two-sample test on the velocity integrated \hii\ optical depths 
finds that the null hypothesis that the low-$z$ and high-$z$ sub-samples are drawn from the 
same distribution is ruled out at $\approx 3\sigma$ significance. We thus obtain 
statistically-significant evidence for redshift evolution in the strength of associated 
\hii\ absorption in the Caltech-Jodrell Flat-spectrum AGN sample.

However, we also found evidence for a significant bias in the intrinsic luminosities 
of the AGNs of our sample, with the high-$z$ AGNs having higher rest-frame 1216~\AA\ UV and 
1.4~GHz radio luminosities. Examining the dependence of the strength of the \hii\ absorption 
on AGN luminosity, the null hypothesis that the velocity-integrated \hii\ optical depths of 
the high-luminosity and low-luminosity AGNs arise from the same distribution is ruled 
out at $\approx (3.3-3.6)\sigma$ significance in a Peto-Prentice two-sample test (for the 
1216~\AA\ UV and 1.4~GHz radio luminosities). 

We also examined the possibility that the lower strength of \hii\ absorption in high-$z$ AGNs
might arise due to a typically lower covering factor for the high-$z$ sub-sample. This 
could occur if the radio continuum at the redshifted \hii\ line frequency for the high-$z$ 
sub-sample is dominated by extended emission. We used the AGN spectral index around the 
redshifted \hii\ line frequency as a proxy for source compactness, since extended emission
is expected to have a steep spectrum. We find no evidence in two-sample tests that the 
strength of the \hii\ absorption depends on AGN spectral index, or that the spectral index 
itself depends on the AGN redshift. It is hence unlikely that the observed weakness of the 
\hii\ absorption in high-$z$ AGNs arises due to a low covering factor.

We find no statistically-significant evidence that the strength of \hii\ absorption 
depends on AGN colour. However, five of the 12 AGNs with \hii\ detections and estimates of 
the (R$-$K) colour show relatively red colours, with (R$-$K)~$>3$. Three of these systems have 
 both high integrated \hii\ optical depths and (R$-$K)~$> 5$; for these, the red colour is 
likely to arise due to dust at the AGN redshift. 

The above results are consistent with those from our pilot GMRT \hii\ absorption survey of the 
CJF sample \citet{aditya16}, but are now based on a significantly larger AGN sample (92 AGNs
against 52 in \citealt{aditya16}), and with more than half the sample at $z > 1$. The median redshift 
of the present sample ($z_{\rm med} = 1.2$) is also significantly higher than that 
($z_{\rm med} = 0.76$) of \citet{aditya16}.

In summary, the strength of associated \hii\ absorption in the CJF AGN sample appears to depend 
on both redshift and AGN luminosity, with weaker \hii\ absorption 
at high redshifts and high luminosities. This could arise due to (1)~lower amounts of neutral
 gas in high-$z$, high-luminosity AGN environments, due to either redshift evolution or ionization 
of the \hi\ by the far-UV AGN radiation, or (2)~higher spin temperatures in high-$z$, high-luminosity 
AGN environments, due to either a preponderance of warm neutral gas around high-$z$ AGNs or 
spin temperatures greater than the kinetic temperature due to the high AGN UV/radio luminosity. 
Unfortunately, the luminosity bias in our sample, with the higher-luminosity AGNs located at higher 
redshifts, implies that the present dataset does not allow us to distinguish between the 
above possibilities to identify whether redshift or AGN luminosity is the primary driving 
factor in determining the strength of the associated \hii\ absorption; this will be the focus 
of future studies.

\section*{Acknowledgements}
We thank the staff of the GMRT who have made these observations possible. The GMRT is run by the National Centre 
for Radio Astrophysics of the Tata Institute of Fundamental Research. NK acknowledges support from the 
Department of Science and Technology via a Swarnajayanti Fellowship (DST/SJF/PSA-01/2012-13). 
JA was supported during a part of this work by the Indo-French Centre for the Promotion of Advanced 
Research under Project 5504-B (PIs: N. Gupta, P. Noterdaeme).


\bibliographystyle{mnras}
\bibliography{ms}

\appendix 

\section{}

\begin{figure*}
\caption[]{The GMRT spectra for the 37 CJF sources with non-detections of \hii\ absorption. All spectra 
	have been Hanning-smoothed and resampled. The shaded channels in the spectra are corrupted by RFI.
	\label{fig:nondetect}}
\begin{tabular}{ccc}

\includegraphics[scale=0.28]{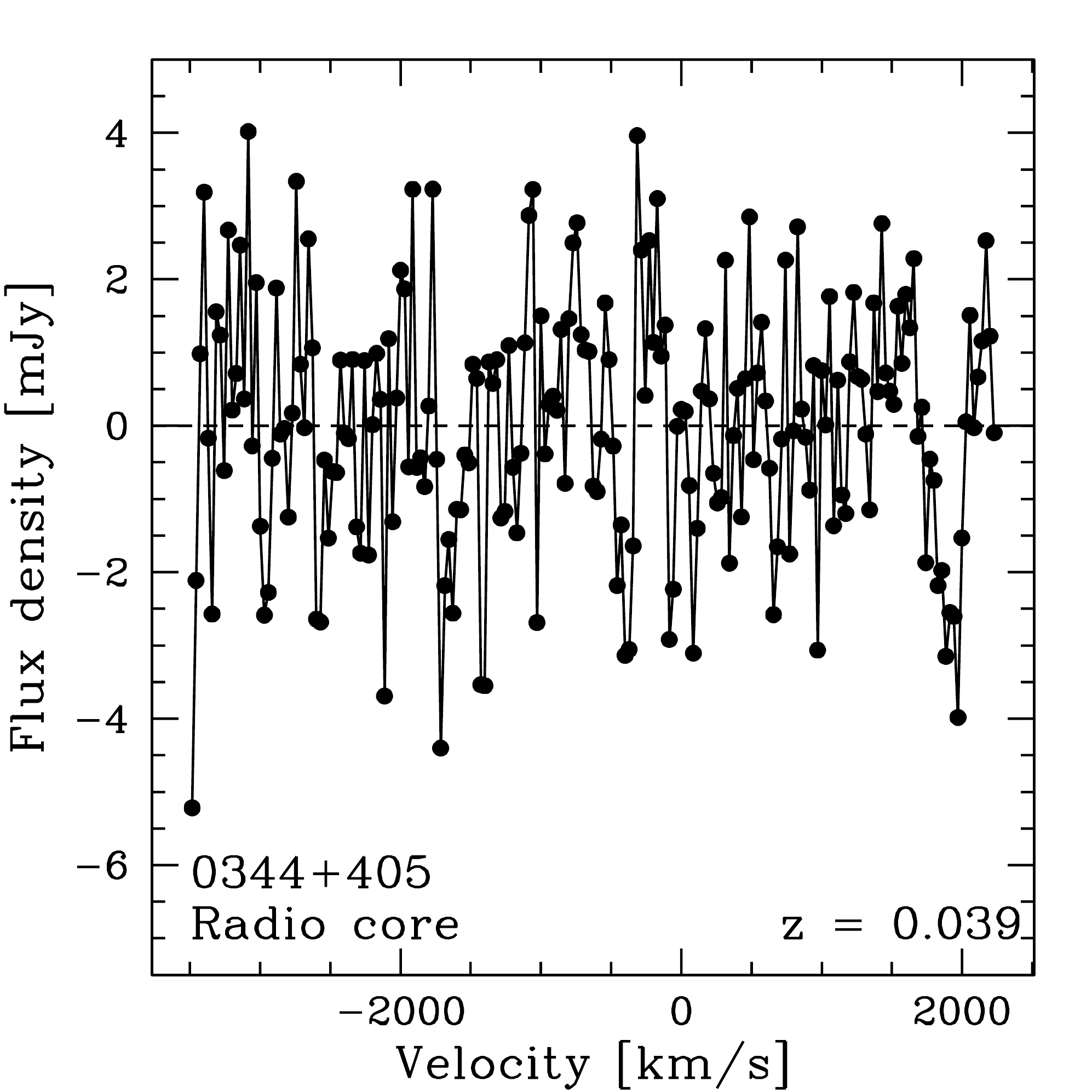} &
\includegraphics[scale=0.28]{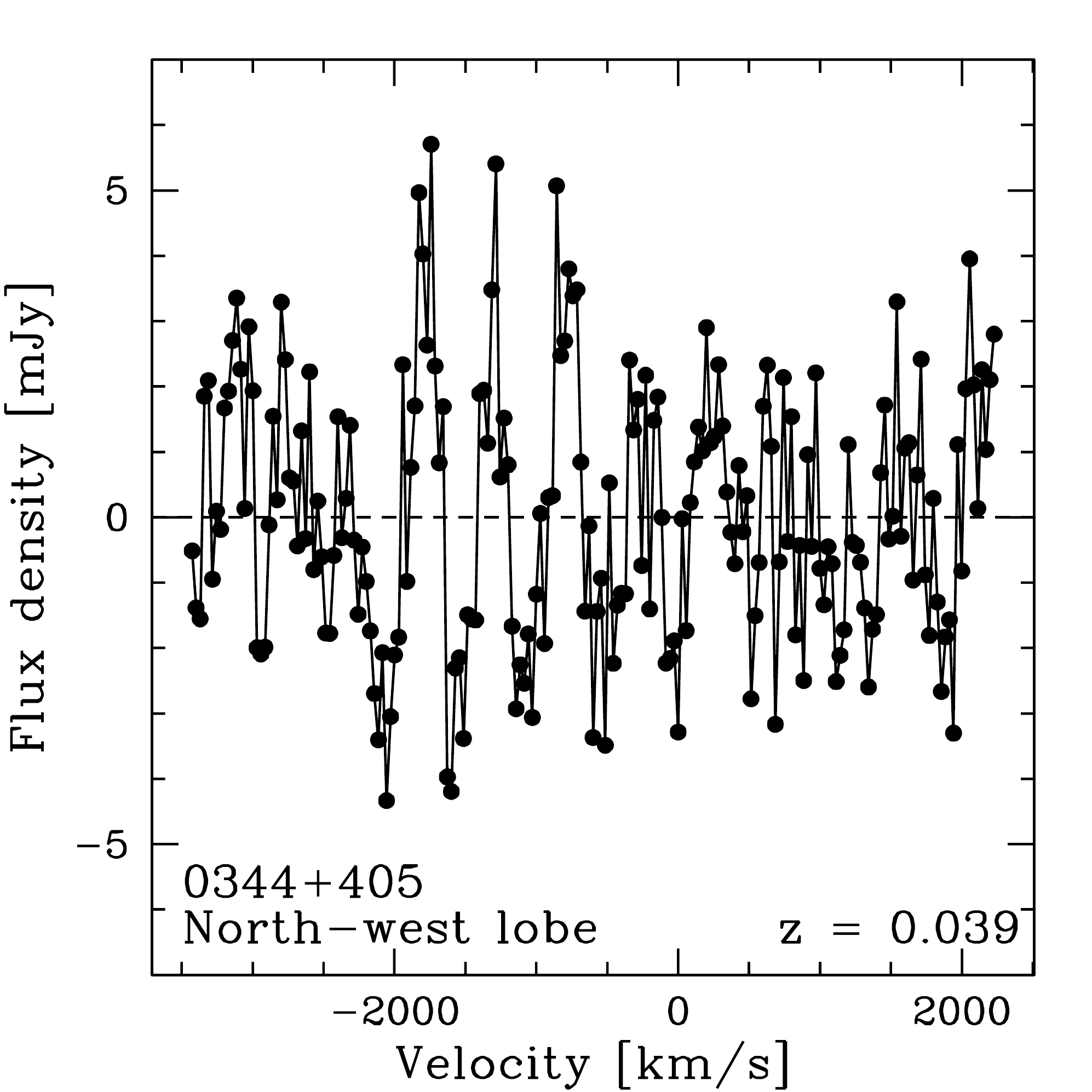} &
\includegraphics[scale=0.28]{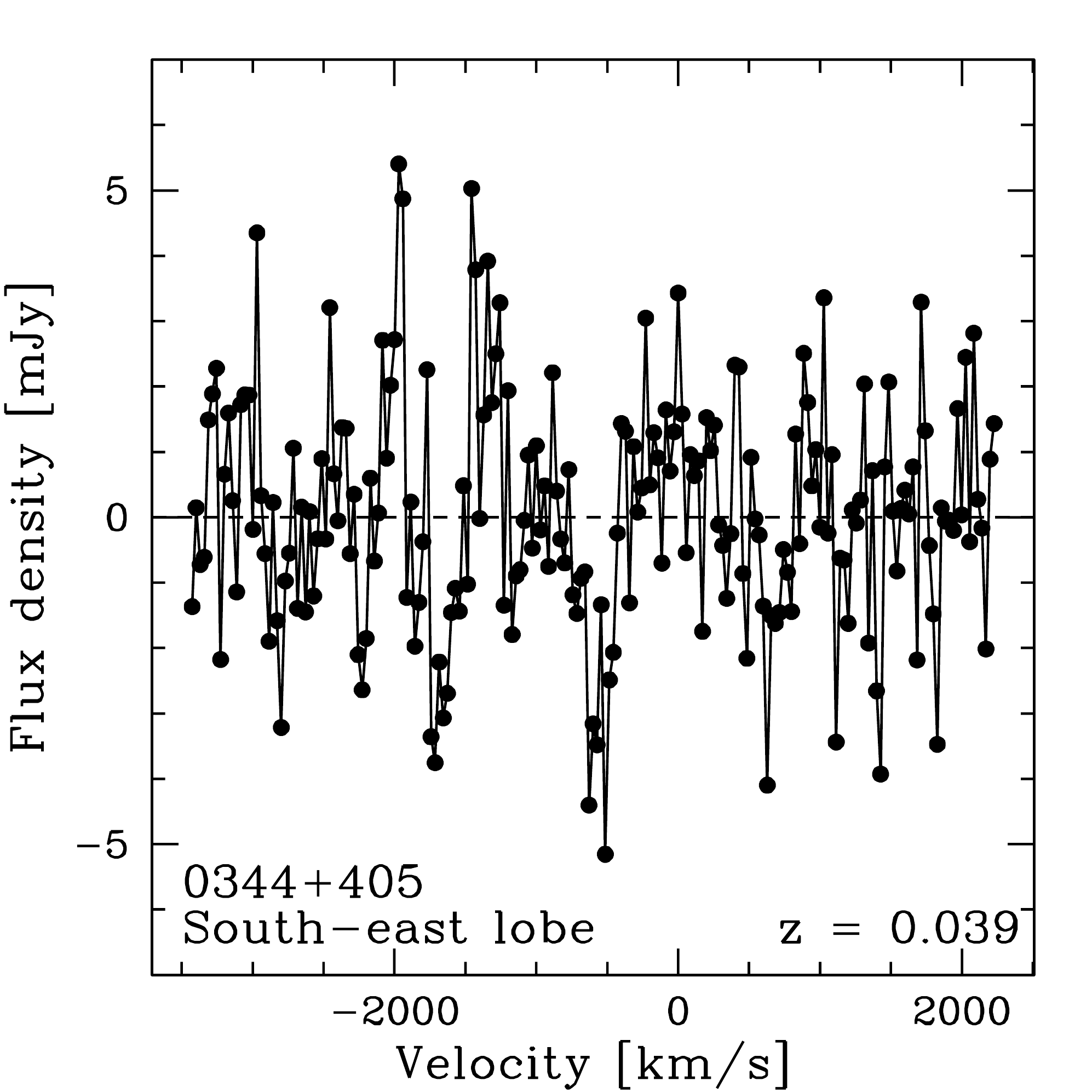} \\

\includegraphics[scale=0.28]{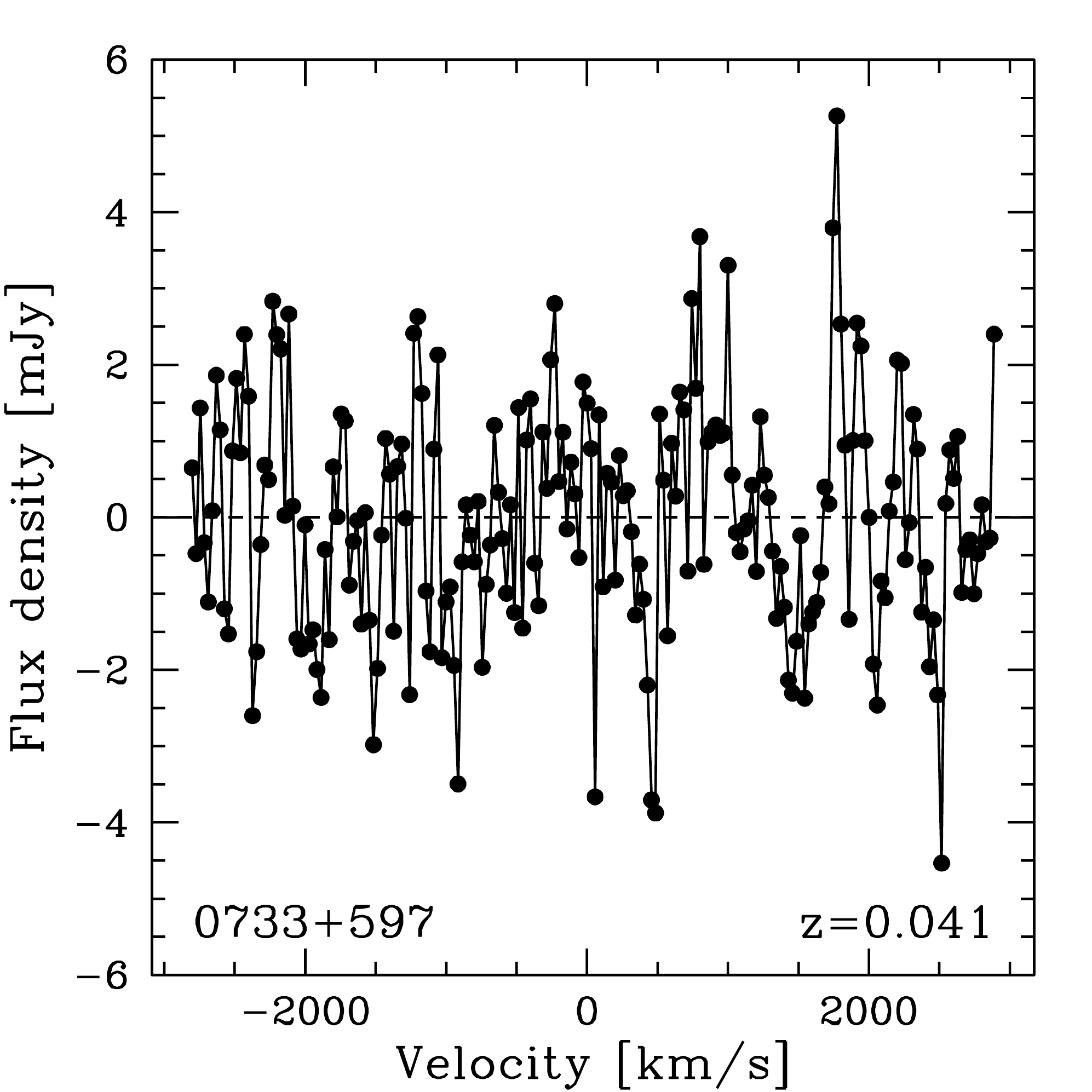} &
\includegraphics[scale=0.28]{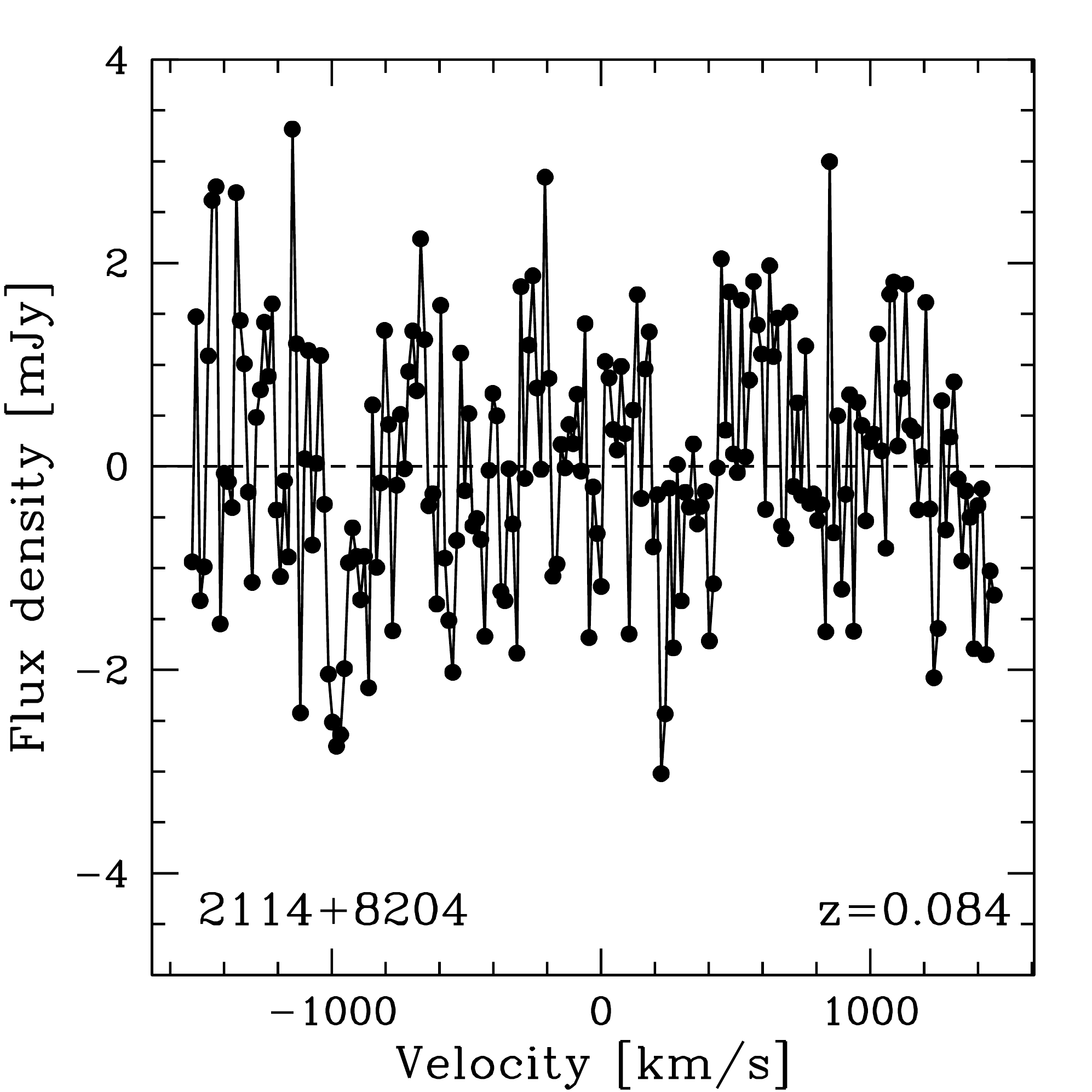}  &
\includegraphics[scale=0.28]{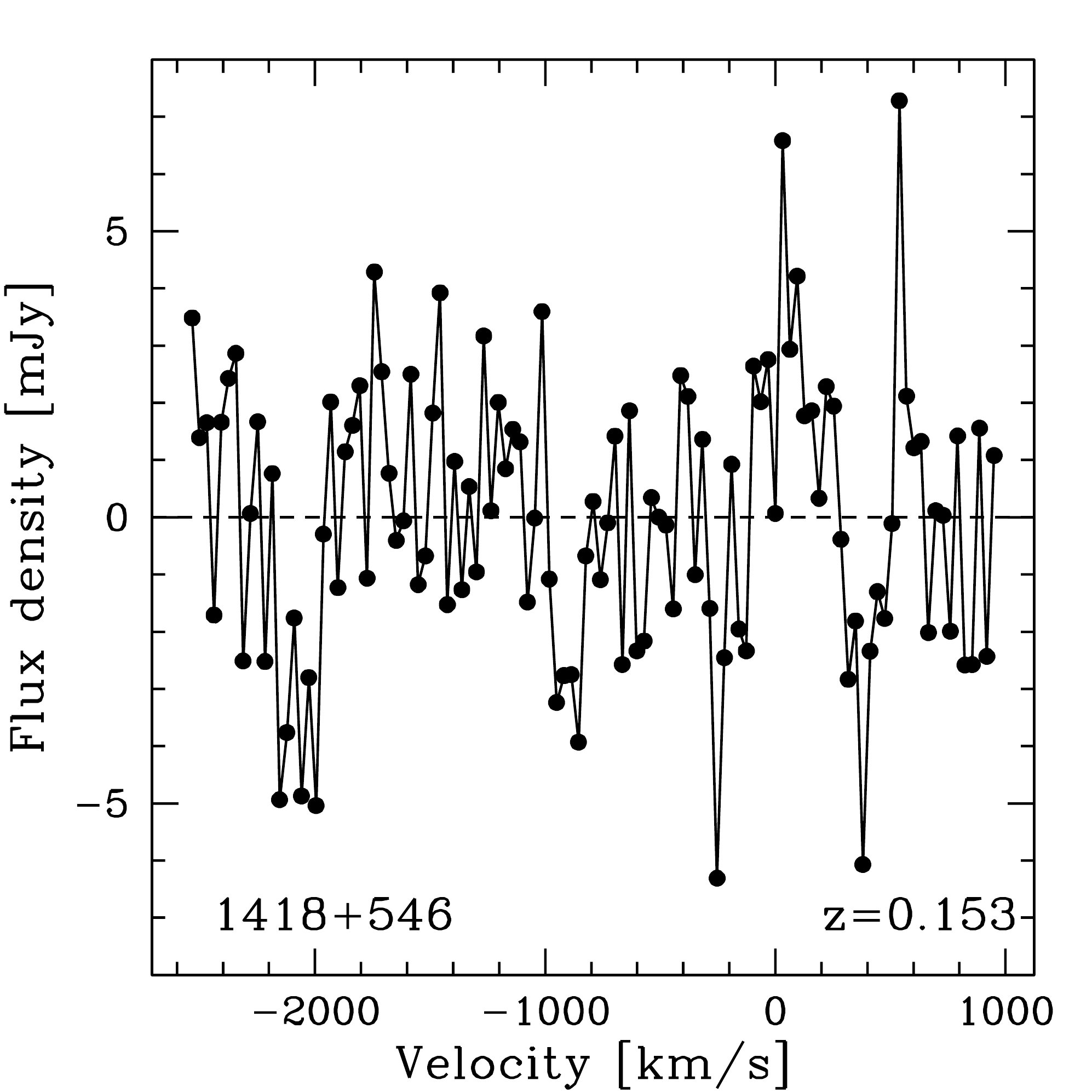} \\

\includegraphics[scale=0.28]{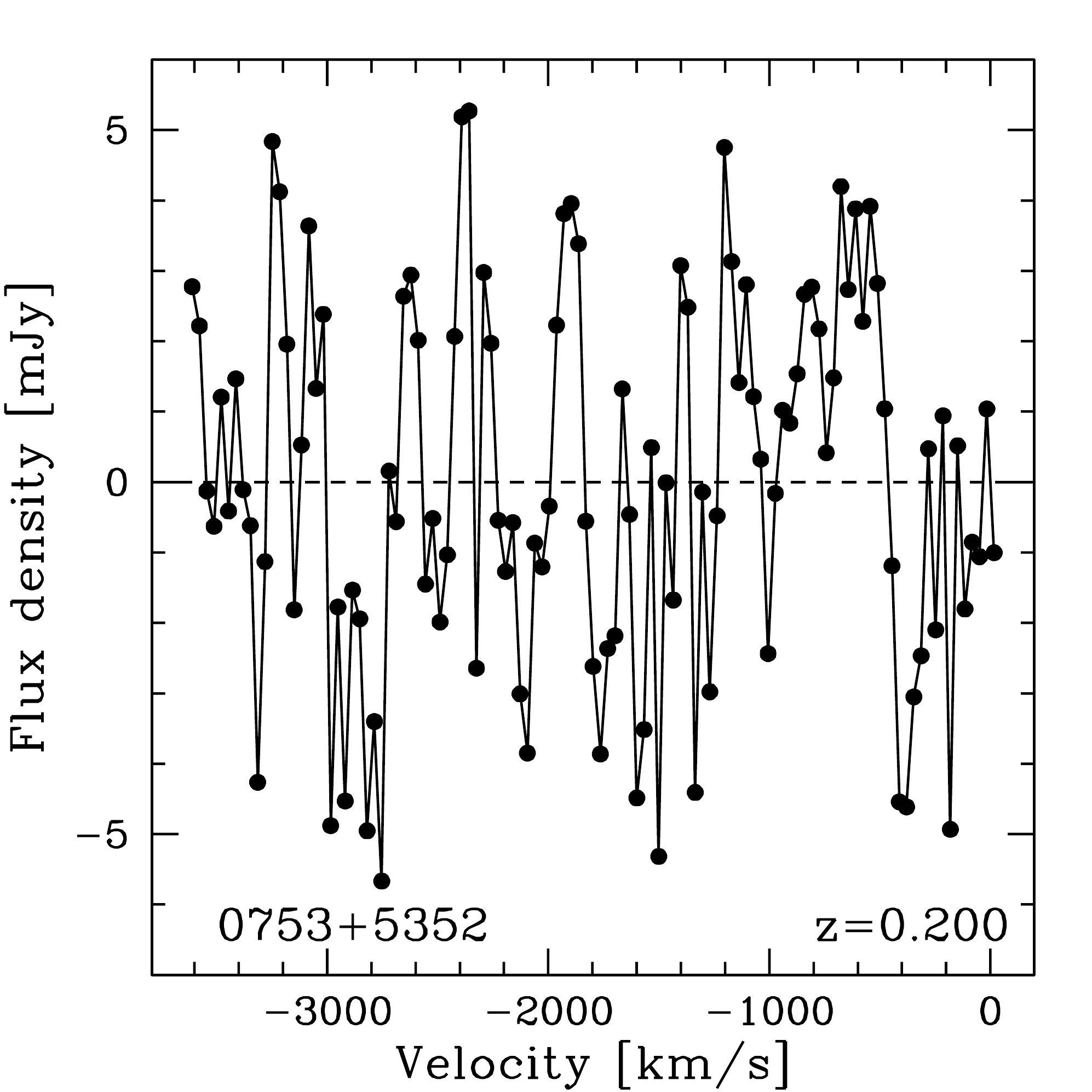} &
\includegraphics[scale=0.28]{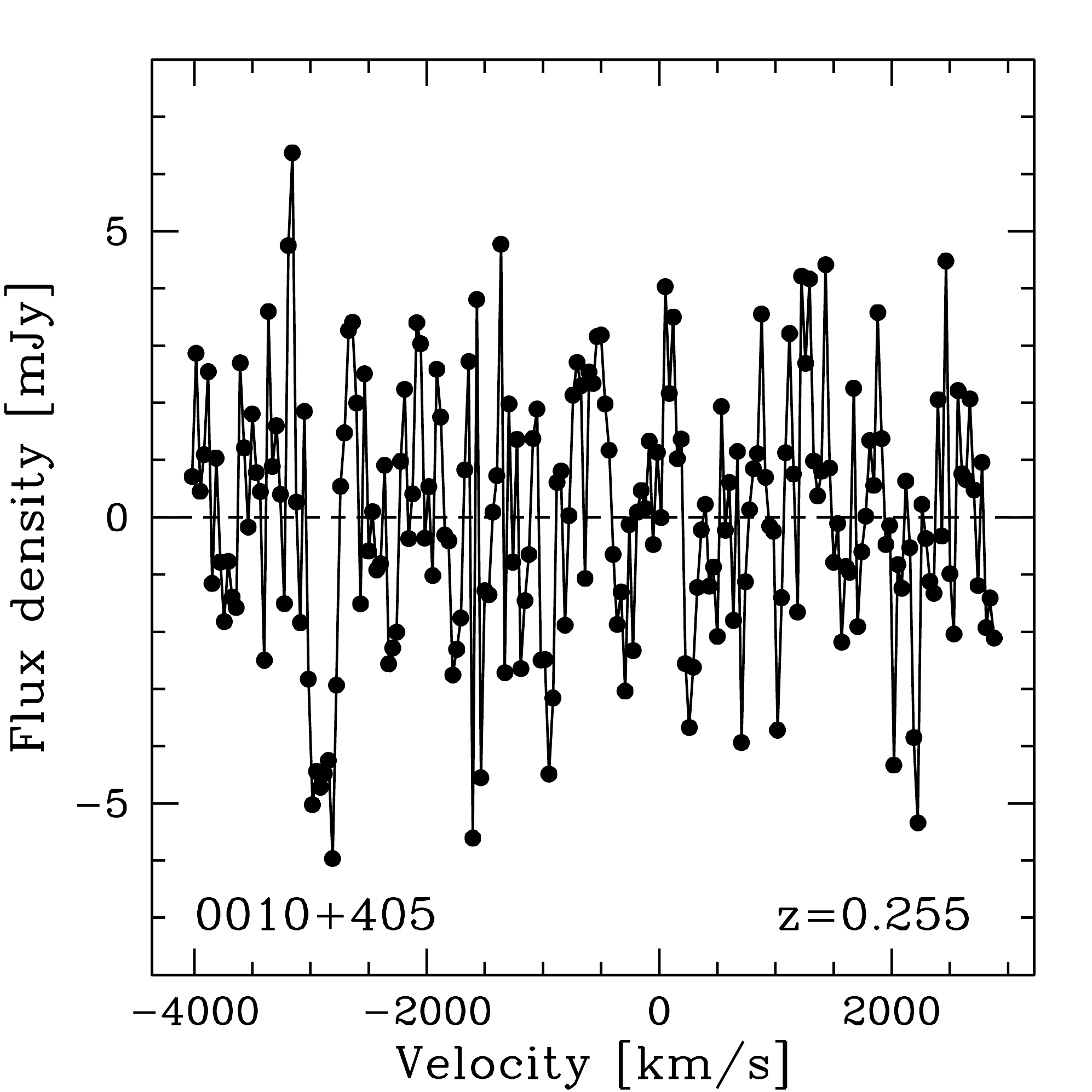}&
\includegraphics[scale=0.28]{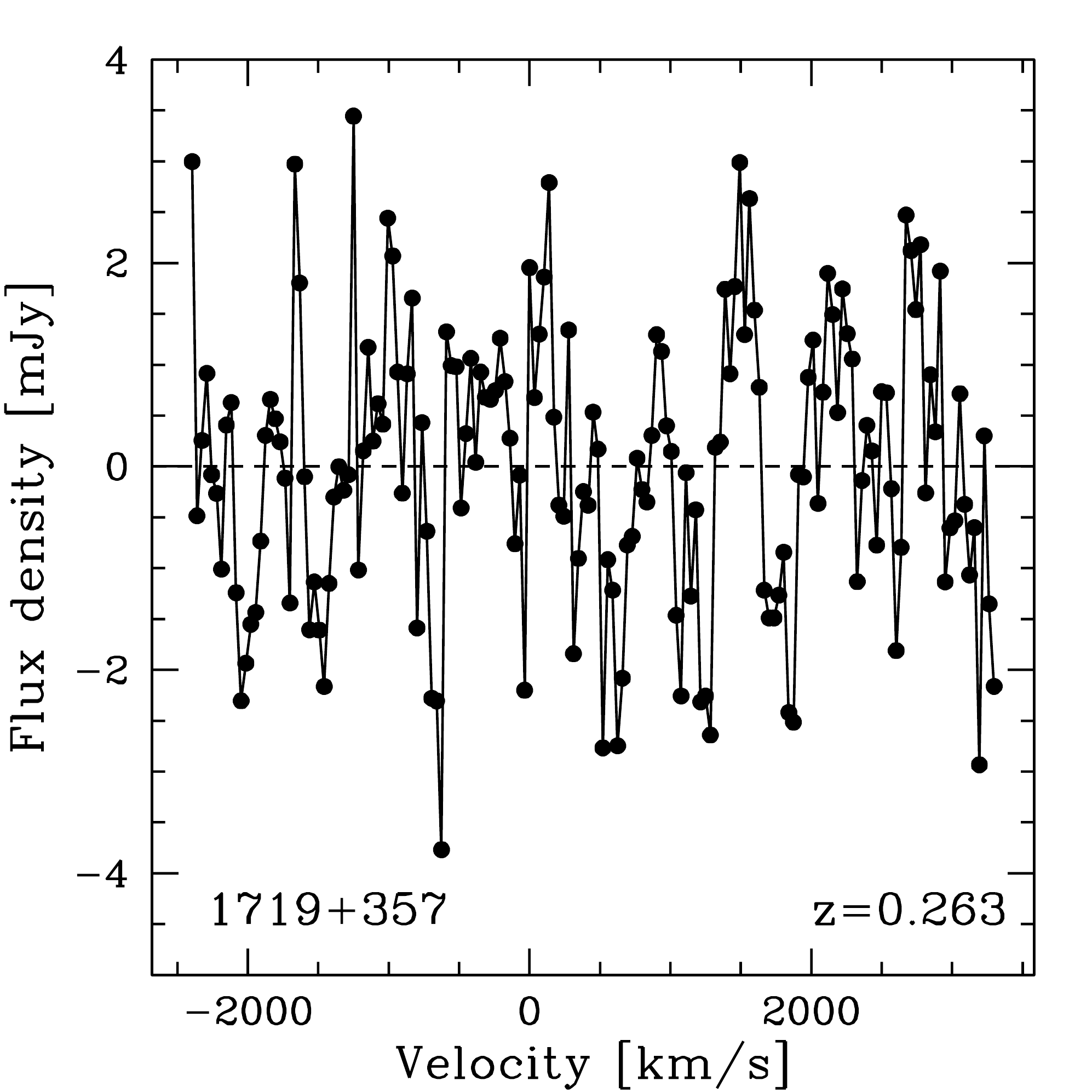}\\

\includegraphics[scale=0.28]{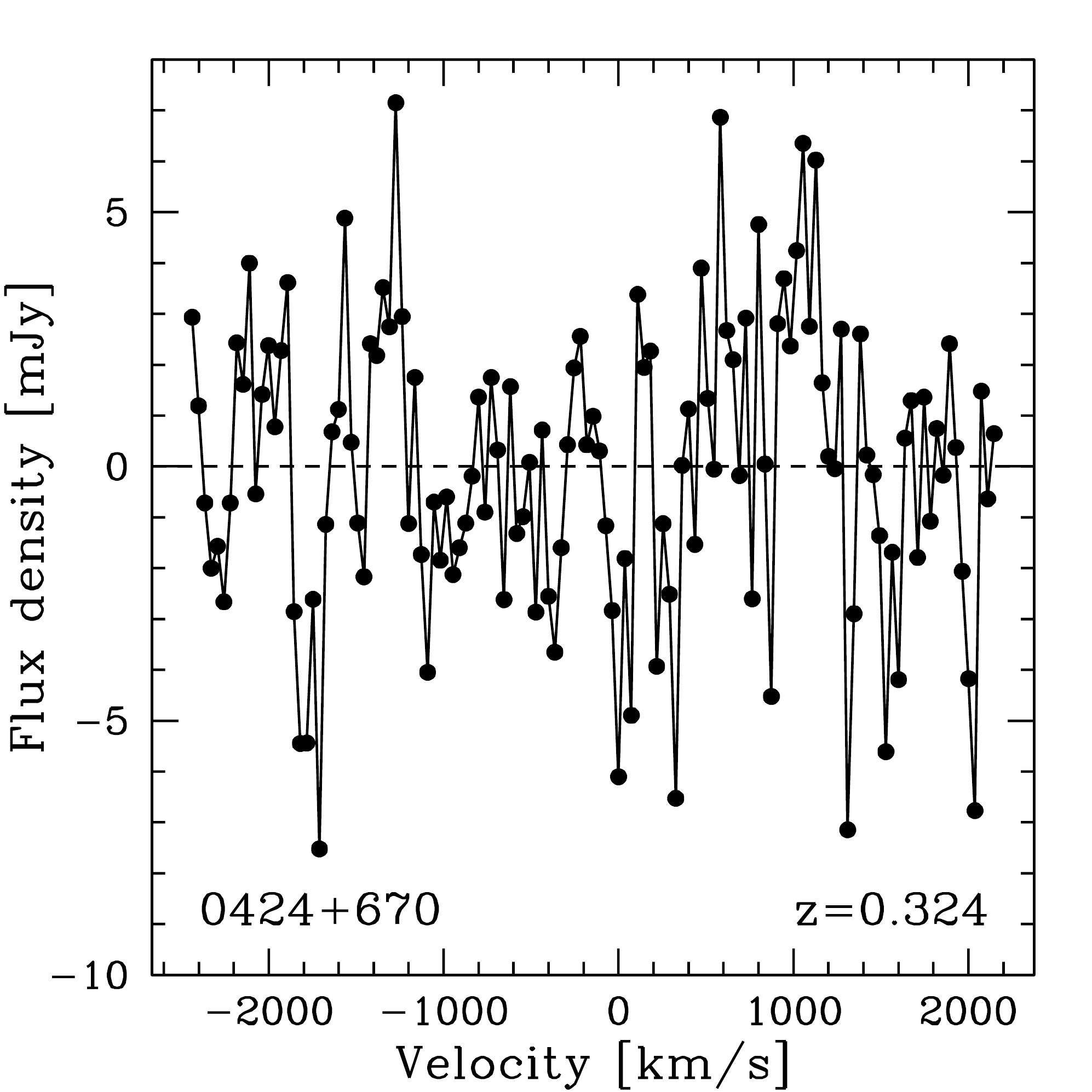} &
\includegraphics[scale=0.28]{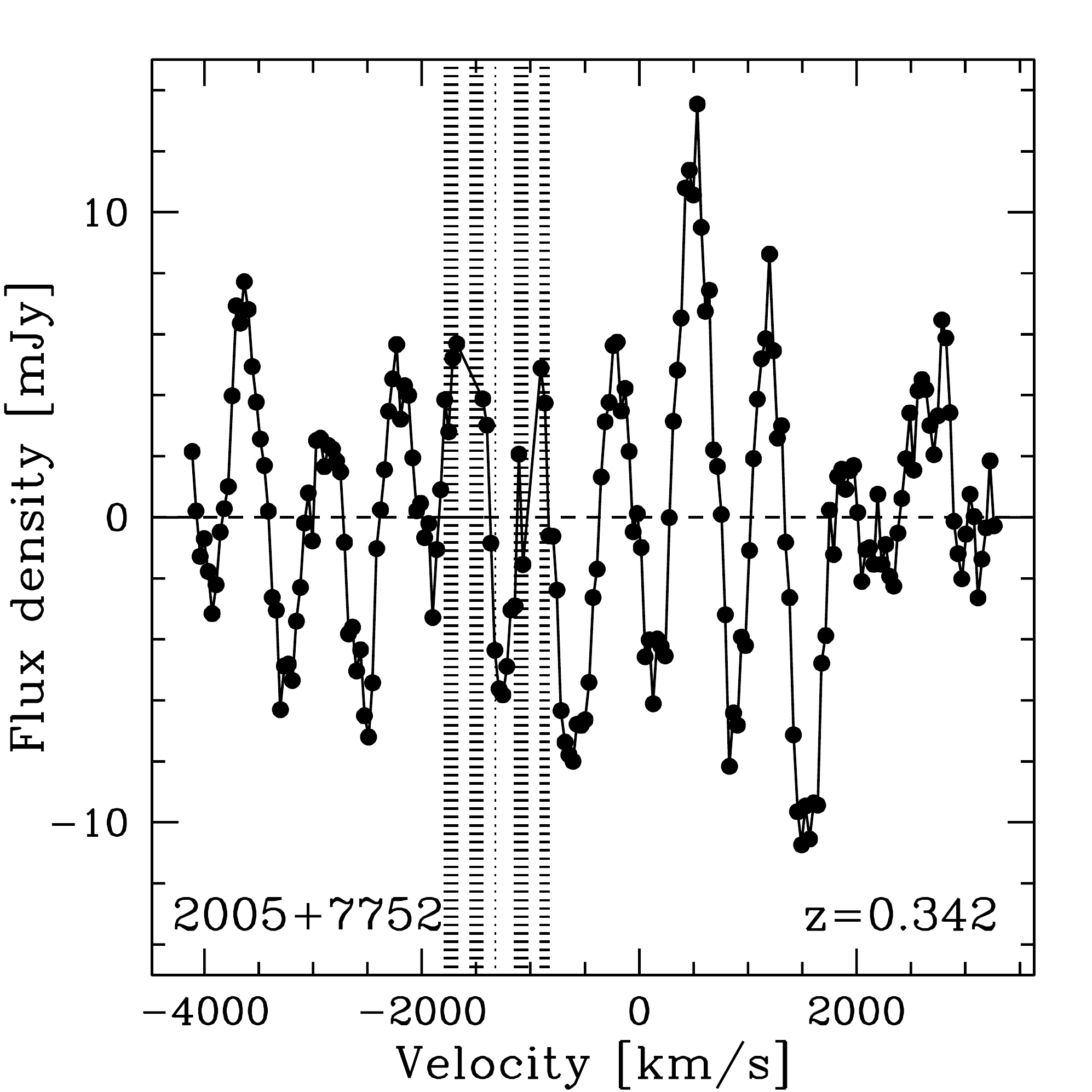}  &
\includegraphics[scale=0.28]{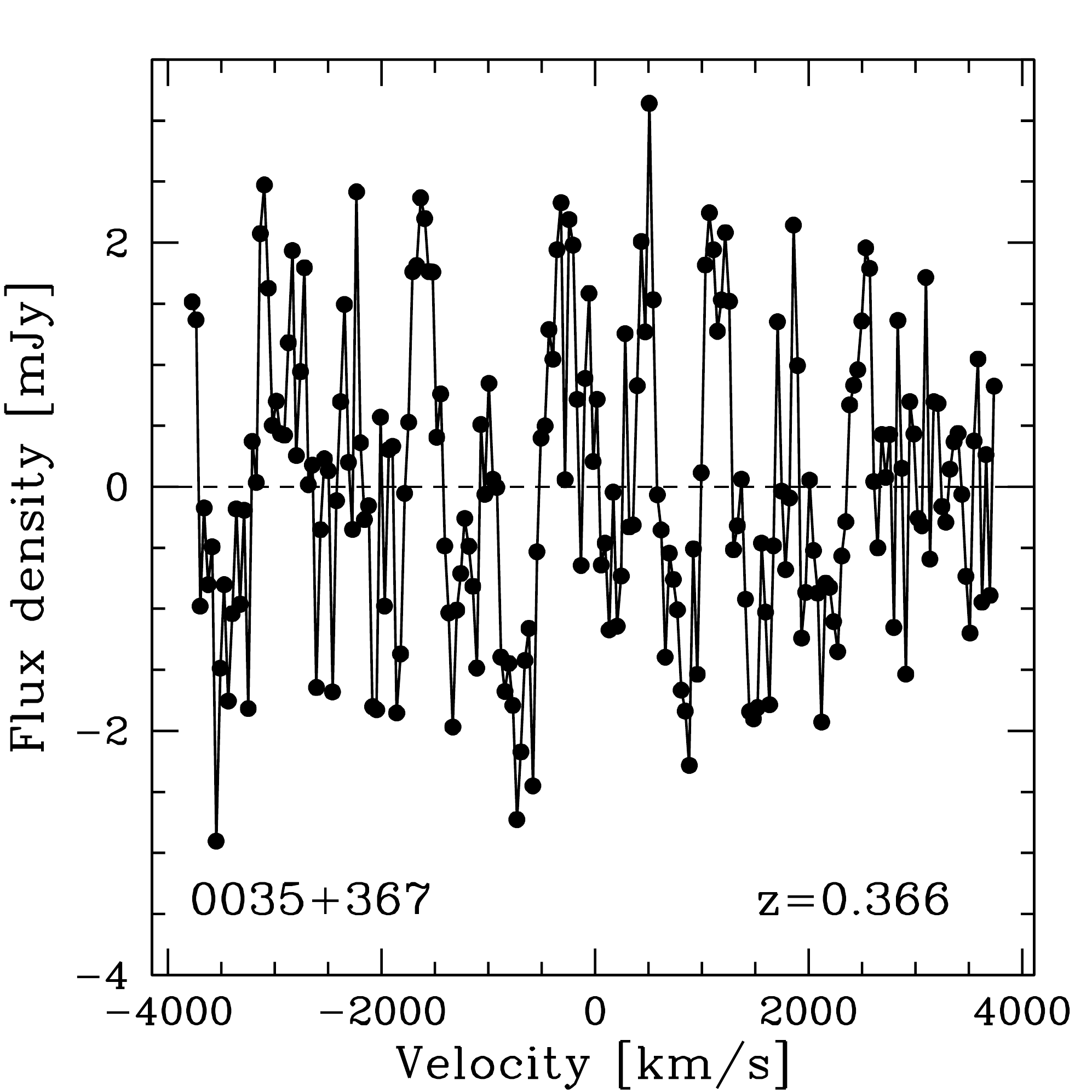} \\

\end{tabular}
\end{figure*}

\begin{figure*}
\begin{tabular}{ccc}
\multicolumn{2}{ c }{{{\bf Figure A1.} (contd.) The GMRT \hii\ spectra for the 37 non-detections}} \\

\includegraphics[scale=0.28]{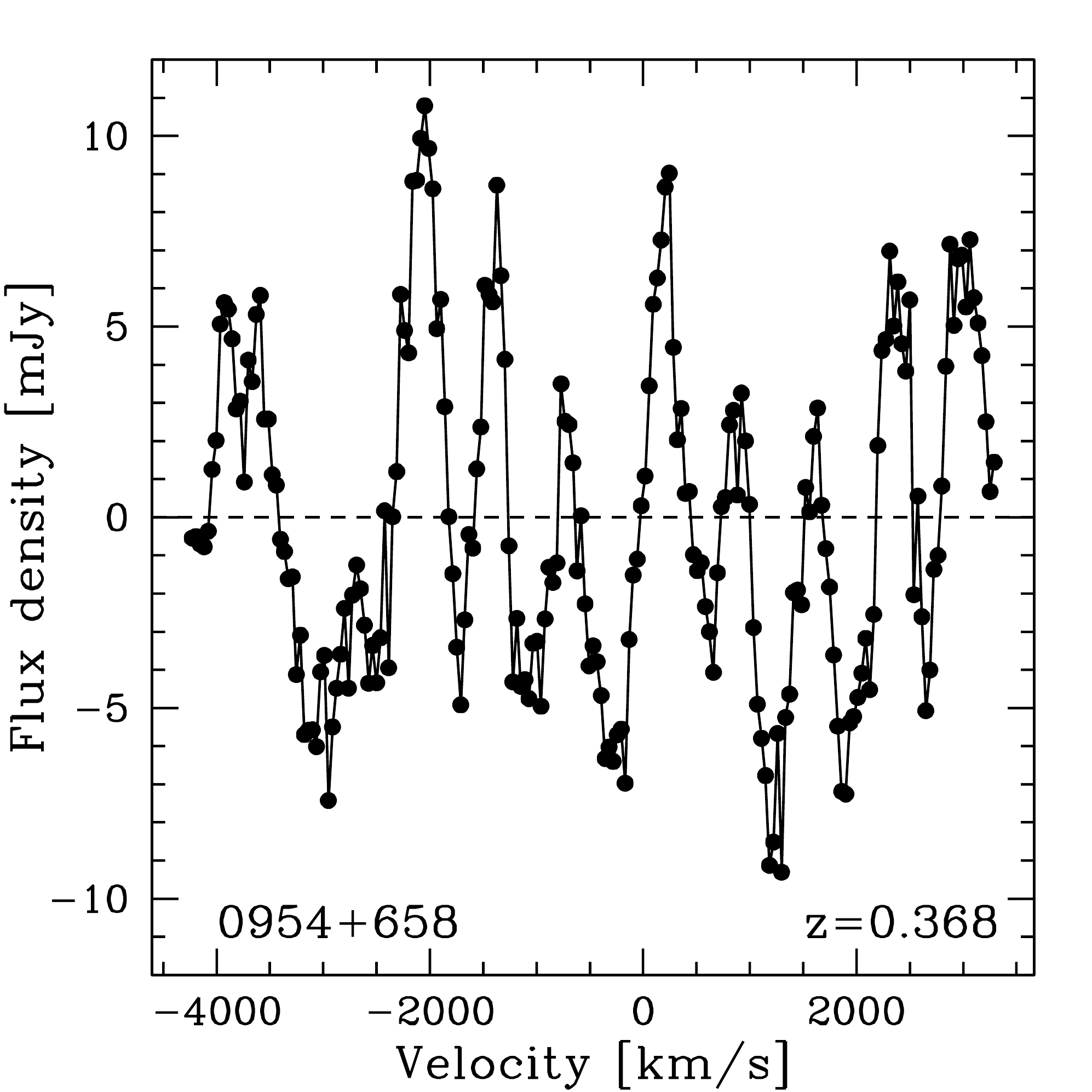} &
\includegraphics[scale=0.28]{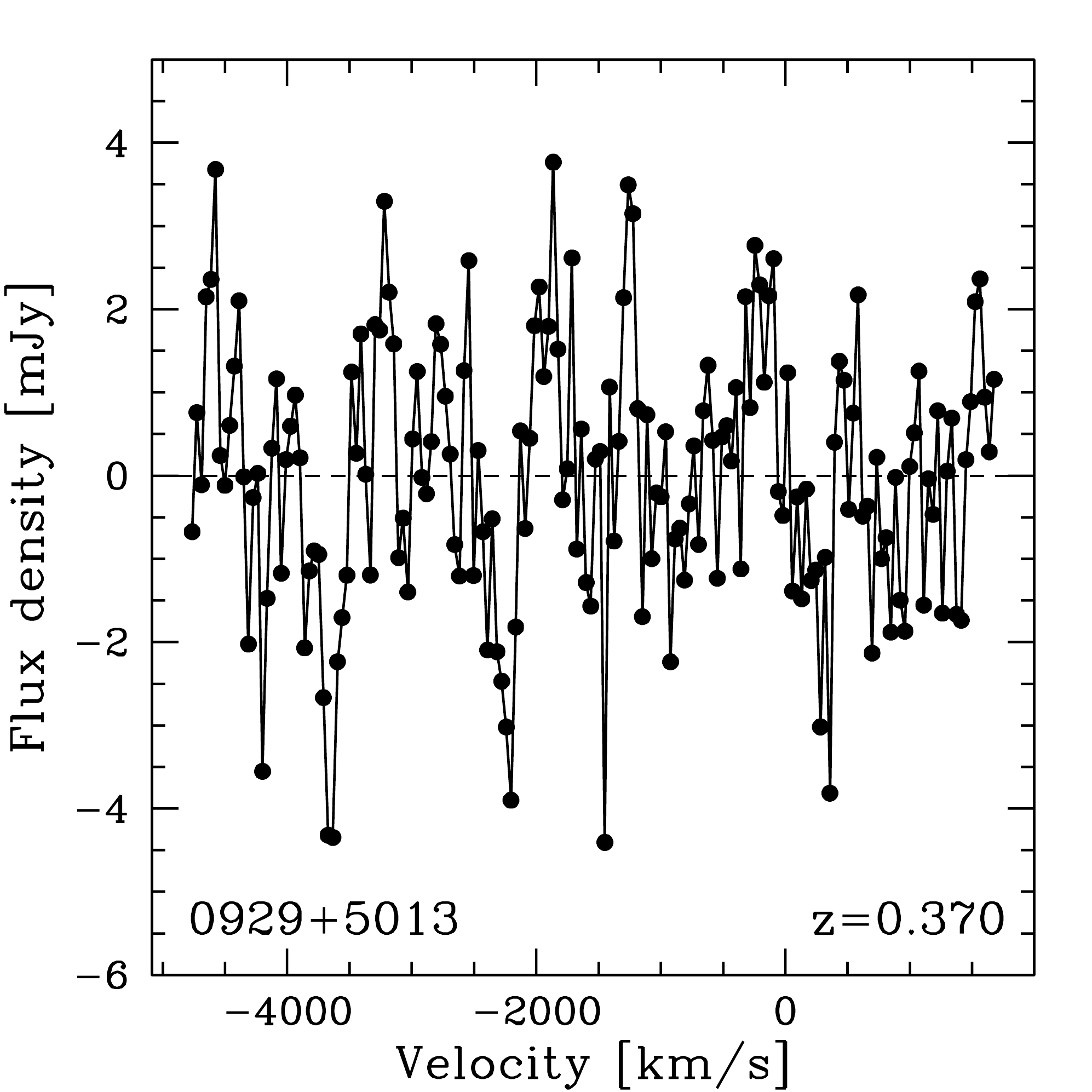}  &
\includegraphics[scale=0.28]{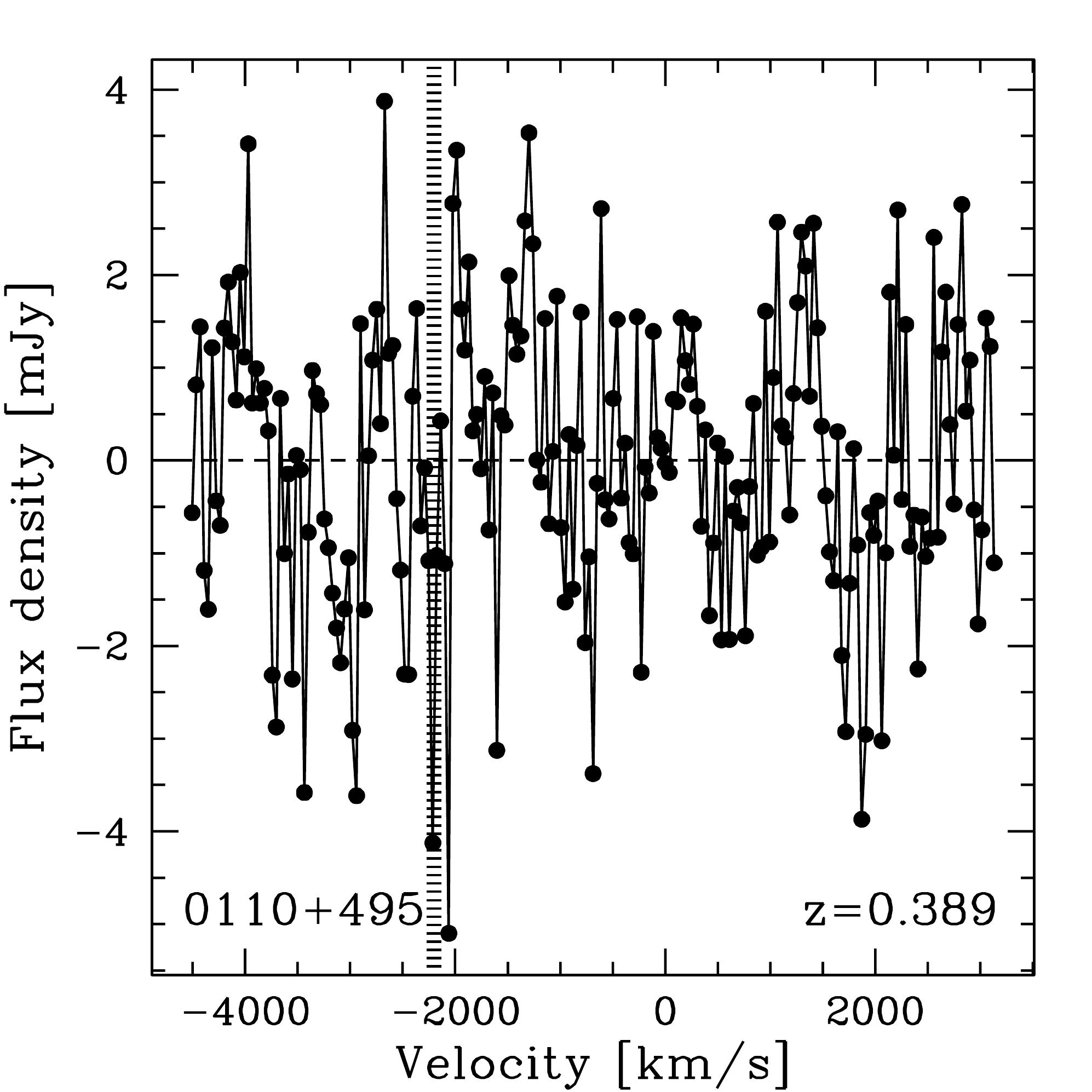} \\

\includegraphics[scale=0.28]{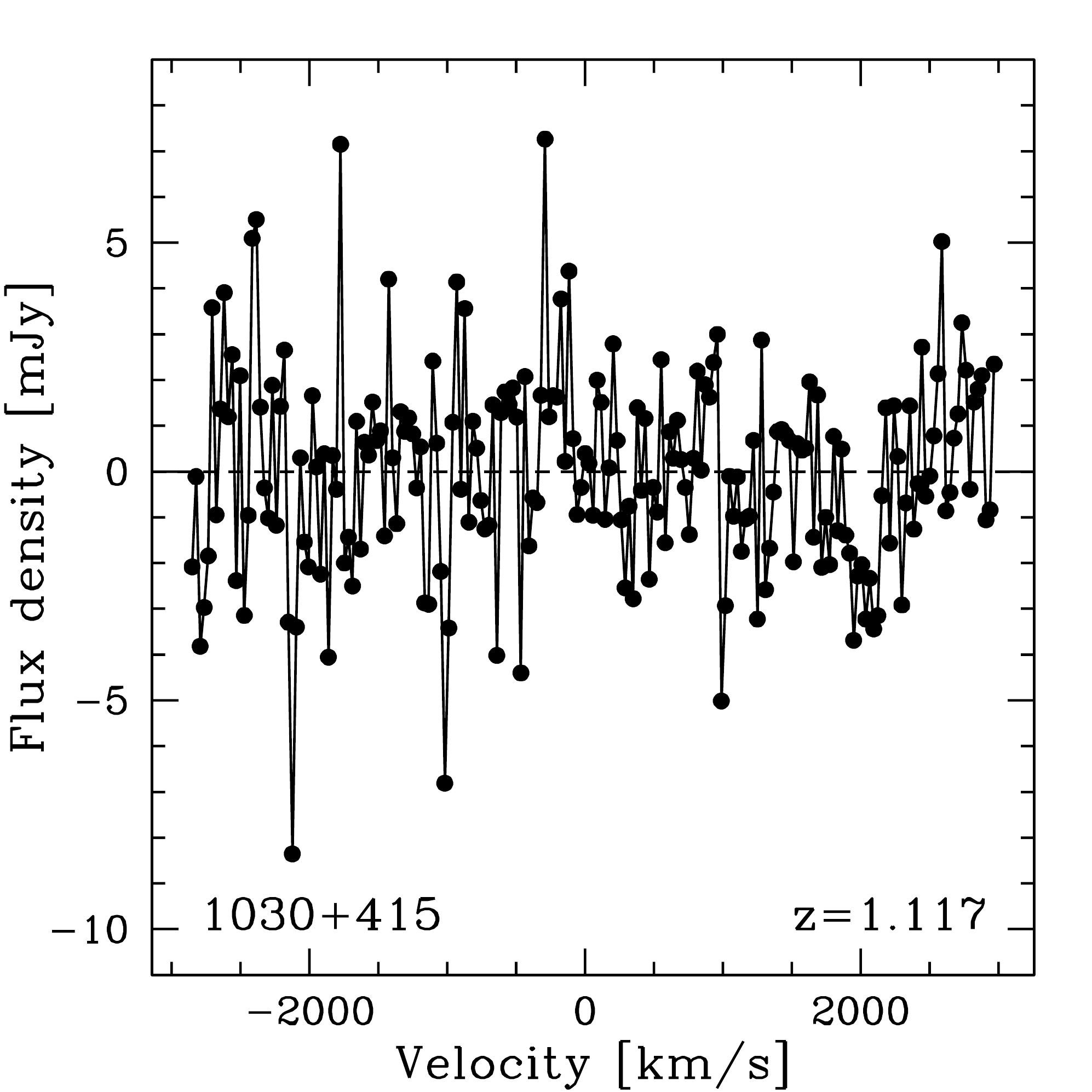} &
\includegraphics[scale=0.28]{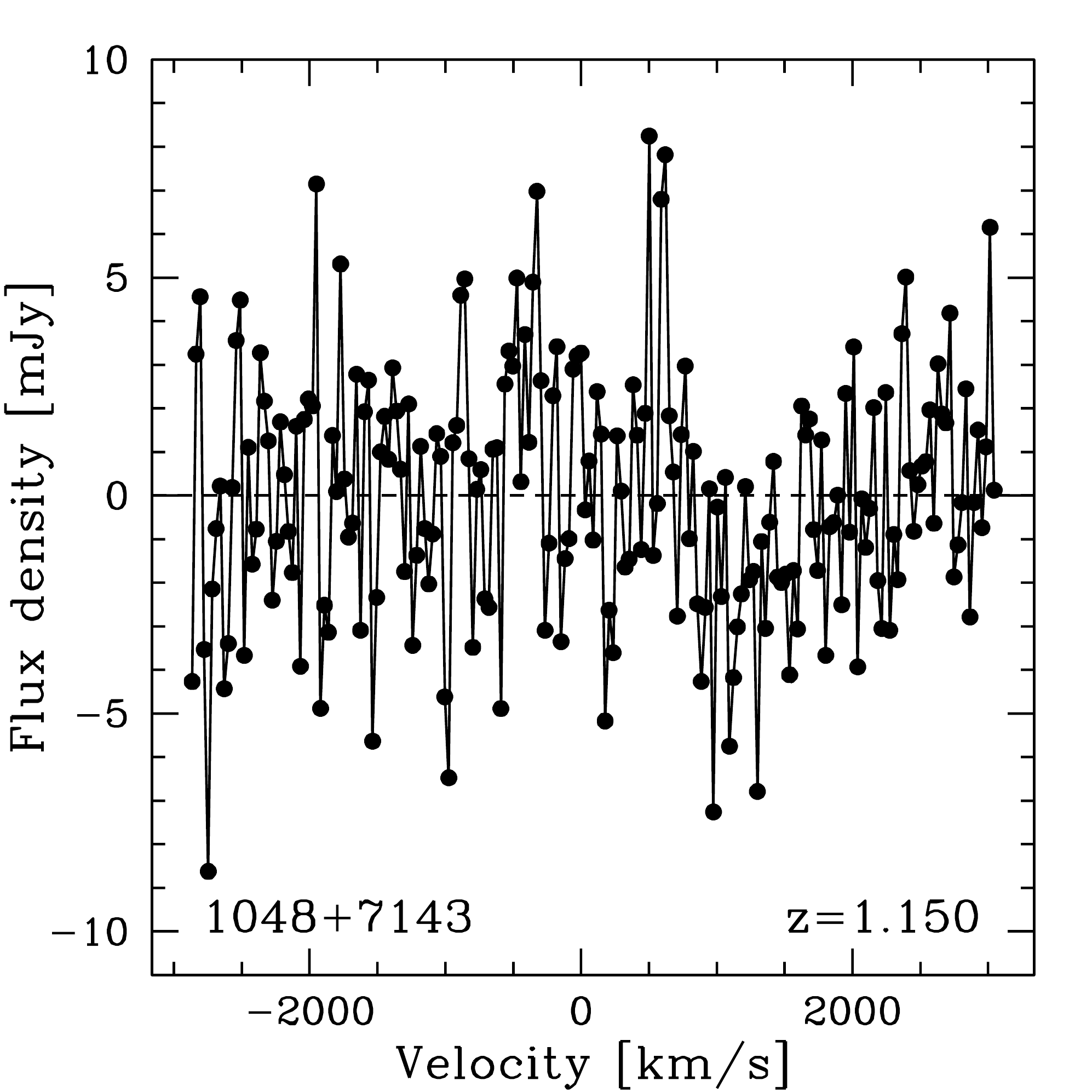}  &
\includegraphics[scale=0.28]{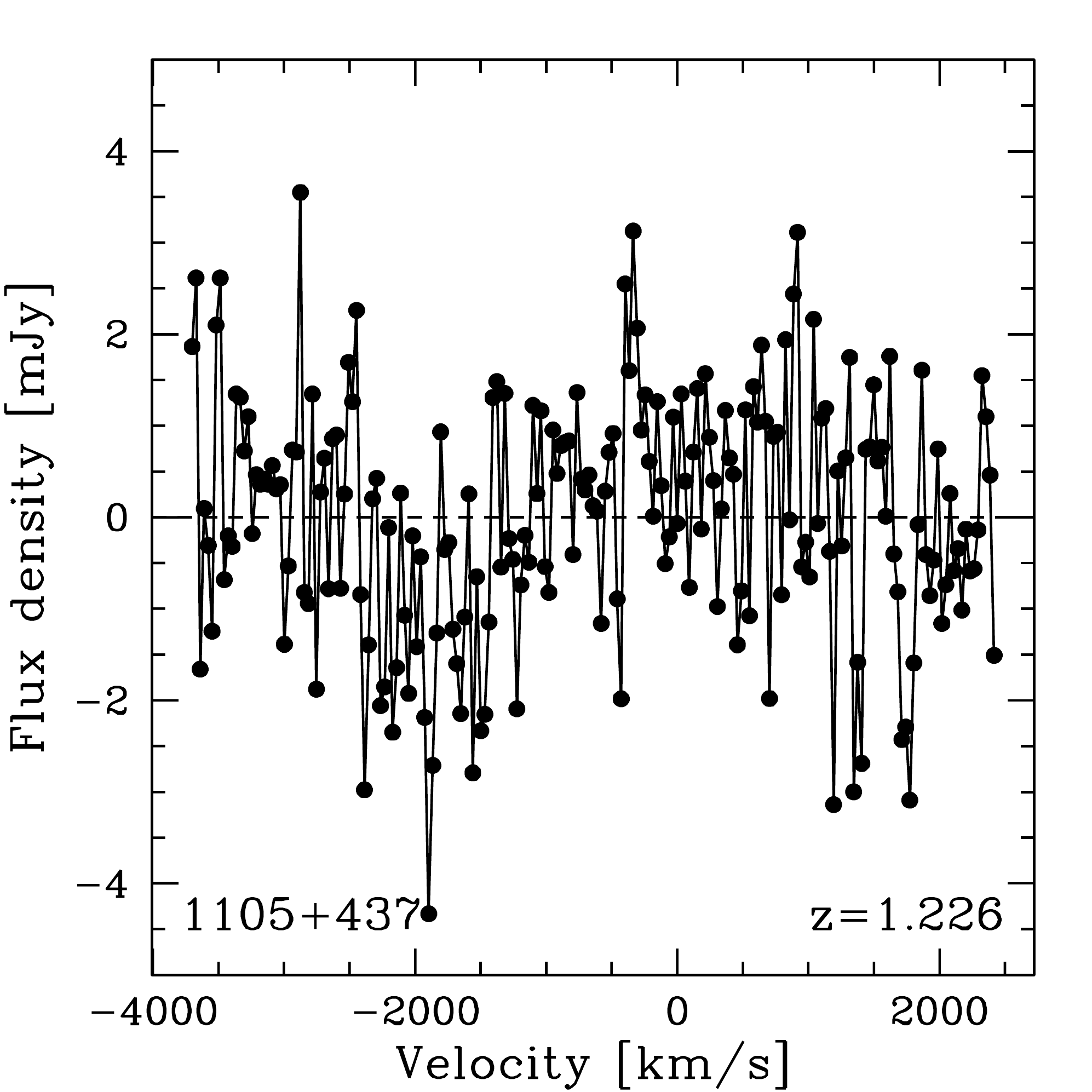} \\

\includegraphics[scale=0.28]{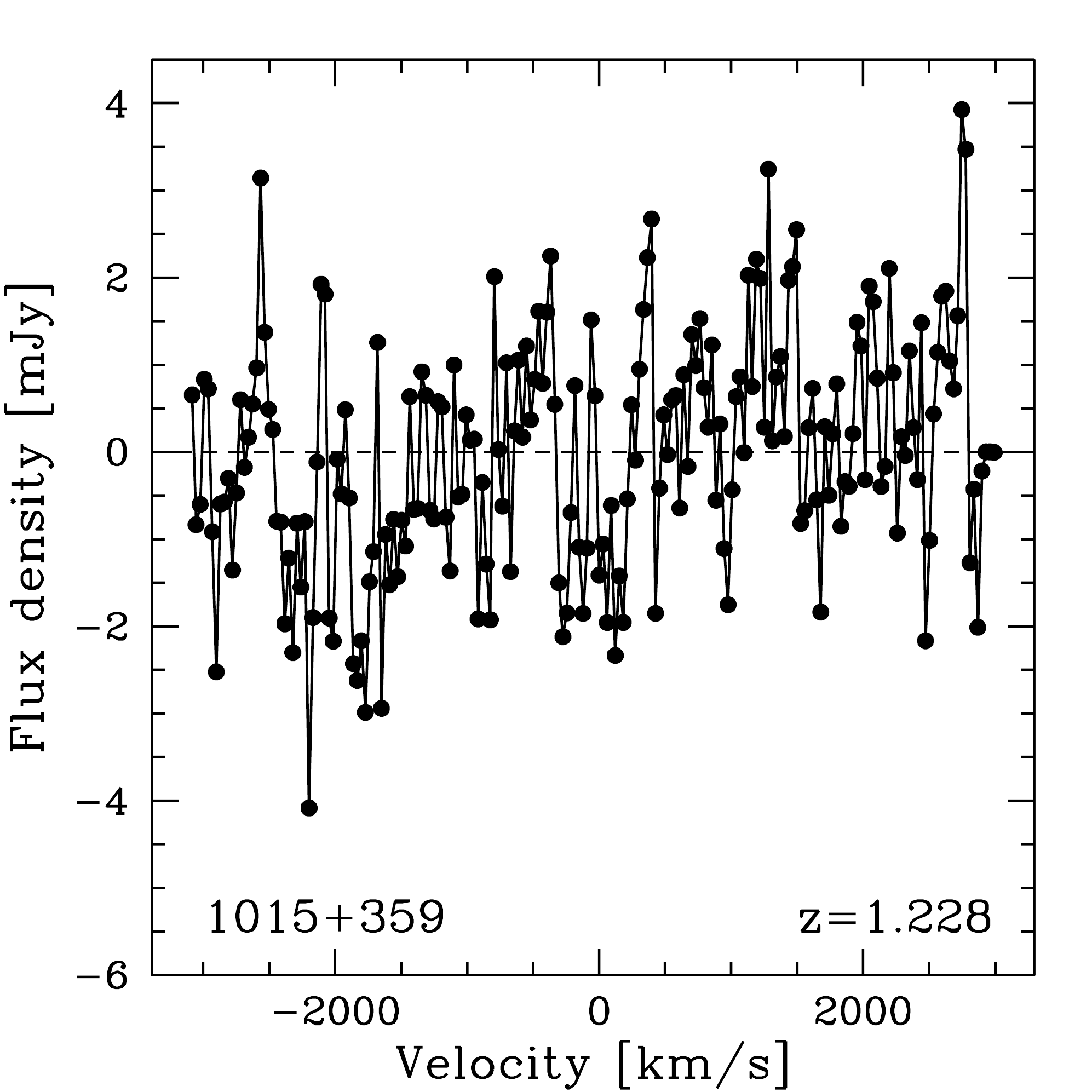} &
\includegraphics[scale=0.28]{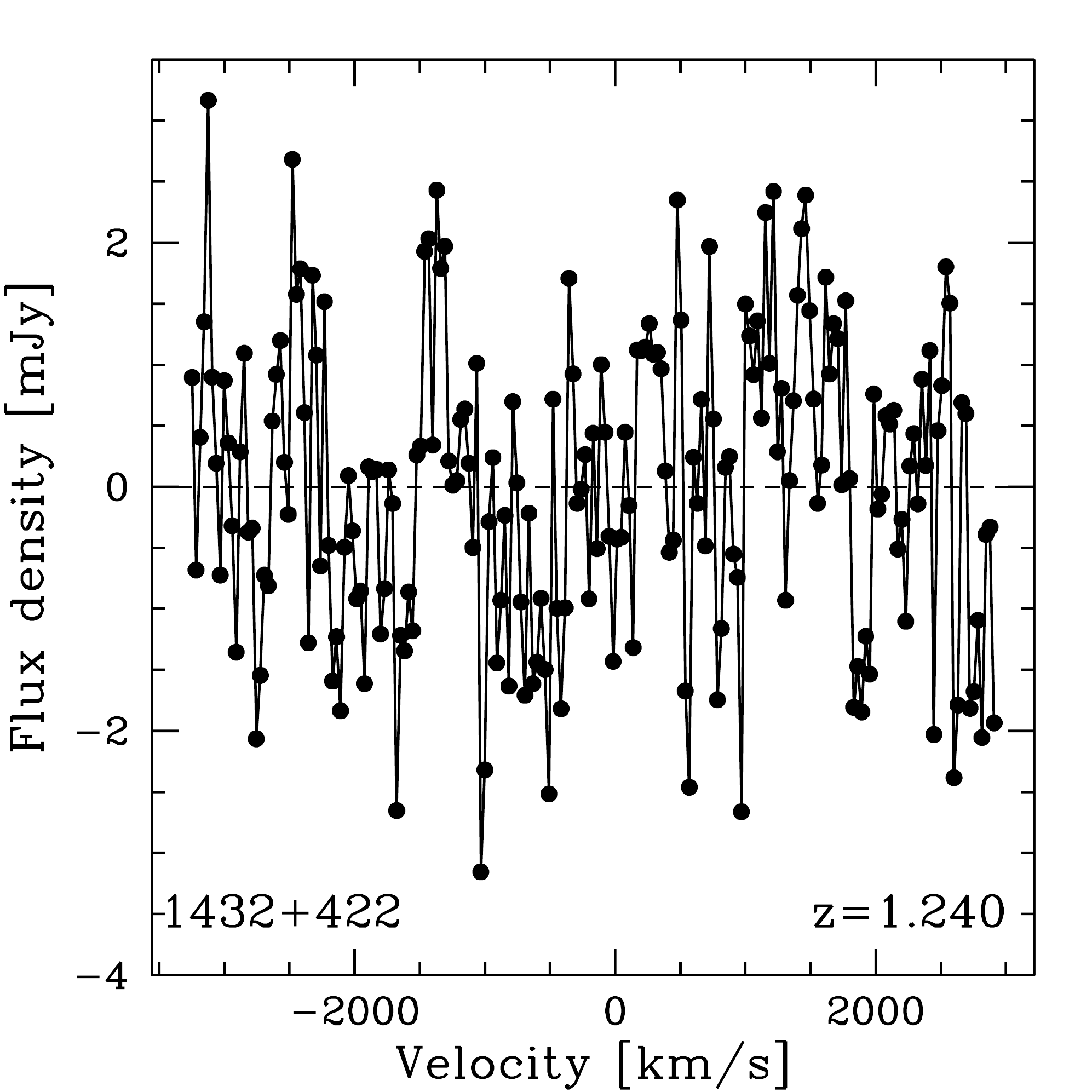} &
\includegraphics[scale=0.28]{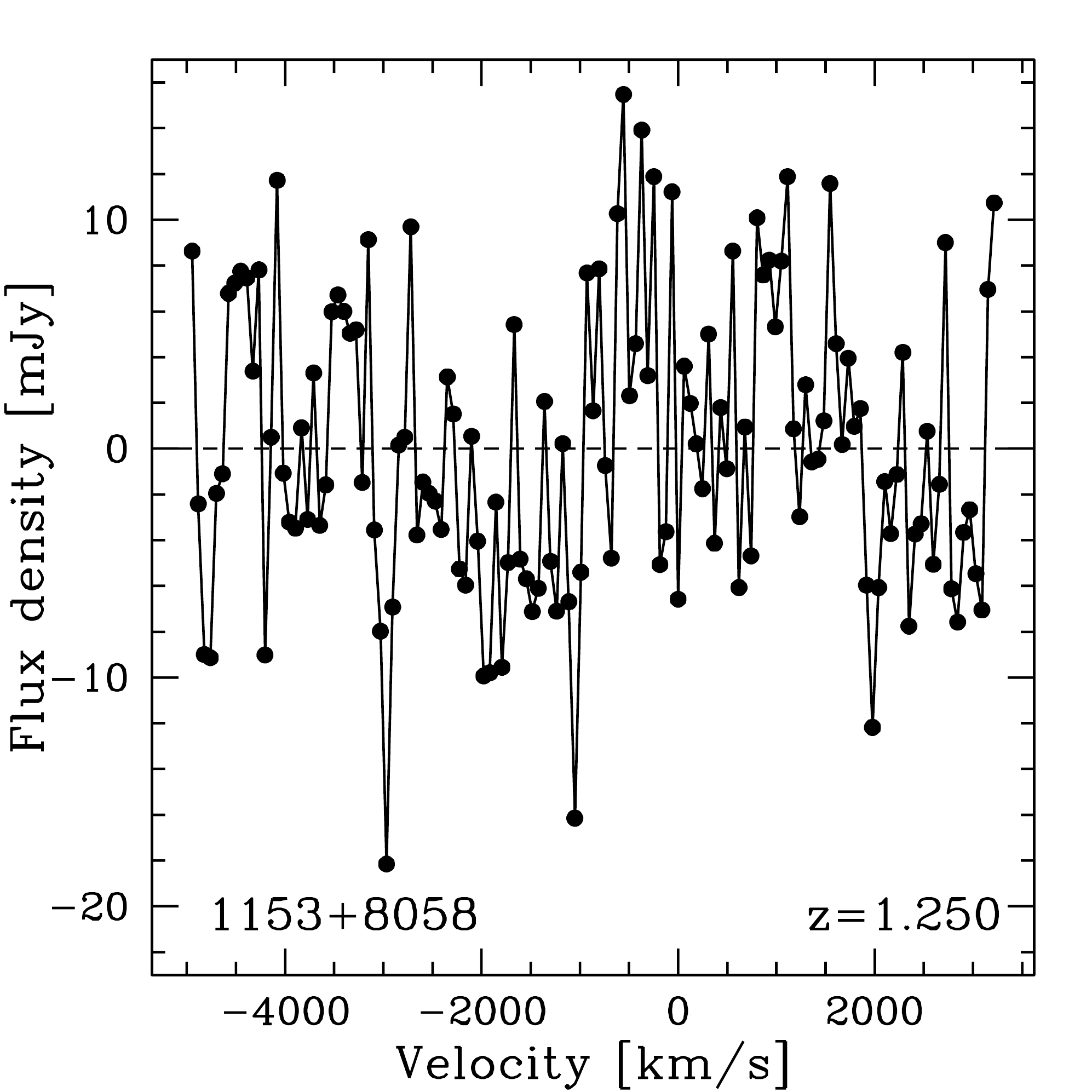}  \\

\includegraphics[scale=0.28]{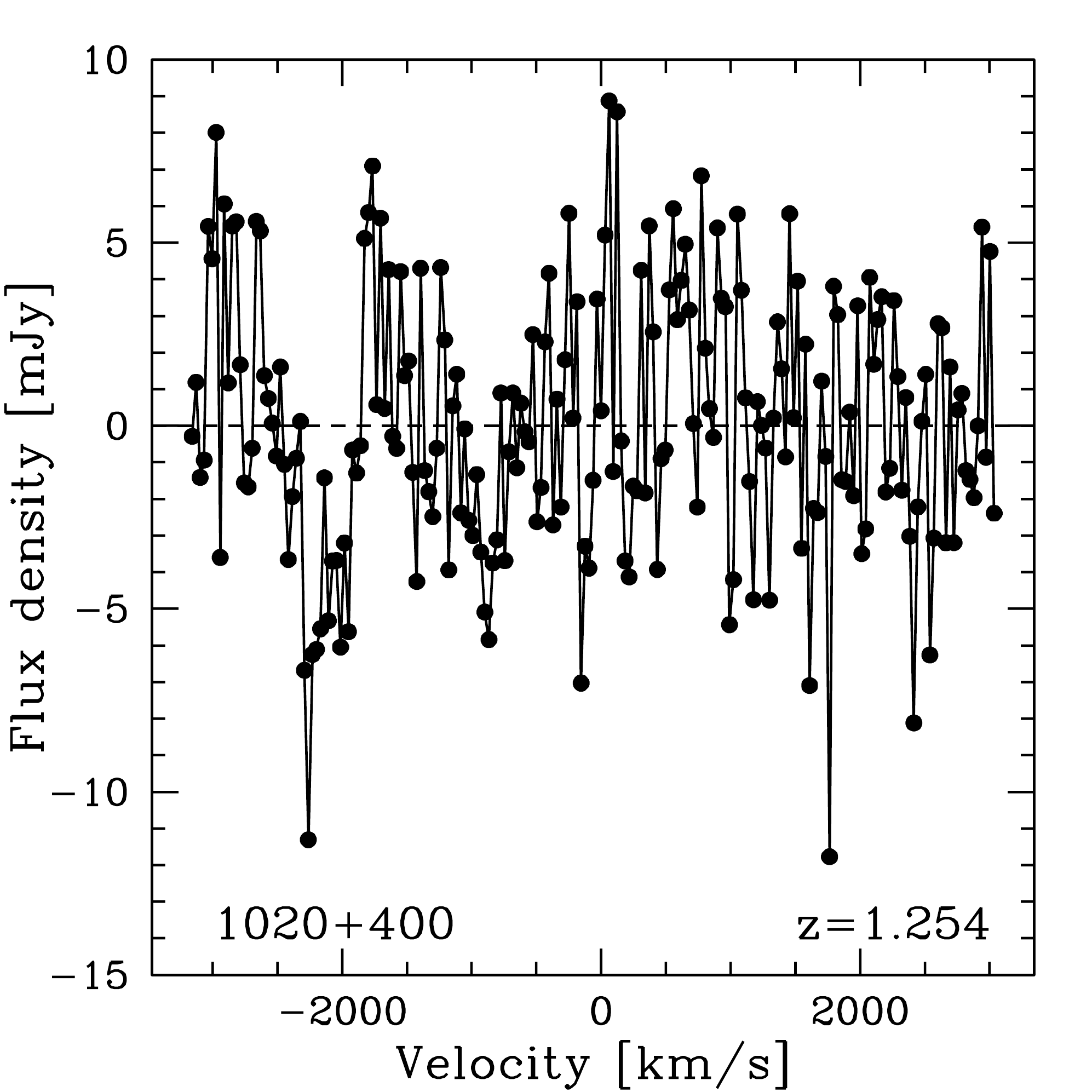} &
\includegraphics[scale=0.28]{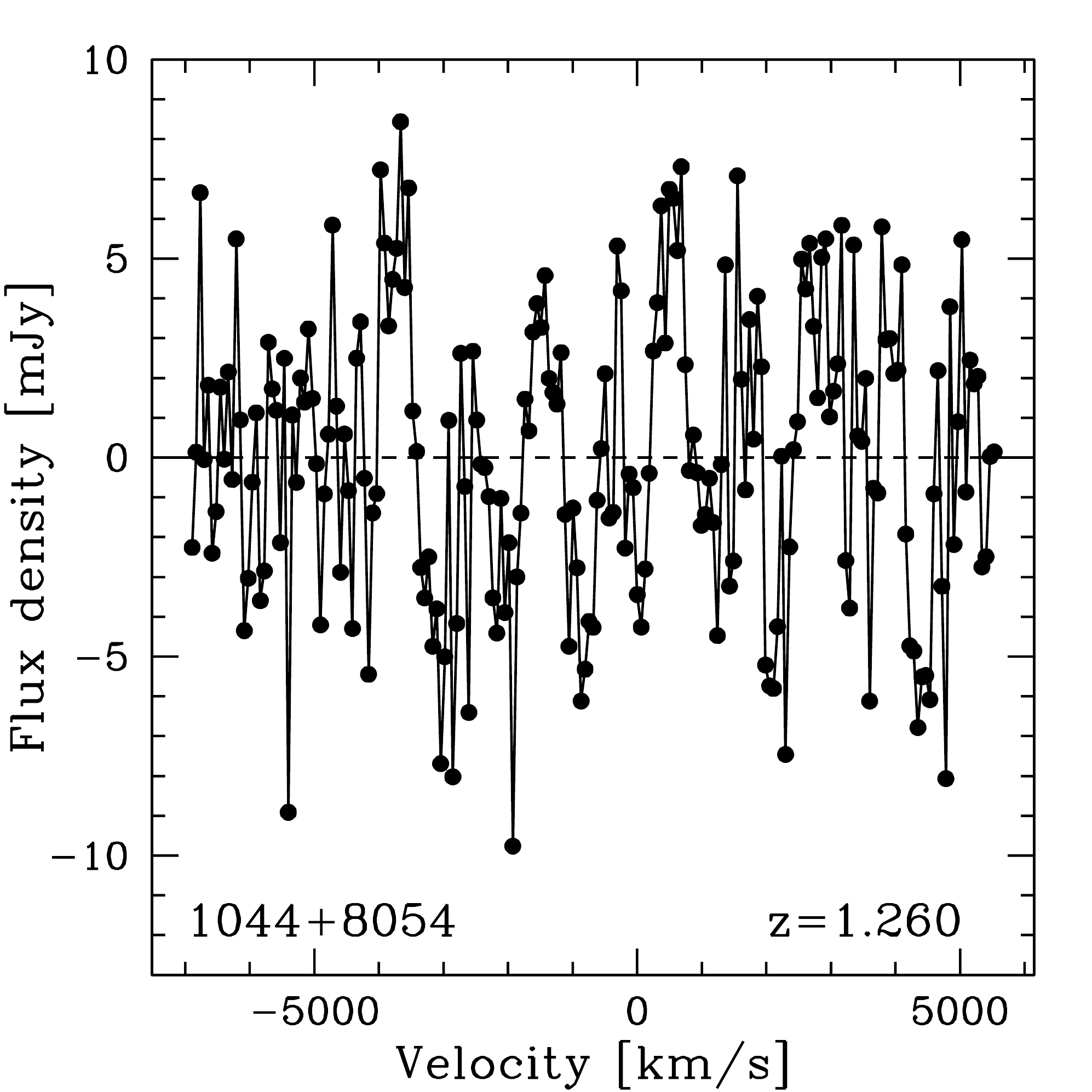}  &
\includegraphics[scale=0.28]{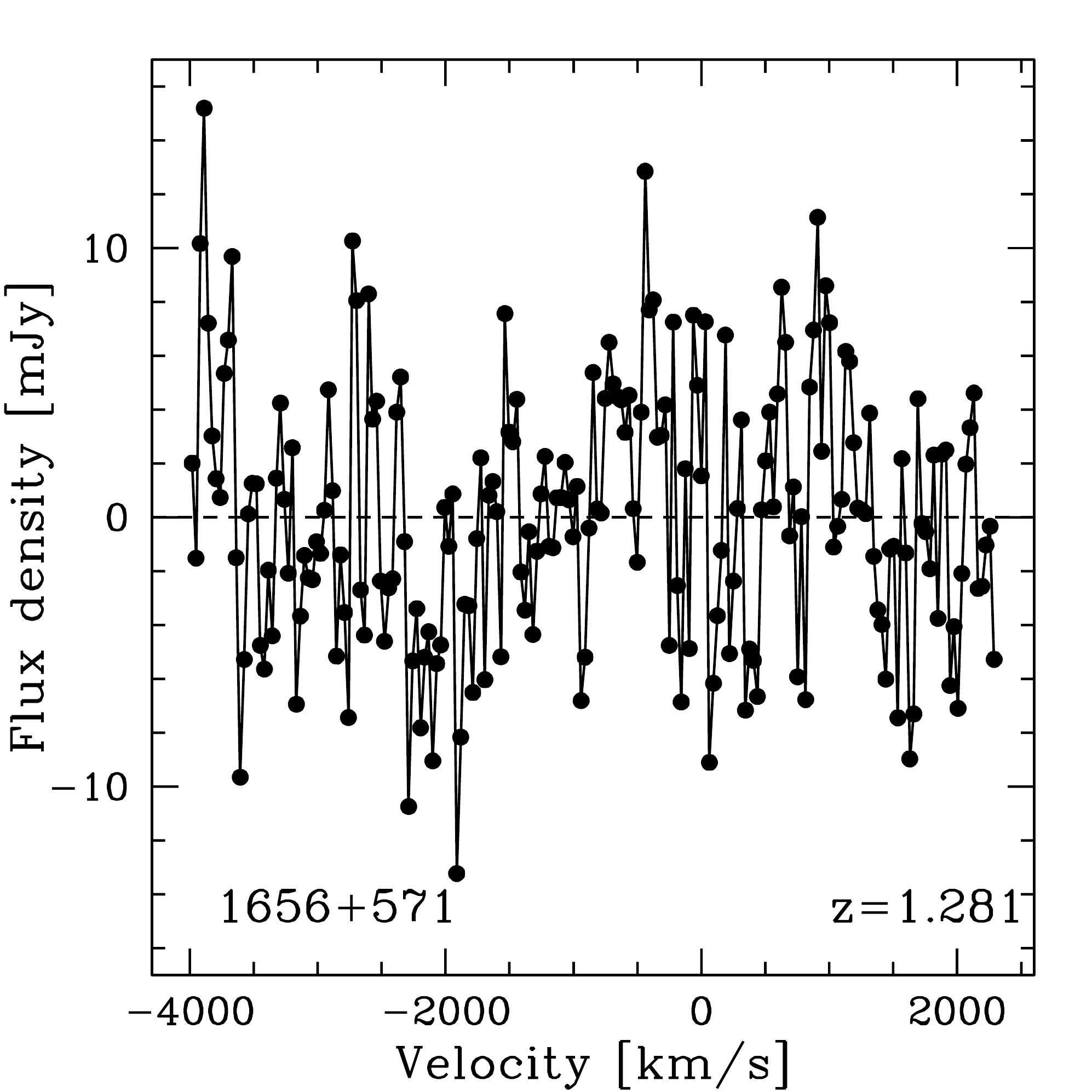}  \\

\end{tabular}
\end{figure*}

\begin{figure*}
\begin{tabular}{ccc}
\multicolumn{2}{ c }{{{\bf Figure A1.} (contd.) The GMRT \hii\ spectra for the 37 non-detections.}} \\

\includegraphics[scale=0.28]{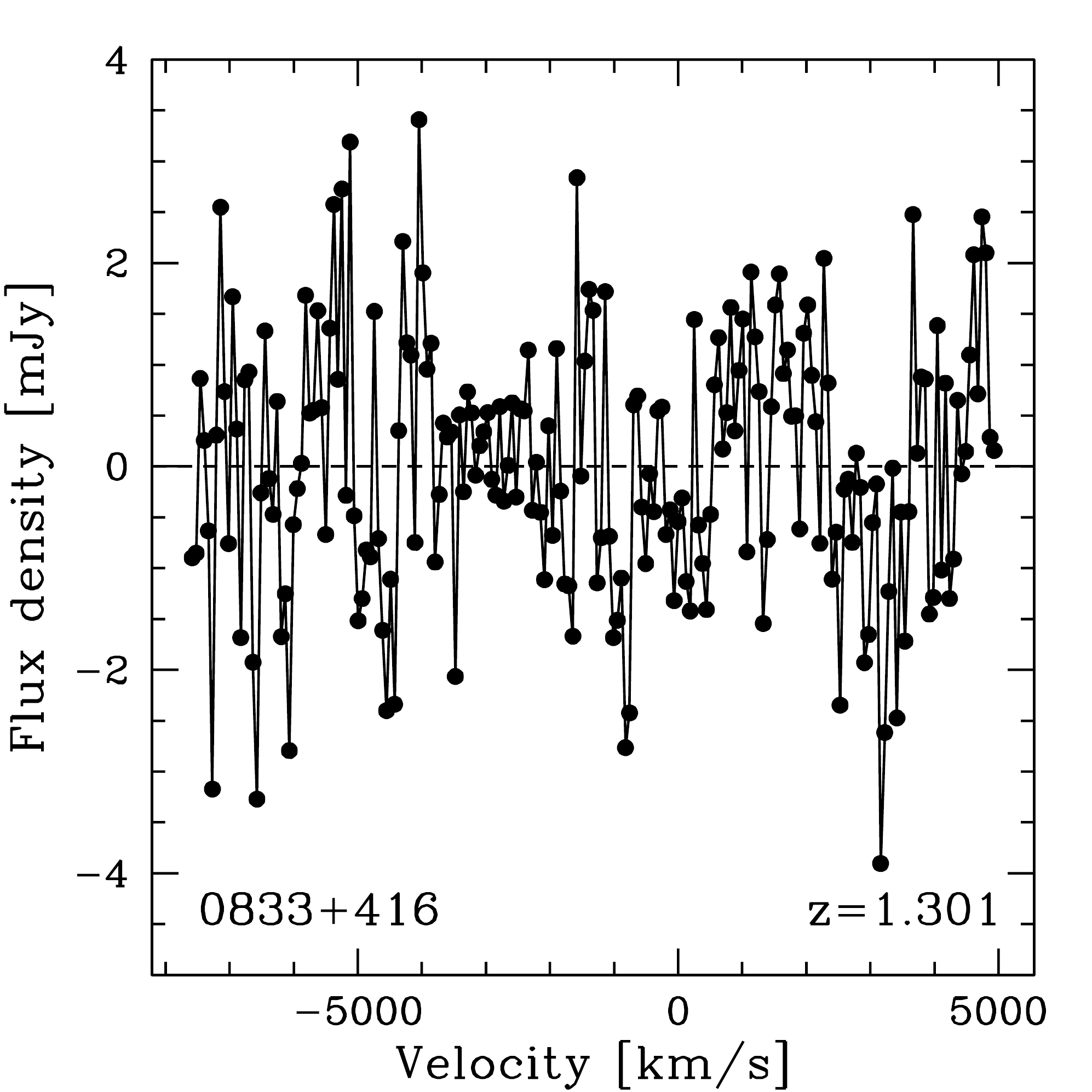} &
\includegraphics[scale=0.28]{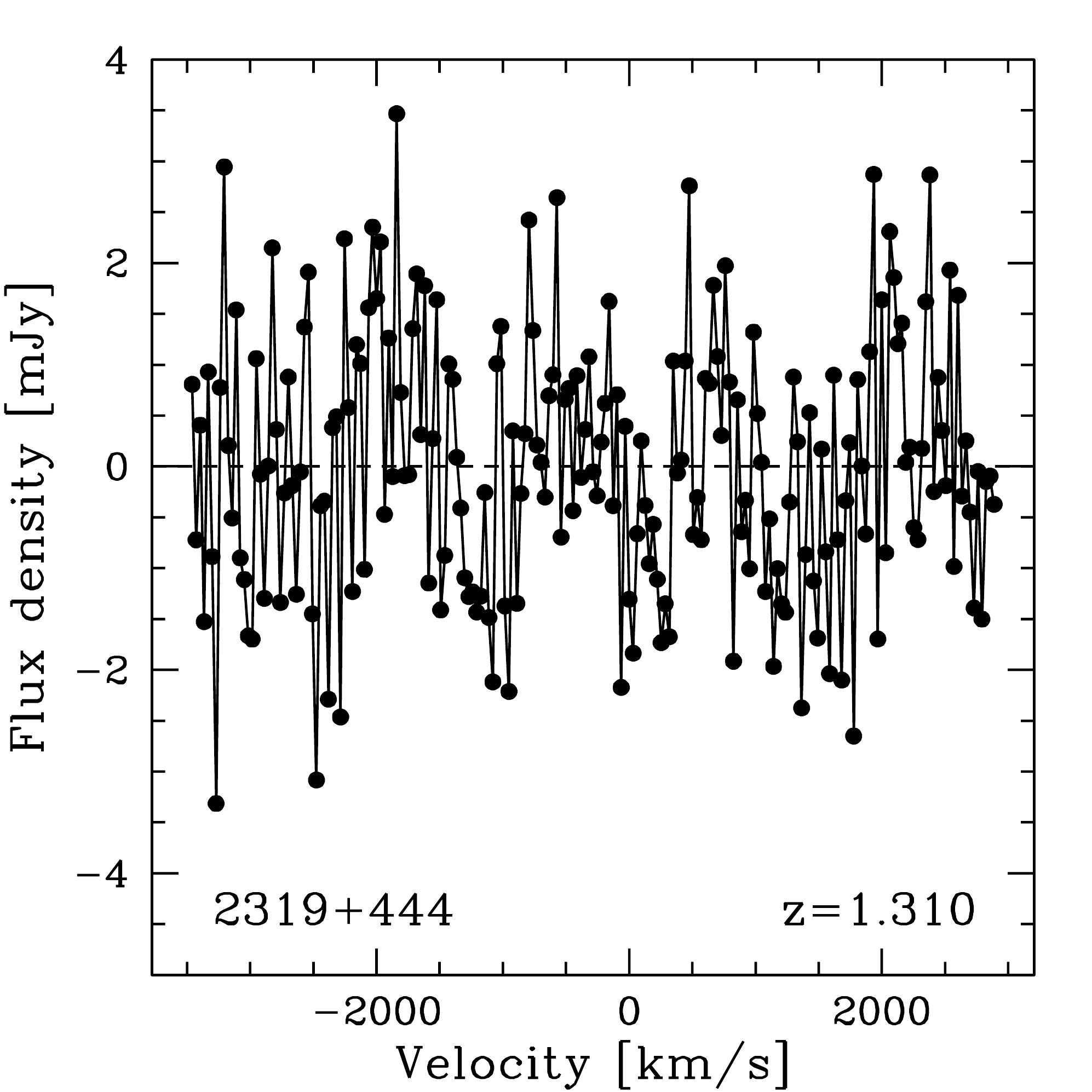} &
\includegraphics[scale=0.28]{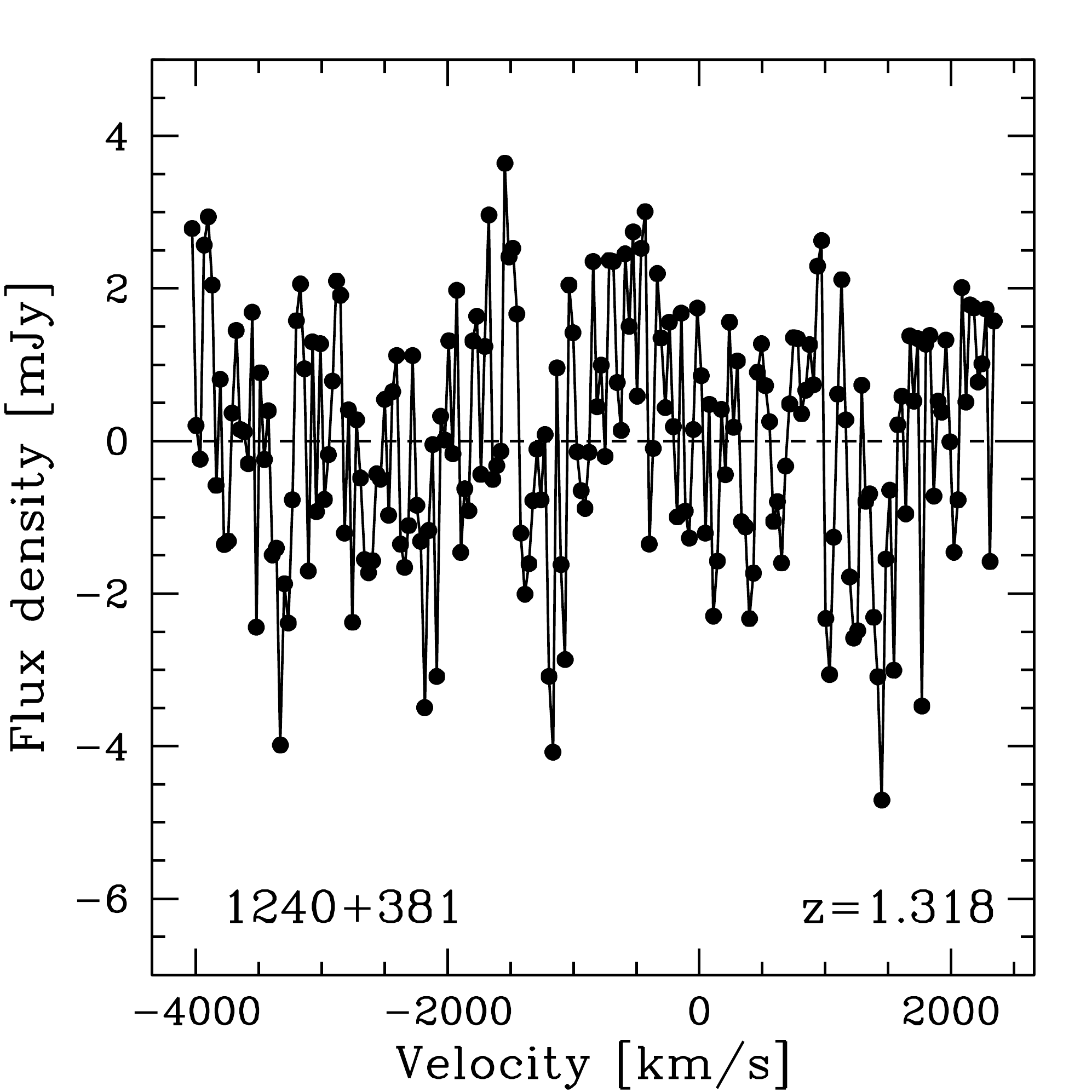} \\

\includegraphics[scale=0.28]{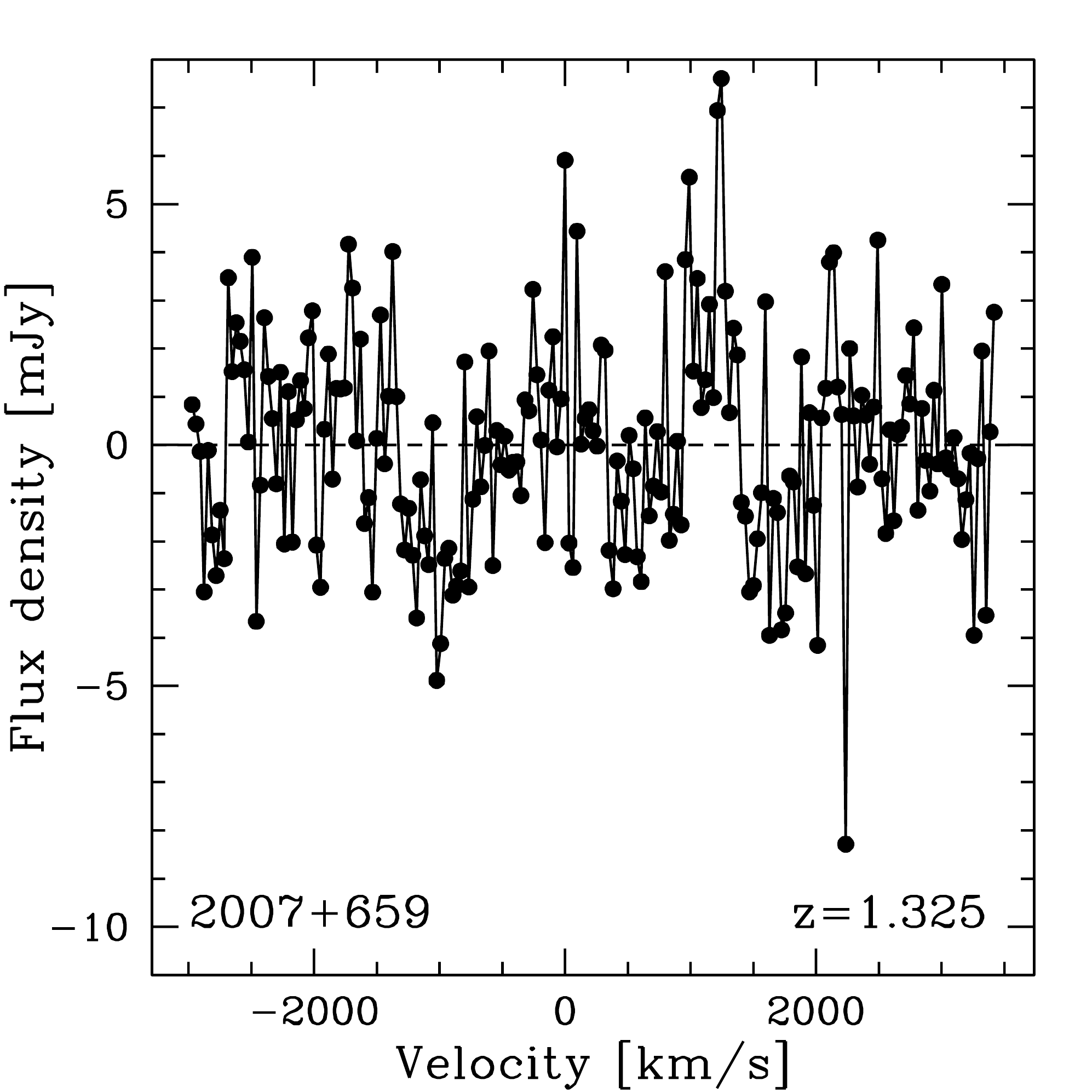} &
\includegraphics[scale=0.28]{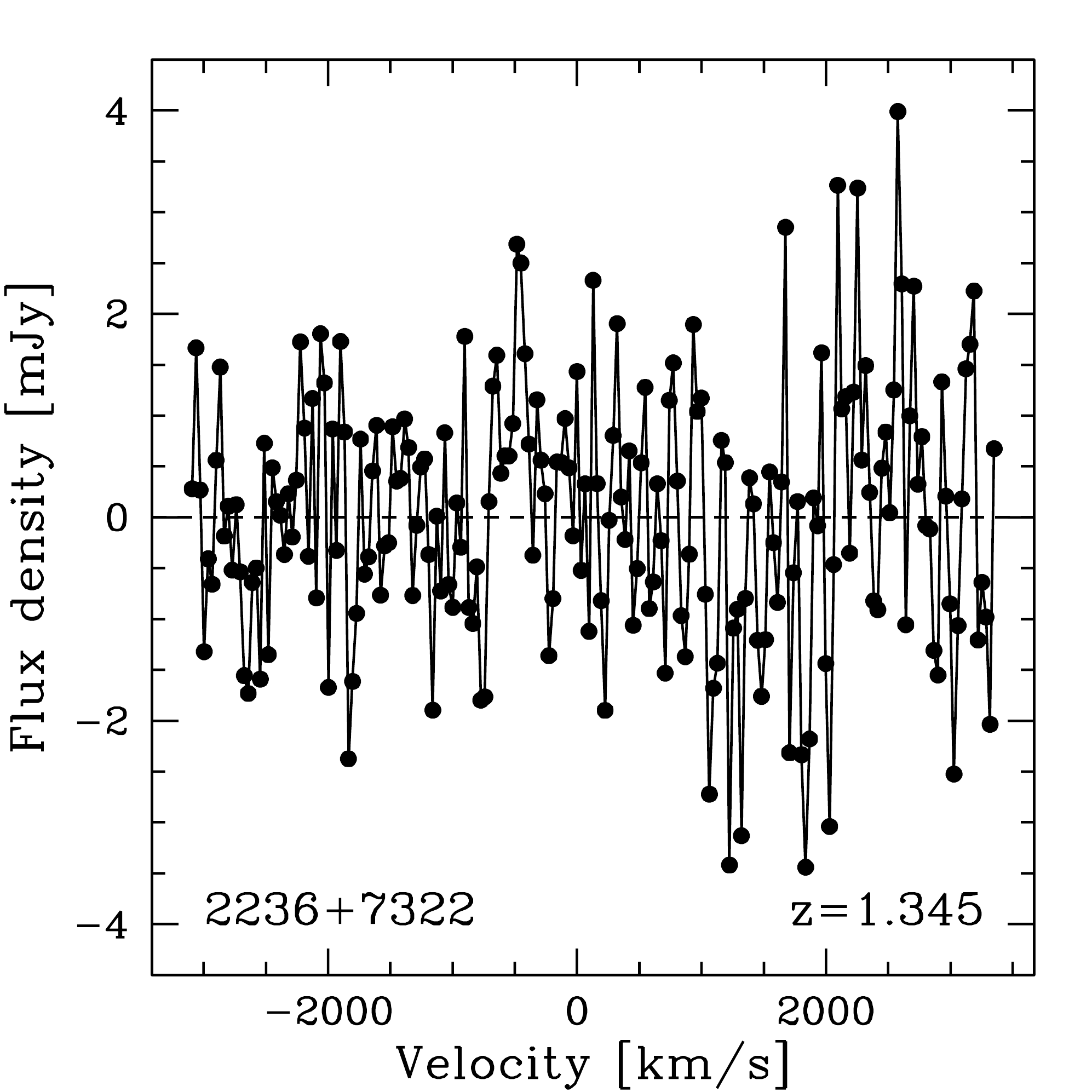}  &
\includegraphics[scale=0.28]{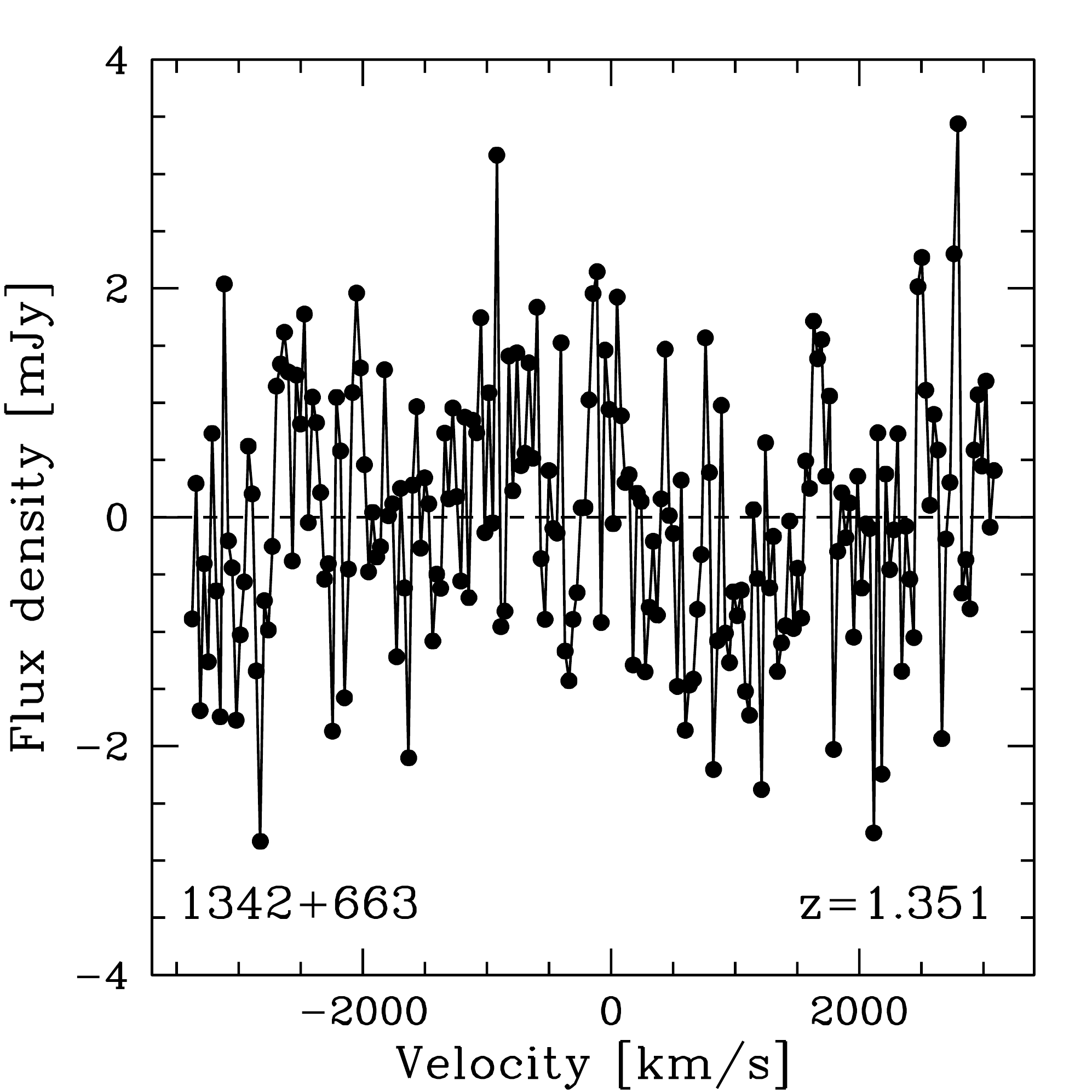} \\

\includegraphics[scale=0.28]{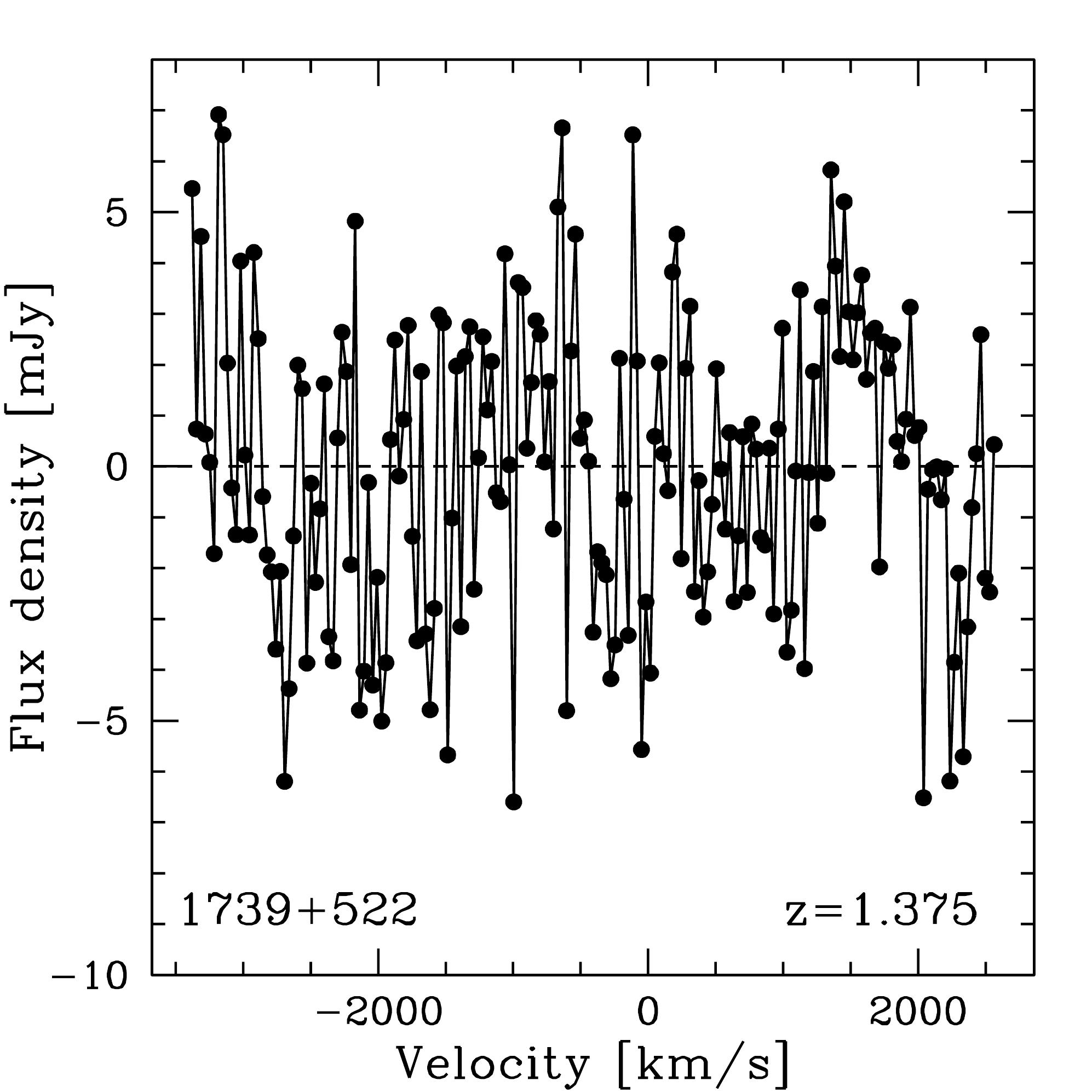} &
\includegraphics[scale=0.28]{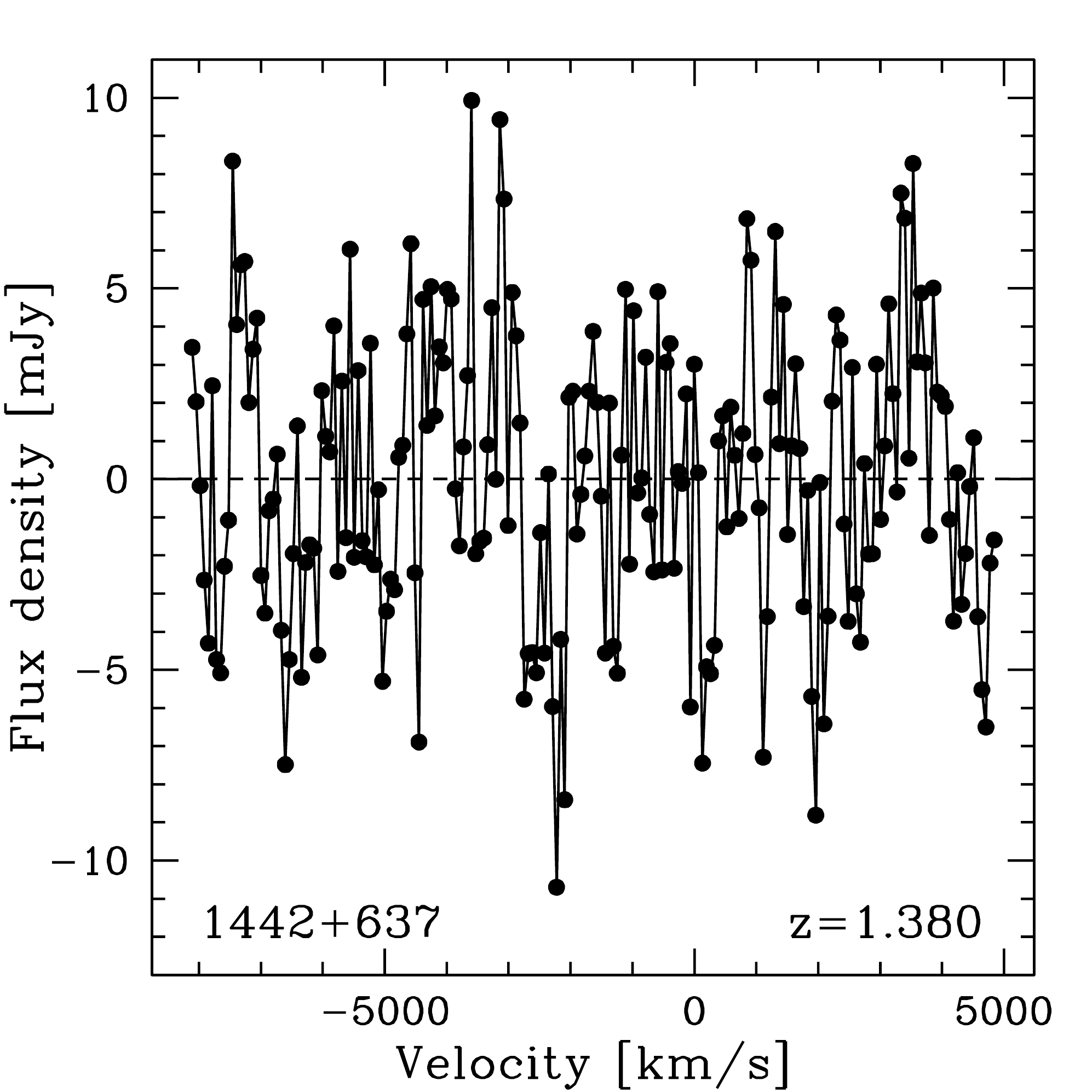} &
\includegraphics[scale=0.28]{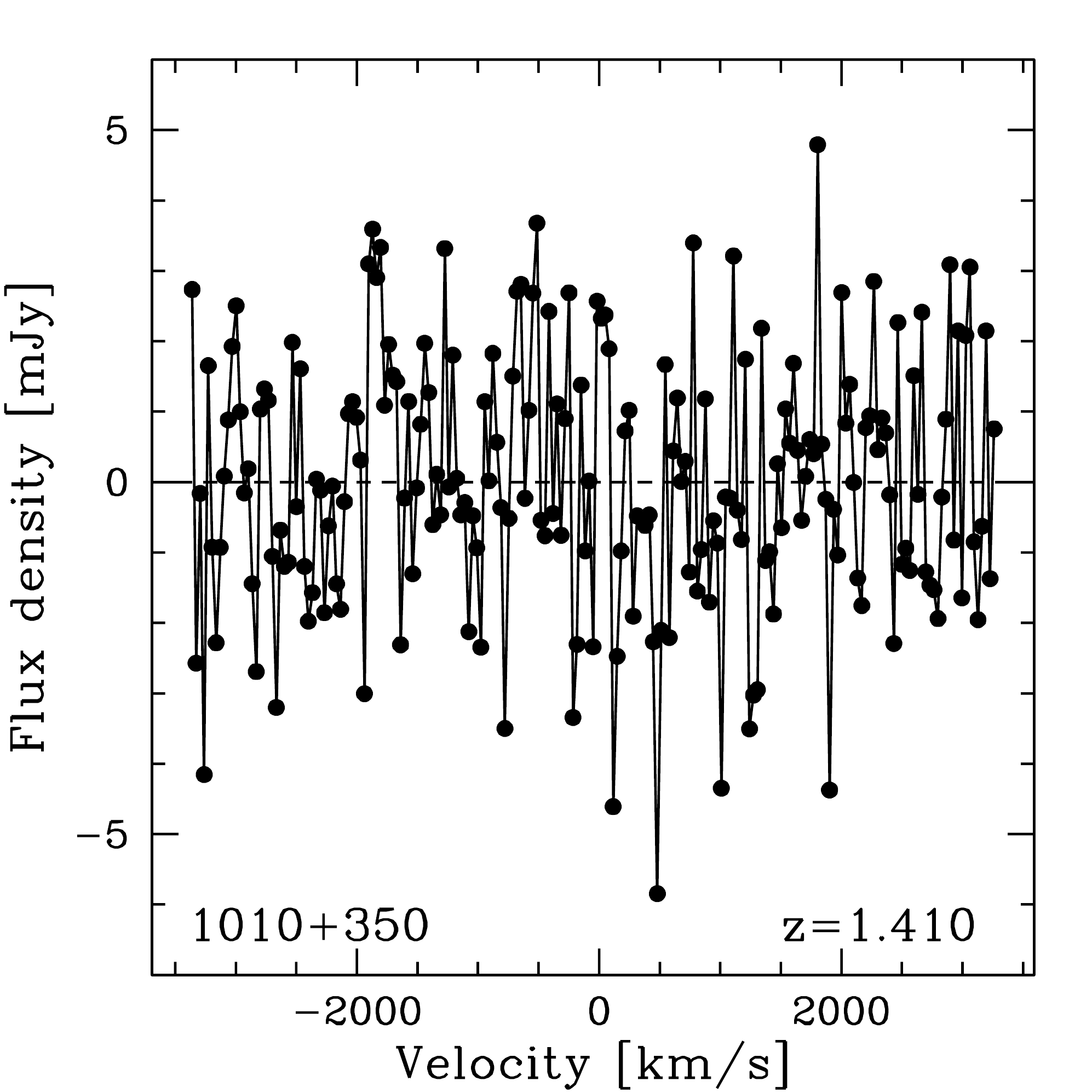}\\

\includegraphics[scale=0.28]{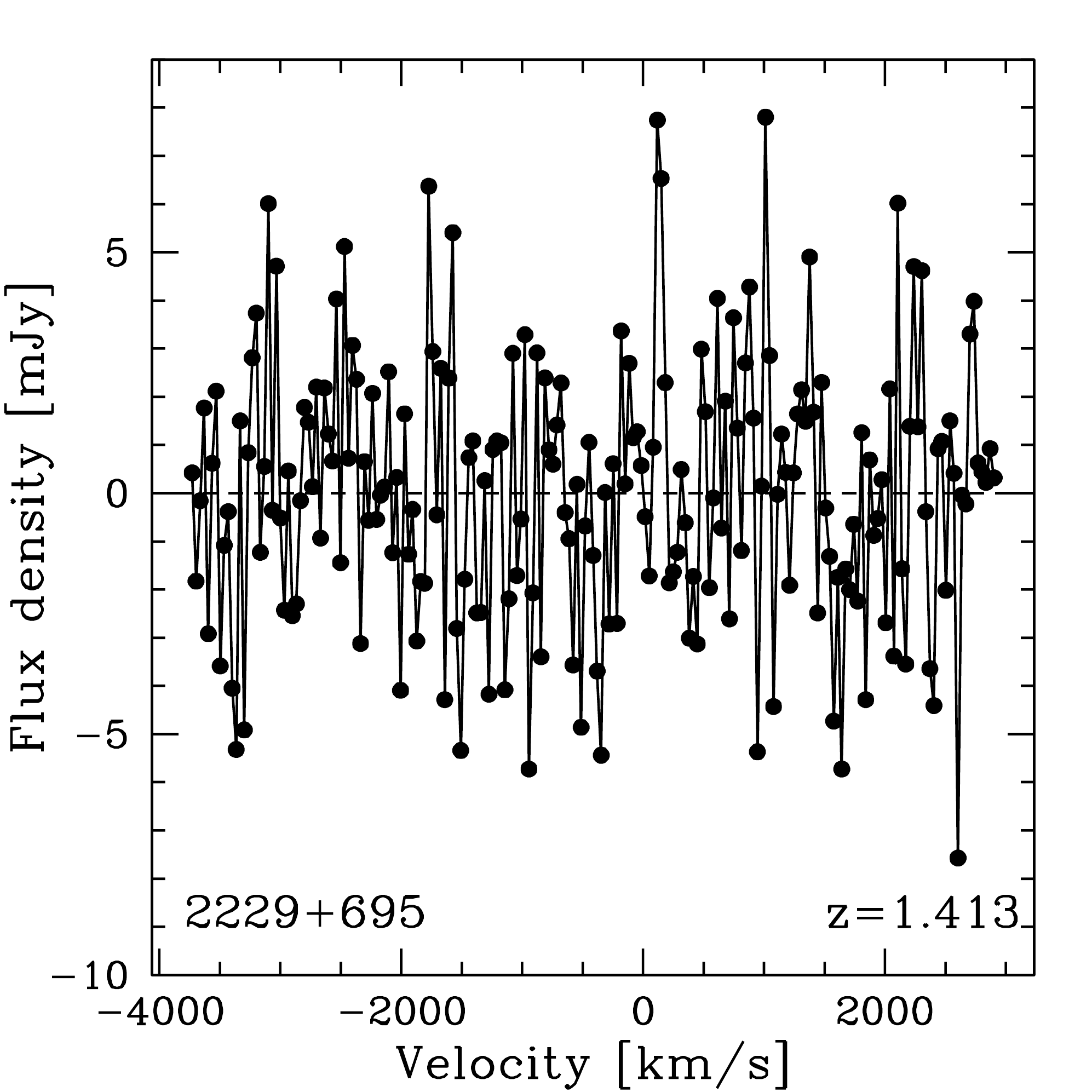} &
\includegraphics[scale=0.28]{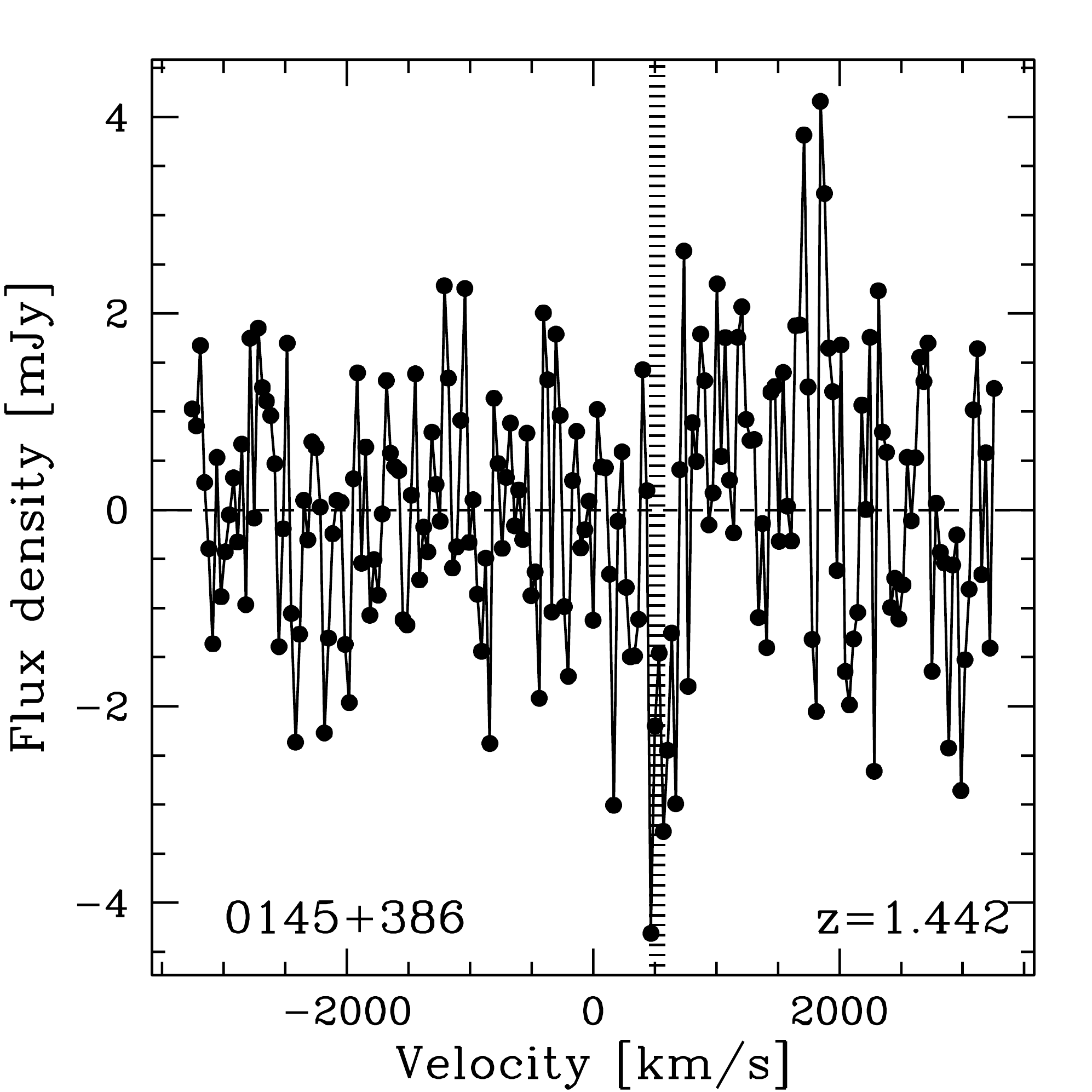} &
\includegraphics[scale=0.28]{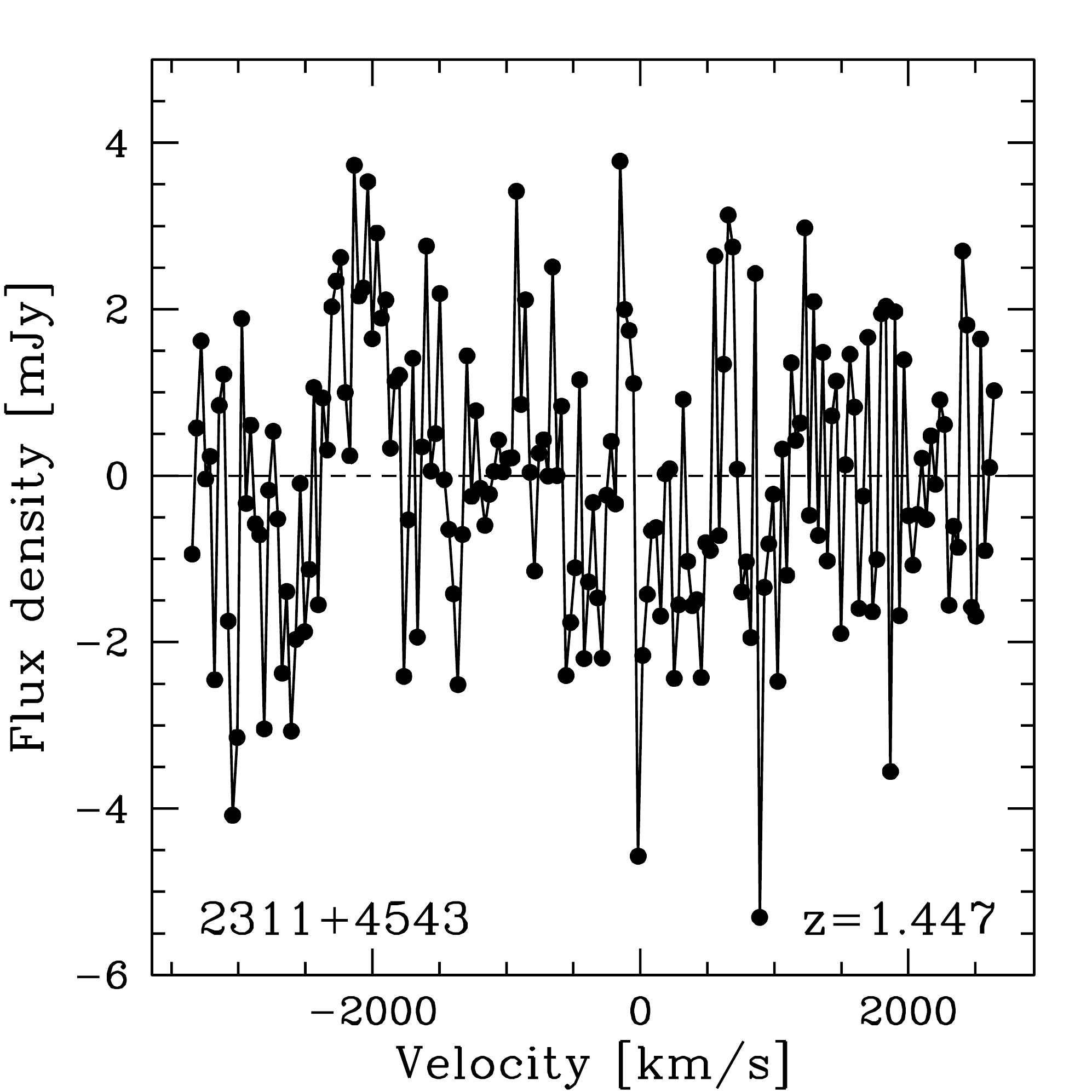}  \\

\end{tabular}
\end{figure*}

\begin{figure*}
\begin{tabular}{ccc}
\multicolumn{2}{ c }{{{\bf Figure A1.} (contd.) The GMRT \hii\ spectra for the 37 non-detections.}} \\

\includegraphics[scale=0.28]{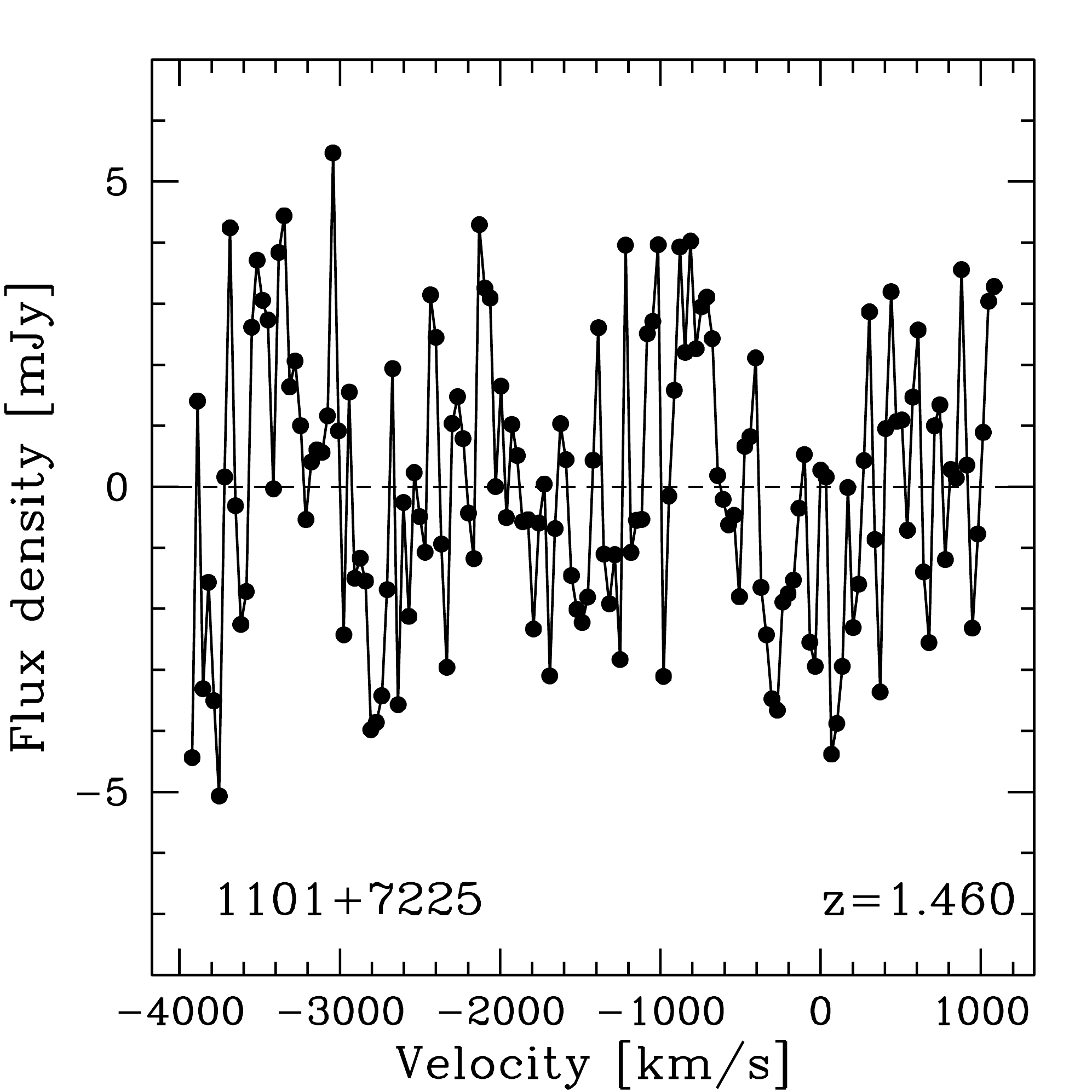} &
\includegraphics[scale=0.28]{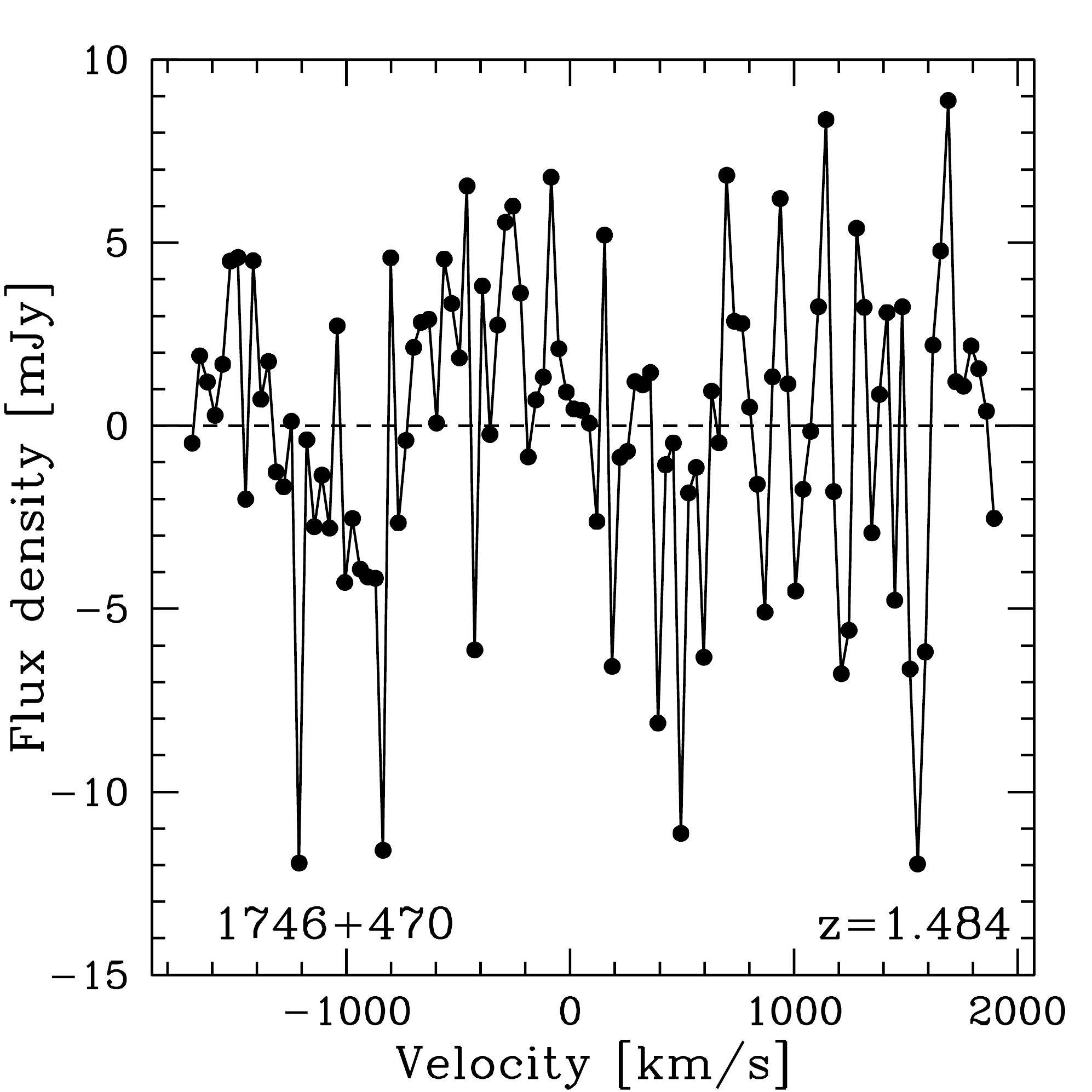} &
\includegraphics[scale=0.28]{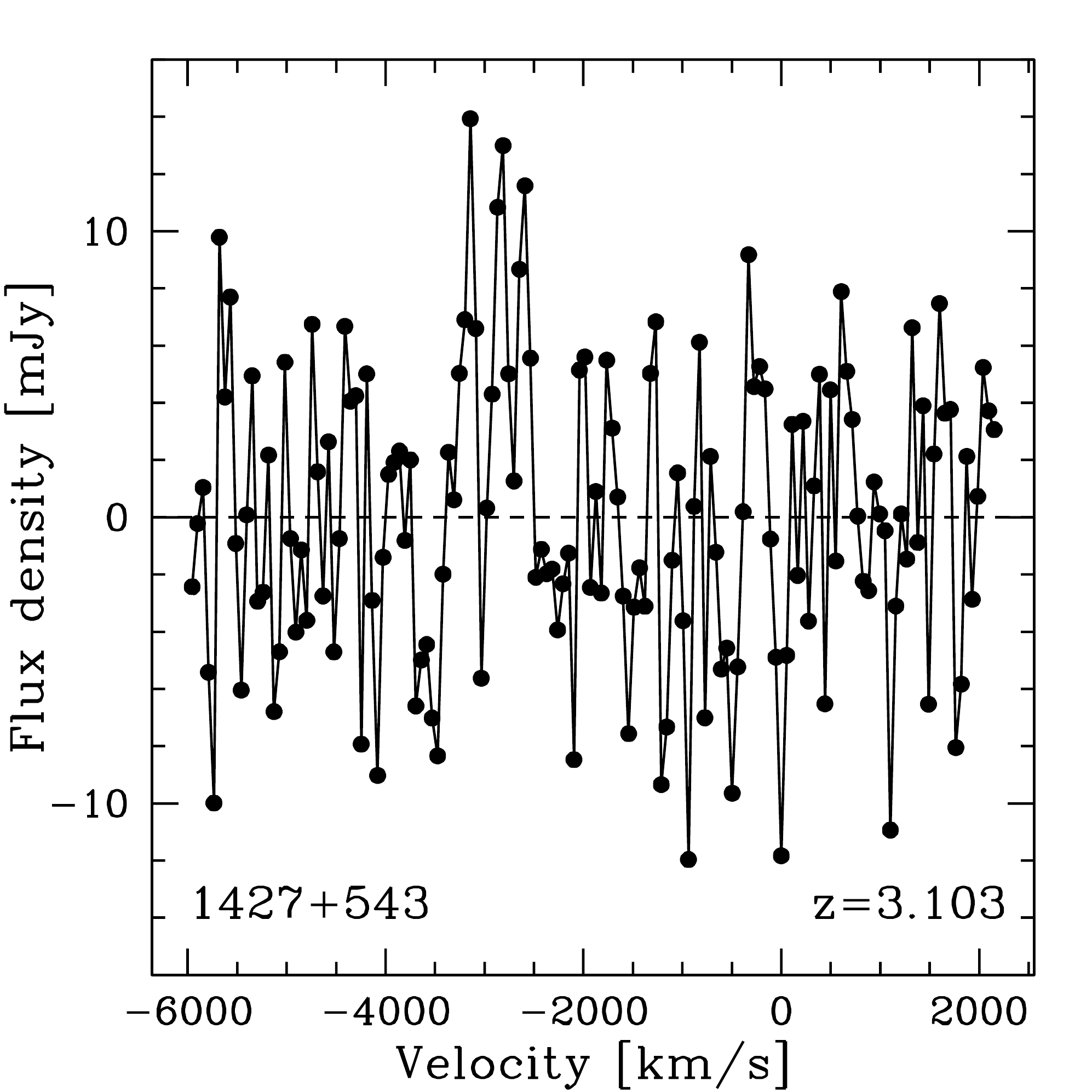} \\

\end{tabular}
\end{figure*}


\bsp	
\label{lastpage}
\end{document}